\documentclass[fleqn,useAMS,usenatbib]{mnras}

\usepackage{newtxtext,newtxmath}
\usepackage[T1]{fontenc}
\usepackage{graphicx}
\usepackage{caption}
\usepackage{tabularx}
\usepackage{pifont}
\newcommand{\MSUN}{\rm {M}_{\sun}}
\newcommand{\RVIR}{{R}_{\rm 200c}}

\newcommand{\MSTARS}{{M}_{*}}

\title[Soft X-ray line emission from the CGM]{X-ray metal line emission from the hot circumgalactic medium: probing the effects of supermassive black hole feedback}
\author[Truong et al.]{Nhut Truong$^{1,2,3}$\thanks{E-mail: nhut.truong@nasa.gov}, Annalisa Pillepich$^{3}$, Dylan Nelson$^{4}$, \'Akos Bogd\'an$^{5}$, Gerrit Schellenberger $^{5}$, \newauthor
Priyanka Chakraborty$^{5}$, William R. Forman$^{5}$, Ralph Kraft$^{5}$, Maxim Markevitch$^1$, Anna Ogorzalek$^{1,6}$, \newauthor Benjamin D. Oppenheimer$^7$, 
Arnab Sarkar$^{8}$, Sylvain Veilleux$^{6,9}$, Mark Vogelsberger$^8$, Q. Daniel Wang$^{10}$,\newauthor Norbert Werner$^{11}$, 
Irina Zhuravleva$^{12}$, and John Zuhone$^{5}$ 
\\
\\
$^{1}$\ NASA Goddard Space Flight Center, Greenbelt, MD 20771, USA\\
$^2$\ Center for Space Sciences and Technology, University of Maryland, Baltimore County, 1000 Hilltop Circle, Baltimore, MD 21250, USA \\
$^{3}$\ Max-Planck-Institut f{\"u}r Astronomie, K{\"o}nigstuhl 17, 69117 Heidelberg, Germany\\
$^{4}$\ Universit\"{a}t Heidelberg, Zentrum f\"{u}r Astronomie, Institut f\"{u}r theoretische Astrophysik, Albert-Ueberle-Str. 2, 69120 Heidelberg, Germany\\
$^5$\ Center for Astrophysics \ding{120} Harvard \& Smithsonian, 60 Garden Street, Cambridge, MA 02138, USA \\
$^6$\ Deparment of Astronomy, University of Maryland, College Park MD 20742-2421, USA \\
$^7$\ CASA, Department of Astrophysical and Planetary Sciences, University of Colorado, 389 UCB, Boulder, CO 80309, USA \\
$^8$\ Department of Physics, Kavli Institute for Astrophysics and Space Research, Massachusetts Institute of Technology, Cambridge, MA 02139, USA \\
$^9$\ Joint Space-Science Institute, University of Maryland, College Park, MD 20742, USA \\
$^{10}$\ Astronomy Department, University of Massachusetts, Amherst, MA 01003, USA \\
$^{11}$\ Department of Theoretical Physics and Astrophysics, Faculty of Science, Masaryk University, Kotl\'a\v{r}sk\'a 2, Brno, 611 37, Czech Republic \\ 
$^{12}$\ Department of Astronomy and Astrophysics, The University of Chicago, Chicago, IL 60637, USA \\
}

\volume{{\rm submitted}}
\date{}
\pubyear{2023}

\begin{document}
\label{firstpage}
\pagerange{\pageref{firstpage}--\pageref{lastpage}}
\maketitle

\begin{abstract}
We derive predictions from state-of-the-art cosmological galaxy simulations for the spatial distribution of the hot circumgalactic medium (CGM, ${\rm [0.1-1]R_{200c}}$) through its emission lines in the X-ray soft band ($[0.3-1.3]$ keV). In particular, we compare IllustrisTNG, EAGLE, and SIMBA and focus on galaxies with stellar mass $10^{10-11.6}\, \MSUN$ at $z=0$. The three simulation models return significantly different surface brightness radial profiles of prominent emission lines from ionized metals such as OVII(f), OVIII, and FeXVII as a function of galaxy mass. Likewise, the three simulations predict varying azimuthal distributions of line emission with respect to the galactic stellar planes, with IllustrisTNG predicting the strongest angular modulation of CGM physical properties at radial range ${\gtrsim0.3-0.5\,R_{200c}}$. This anisotropic signal is more prominent for higher-energy lines, where it can manifest as X-ray eROSITA-like bubbles. Despite different models of stellar and supermassive black hole (SMBH) feedback, the three simulations consistently predict a dichotomy between star-forming and quiescent galaxies at the Milky-Way and Andromeda mass range, where the former are X-ray brighter than the latter. This is a signature of SMBH-driven outflows, which are responsible for quenching star formation. Finally, we explore the prospect of testing these predictions with a microcalorimeter-based X-ray mission concept with a large field-of-view. Such a mission would probe the extended hot CGM via soft X-ray line emission, determine the physical properties of the CGM, including temperature, from the measurement of line ratios, and provide critical constraints on the efficiency and impact of SMBH feedback on the CGM.
\end{abstract}

\begin{keywords}
galaxies: evolution -- galaxies: formation -- galaxies:haloes -- X-ray: galaxies -- (galaxies) quasars: supermassive black holes
\end{keywords}

%%%%%%%%%%%%%%%%%%% INTRODUCTION %%%%%%%%%%%%%%%%%%%

\section{Introduction}
\label{sec:intro}

In the standard $\Lambda$CDM model of structure formation, dark matter (DM) haloes form first and the baryonic component follows into these already established gravitational potential wells. Unlike DM, gas component is collisional, and cosmic gas infalling into high-mass halos is shock-heated to approximately the virial temperature, leading to the formation of hot gaseous atmospheres (\citealt{white.rees.1979, white.frenk.1991, kauffmann.etal.1993}, see \citealt{werner.etal.2019} for a recent review). These hot atmospheres -- the circumgalactic medium (CGM) -- play an essential role in regulating the growth and evolution of galaxies. However, despite being predicted long ago, observational detection of the extended hot CGM at galactic mass scales (${\rm M_{200c}\gtrsim10^{12}\, \MSUN}$)\footnote{${\rm M_{200c}}$ is defined as the total mass within the radius of ${\rm R_{200c}}$, within which the average matter density is 200 times the cosmological critical density at a given redshift, namely it is defined by the relation: ${\rm M_{200c}=\frac{4}{3}\pi\times200\rho_c(z)R_{200c}^3}$.} remains difficult, if not elusive. 
  
Observationally, a broad, multi-wavelength view of the multi-phase CGM has been established (see e.g. \citealt{tumlinson.etal.2017} for a general review). For individual objects, several works have reported the detection of diffuse, emitting gas in the X-ray, observed by Chandra and XMM-Newton. These observations have targeted both early-type galaxies (e.g. \citealt{goulding.etal.2016,babyk.etal.2018, bregman.etal.2018, lakhchaura.etal.2018}) and late-type galaxies, at the Milky-Way (MW) mass range or below (e.g. \citealt{strickland.etal.2004, tullmann.etal.2006,yamasaki.etal.2009, mineo.etal.2012, bogdan.etal.2013, bogdan.etal.2015, bogdan.etal.2017, li.wang.2013, li.etal.2017}), and a few more massive merger galaxies (e.g. \citealt{nardini.etal.2013,veilleux.etal.2014, liu.etal.2019}). Nonetheless, due to the rapid drop of density and thus emissivity with distance, existing observations with CCD-based instruments such as Chandra/XMM-Newton primarily probe the X-ray emission within the central regions of halos (up to 60-70 kpc). Moreover, these CCD-based observations unable to yield substantial constraints on the thermodynamcis and metal content of the CGM.
 
Emission from hot gas at larger radii is measurable by stacking X-ray photons at the positions of galaxies from large-scale surveys. \cite{anderson.etal.2013,anderson.etal.2015, Bogdan.Goulding.2015} demonstrate the potential of this technique by combining hundreds of thousands of galaxies from local ($z < 0.5$) galaxy surveys, including the Sloan Digital Sky Survey (SDSS), and X-ray data from the ROSAT All-Sky Survey, finding extended X-ray emission down to galaxy stellar masses as low as $\MSTARS\sim10^{10.8}\, \MSUN$.

More recently, \cite{comparat.etal.2022} and \cite{chadayammuri.etal.2022} use X-ray data from the eROSITA Final Equatorial Depth Survey (eFEDS). By stacking X-ray events around thousand of galaxies identified in the Galaxy And Mass Assembly (GAMA) and SDSS surveys, respectively, they characterize the radial profile of extended X-ray emission from the CGM. They both detect X-ray emission in the CGM out to galactocentric distances of $\sim100$ kpc for galaxies with $\MSTARS\lesssim10^{11}\, \MSUN$. However, the quantitative result depends on details of the stacking methodology, including the selected galaxy sample, removal of point sources including background AGN, contaminating X-ray photons from PSF effects, and so on. Regardless, stacking can only measure the mean (averaged) profile of a galaxy population, which can be biased high by a few brightest galaxies. Additionally, these stacking experiments cannot directly constrain the physical properties of the CGM such as temperature or metal abundance. These limitations in current observations with X-ray CCD detectors of the hot CGM preclude a comprehensive understanding of its crucial role in galaxy formation and evolution. 

Theoretically, recent cosmological hydrodynamical simulations of galaxies predict that the CGM of massive galaxies ($\MSTARS\sim 10^{10-11.5}\,\MSUN$) is strongly affected by feedback activity, especially due to energy released from supermassive black hole (SMBH) accretion. Three key phenomenological CGM features have been recently predicted and further quantified with modern galaxy simulations:

\begin{enumerate}
    \item {\it Diversity between star-forming versus quiescent galaxies in hot gas content.} Recent studies based on the IllustrisTNG and EAGLE simulations suggest that central massive galaxies are quenched, at least in part, by ejective outflows produced by SMBH feedback (e.g. \citealt{terrazas.etal.2020, zinger.etal.2020,davies.etal.2020}). As a consequence, these simulations predict that star-forming galaxies are significantly more X-ray luminous compared to their quiescent counterparts at the transitional mass scale of $\sim 10^{10.5-11.2}\,\MSUN$ in stars (\citealt{truong.etal.2020, oppenheimer.etal.2020}), and overall have different CGM properties \citep{nelson.etal.2019a, ramesh.etal.2023a}. Chandra observations of nearby galaxies tentatively support this finding (although see \citealt{bogdan.etal.2011} for a different result). However, \cite{truong.etal.2020} show that, as current X-ray observations are limited to the central region of galaxies (within at most a few times ${\rm R_e}$), it is challenging to distinguish the X-ray emission from diffuse halo gas versus contamination from the hot interstellar medium and point-like sources such as X-ray binary stars, especially in star-forming galaxies. Simultaneously, recent stacking studies using eFEDS \citep{comparat.etal.2022,chadayammuri.etal.2022} find discrepant results for the X-ray surface brightness profiles of star-forming versus quiescent galaxies, highlighting the difficulty of this measurement.  \cite{comparat.etal.2022} reports robust detection of the stacked signal only in quiescent galaxies at the MW-mass range, whereas \cite{chadayammuri.etal.2022} reports the detection in both star-forming and quiescent populations, with the former being mildly more X-ray luminous than the latter. The discrepancy could be partly explained by different stacking methods adopted in the two studies, including factors such as the selection of stacking samples and the removal of X-ray contaminants, e.g. emission from AGN and galaxy groups/clusters, from the CGM emission.    
    \item {\it Angular/azimuthal distribution of the CGM.} SMBH feedback can cause an angular dependence in properties of the CGM around massive galaxies, such as temperature, density, and metallicity, with respect to the orientation of the central galaxy (\citealt{truong.etal.2021b,nica.etal.2022}). According to the IllustrisTNG simulations, this produces galaxy population-wide observable signatures such as broad-band X-ray luminosity or X-ray hardness, which could be detectable via stacking of eROSITA all-sky data. At the individual galaxy level, this angular dependence of the CGM can manifest itself even in the form of X-ray emitting bubbles, shells, and cavities in the CGM above and below the stellar disks of MW and Andromeda-like simulated galaxies, similar in morphology to the eROSITA/Fermi bubbles in our own Galaxy \citep{pillepich.etal.2021}.
    \item {\it Thermal content of the CGM.} Observations (e.g. \citealt{kim.fabbiano.2015,li.etal.2017,goulding.etal.2016,babyk.etal.2018}) and simulations (e.g. \citealt{truong.etal.2021}) both suggest that the CGM temperature in ${\rm M_{200c}\sim10^{12}\, \MSUN}$ halos is higher than the self-similar prediction. This indicates that in addition to gravitational heating there is a contribution from other energy sources, i.e. SMBH feedback. However, current measurements of CGM temperature are mainly available for massive early-type galaxies, and limited to the hot ISM within their central regions (e.g. \citealt{kim.fabbiano.2015}). 
\end{enumerate}

Extended emission from the hot CGM can also be observed in narrow band X-ray emission. The warm-hot phase of the CGM has a temperature ${T \sim 10^{5-7}\, \rm{K}}$ such that it emits abundantly in the soft X-ray band ([0.3-2.0] keV), where the emission is expected to be dominated by individual metal lines (\citealt{bertone.etal.2010, vandeVoort.schaye.2013,wijers.schaye.2022}). Metal line emission has the potential to probe the extended hot CGM without significant confusion due to foreground emission from the CGM of our Milky Way, unlike current low spectral resolution observations. With a high spectral resolution instrument ($\sim$ eV), line emission from a distant galaxy is observable as it redshifts out of the energy range occupied by the same lines in the MW foreground (see Section~\ref{sec:narrow_band_obs}). This technique will be possible with the next generation of X-ray instruments equipped with microcalorimeters, such as XRISM Resolve (\citealt{xrism.whitepaper.2020}) and the mission concepts ATHENA X-IFU (\citealt{Barret.etal.2016, barret.etal.2018}), HUBS (\citealt{cui.etal.2011}), and Line Emission Mapper (LEM, \citealt{kraft.etal.2022}).  

In order to effectively study the CGM in nearby galaxies through line emission, we need both high ($\sim$\,eV) spectral resolution and a large ($\sim$\,10s of arcmin) field of view. High spectral resolution is necessary to resolve individual lines, while a large field of view is required to map the CGM out to the virial radius with a single pointing. In this regards, ATHENA X-IFU and XRISM Resolve are limited by their small field of view, which are of 5 and 3 arcminutes, respectively (see Table 1 in \citealt{kraft.etal.2022} for a detailed comparison between future X-ray spectrometers). Although this limitation can be overcome by observing galaxies with multiple pointings or at higher redshifts, it will be challenging to observe such galaxies given the needed observing time or the large cosmological dimming factor. 

This paper explores observational opportunities to uncover the connection between galactic feedback and CGM properties with the aforementioned Line Emission Mapper (LEM), an X-ray mission concept that has been recently developed with the main scientific goal to study the formation of cosmic structures (\citealt{kraft.etal.2022}). LEM has 1-2 eV spectral resolution and a large field of view (30'x30'), making it an ideal instrument to observe the CGM in emission (\citealt{schellenberger.etal.2023, zuhone.etal.2023}) and in absorption (\citealt{bogdan.etal.2023}). We especially provide quantitative theoretical predictions to probe the effects of SMBH-driven feedback on the spatial distribution and azimuthal anisotropy of the CGM, via narrow-band X-ray imaging and high-resolution X-ray spectroscopy. This work is based on the outcome of current state-of-art cosmological simulations including IllustrisTNG (\citealt{nelson.etal.2018,springel.etal.2018, marinacci.etal.2018,naiman.etal.2018,pillepich.etal.2018b}), EAGLE (\citealt{schaye.etal.2015,crain.etal.2015, Mcalpine.etal.2016}), and SIMBA (\citealt{dave.etal.2019}), which implement different models for SMBH feedback (see Section \ref{sec:sims}). Our main objective is therefore to examine the impact of SMBH feedback on the spatial distribution of line emission in the CGM, as predicted by these three simulations. In doing so we explore the potential that future microcalorimeter-based missions such as LEM have in constraining the physics of galactic formation, feedback, and quenching.

The paper is arranged as follows. In Section~\ref{sec:method} we introduce the three simulations as well as our methodology for computing CGM observables. Section~\ref{section3} presents our main results and the comparison across the three models. In Section~\ref{sec:obs} we present the key observational considerations for the CGM science case of LEM. Finally, we summarize and conclude in Section~\ref{sec:conclusion}. 

%%%%%%%%%%%%%%%%%%%%%%%%%%%%%%%%% METHODS %%%%%%%%%%%%%%%%%%%%%%%%%%%%%%%%%%%%%

\section{Methodology}
\label{sec:method}

\subsection{Cosmological galaxy formation simulations}
\label{sec:sims}

In this paper, we extract quantitative predictions from three current state-of-the-art cosmological simulation models for the formation and evolution of galaxies: TNG100 of the IllustrisTNG suite (TNG hereafter), EAGLE, and SIMBA.\footnote{The data of these simulation are all publicly available. Herein we use versions of EAGLE and SIMBA that have been re-processed to enable an apples-to-apples comparison with TNG. For all three simulations we therefore identify structures (i.e. haloes, subhaloes, and hence galaxies) and their properties in the same manner, with the data of EAGLE and SIMBA having been rewritten exactly in the TNG format \citep{nelson.etal.2019a}, enabling identical analysis routines.} These three simulations assume the $\Lambda$CDM cosmological model with similar parameters, and have similar simulated volumes.

Below are brief descriptions of these simulation projects, which have been presented, analyzed and discussed in a large number of previous scientific works. Of relevance for this paper, all the considered simulation models evolve cosmic gas, cold dark matter, stellar populations and SMBHs from high redshift ($z > 100$) to the current epoch. TNG follows also the dynamics and amplification of magnetic fields. They all account for the heating and cooling of gas, assume a uniform UV/X-ray background arising since the epoch of Reionization, include numerical subgrid implementations for star formation, stellar evolution and chemical enrichment through SNIa, SNII and AGB channels, stellar feedback and AGN feedback, and are fully cosmological. The result of solving the coupled equations of gravity, (magneto)hydrodynamics, and galaxy astrophysics in a large cosmological volume is thousands of galaxies. These span a wide range of halo and stellar masses, stellar morphologies and cosmic environments, including their surrounding dark matter and gaseous haloes. As a cautionary note, these models do not simulate gas colder than $10^4$ K (see e.g. \citealt{veilleux.etal.2020} for a review on this topic) and, among other processes, do not explicitly account for photon propagation and for the explicit radiation-gas interaction.  

\begin{itemize}
    \item \textbf{IllustrisTNG (hereafter TNG)}\footnote{\url{https://www.tng-project.org/}} is a project of cosmological magneto-hydrodynamical simulations of galaxies and their evolution over cosmic time (\citealt{nelson.etal.2018,nelson.etal.2019b, springel.etal.2018,marinacci.etal.2018,naiman.etal.2018,pillepich.etal.2018b, pillepich.etal.2019, nelson.etal.2019a}). The TNG simulations were performed with the moving-mesh \texttt{AREPO} code (\citealt{springel.2010}) and with a galaxy formation model that includes a variety of astrophysical processes: see \citealt{weinberger.etal.2017,pillepich.etal.2018} for a detailed description. TNG includes three main flagship simulations: TNG50, TNG100 and TNG300. Here we mostly use the TNG100 run (aka TNG100-1). TNG100 has a simulated volume of ${\rm \sim(110.7\ cMpc)^3}$ and baryon mass resolution ${m_{\rm baryon}=1.4\times10^6 \, \MSUN}$. In Section~\ref{sec:obs}, we also showcase galaxies simulated with TNG50, with 16 (2.5) times better mass (spatial) resolution than TNG100.
    
    \item \textbf{EAGLE}\footnote{\url{https://eagle.strw.leidenuniv.nl/}} is also a suite of cosmological hydrodynamical simulations of galaxy formation and evolution (\citealt{schaye.etal.2015,crain.etal.2015,schaller.etal.2015, Mcalpine.etal.2016}). The simulations are run with a modified version of the \texttt{GADGET-3} smoothed particle hydrodynamics (SPH) code (\citealt{springel.2005}). Here we employ the flagship run referred as 'Ref-L0100N1504', which encompasses a comoving volume of about 100 Mpc per side and has a baryonic mass resolution of ${m_{\rm baryon}=1.81\times10^{6}\, \MSUN}$. 
    
    \item \textbf{SIMBA}\footnote{\url{http://simba.roe.ac.uk/}} is the next generation of the MUFASA cosmological galaxy formation simulations (\citealt{dave.etal.2019}) and it is performed with GIZMO's meshless finite mass hydrodynamics (\citealt{hopkins.2015}). We employ the main run of the SIMBA simulations that has a simulated box of ${\rm \sim(147\ Mpc)^3}$ and baryon mass resolution ${m_{\rm baryon}=1.82\times10^{7}\, \MSUN}$. 
\end{itemize}

Although they are run with different numerical codes for the solution of gravity and hydrodynamics, the key aspect making these three simulation projects highly complementary is their vastly different underlying implementations of the galaxy astrophysics processes. In particular and of relevance here, TNG, EAGLE and SIMBA rely on different physical models and numerical implementations of SMBH feedback. For example, EAGLE employs a single thermal channel, whereas in TNG and SIMBA different energy injection modes occur depending on SMBH accretion rate, including in both cases a kinetic energy injection. In EAGLE and TNG such energy is always distributed isotropically at the injection scales ($\lesssim$ kpc) whereas SIMBA adopts bipolar jet-like kicks. Despite these differences, feedback from SMBHs is responsible for quenching star formation in massive galaxies in all the three simulation suites. It also drives gas out of the central regions of haloes \citep{davies.etal.2020, truong.etal.2021b}, and even far beyond haloes \citep{ayromlou.nelson.pillepich.2022}, while heating gaseous halos and offsetting the cooling times of these CGM reservoirs \citep{zinger.etal.2020}.

In fact, SMBH-related model choices in the underlying galaxy formation models have been calibrated, either quantitatively or approximately, to reproduce reasonably realistic samples of galaxies, particularly with respect to observations of their stellar component in the local Universe. The degree to which such a model tuning has been performed varies enormously across the three simulation projects, which in fact produce, in detail, different galaxy populations \citep[e.g. different galaxy stellar mass functions or quenched fractions,][]{donnari.etal.2021b, schaye.etal.2015, furlong.etal.2015,dave.etal.2019}. On the other hand, none of these models have been calibrated with respect to detailed properties of the CGM or IGM, making their outcomes in this regime highly predictive \citep[e.g.][]{byrohl.2022}.

\subsection{Galaxy selection and definition of CGM}
\label{sec:cgm}

\begin{table}
    \centering
    \renewcommand{\arraystretch}{1.1}
    \begin{tabular}{ccc}
       Lines  & Energy [eV] & Wavelength (${\rm A}$) \\
       \hline
        CV & 298.97 & 41.470  \\
        CVI & 367.47 & 33.740 \\
        NVIf & 419.86 & 29.530 \\
        NVII & 500.36 & 28.790 \\
        OVIIf & 560.98 & 22.101 \\
        OVIIr & 573.95 & 21.602 \\
        OVIII & 653.49 & 18.973 \\
        FeXVII & 725.05 & 17.100 \\
        NeX & 1021.5 & 12.137
    \end{tabular}
    \caption{The energy and wavelength of the X-ray emission lines studied in this paper, ordered by increasing energy (decreasing wavelength).}
    \label{tab:my_lines}
\end{table}

\begin{figure}
  \centering
  \includegraphics[width=0.48\textwidth]{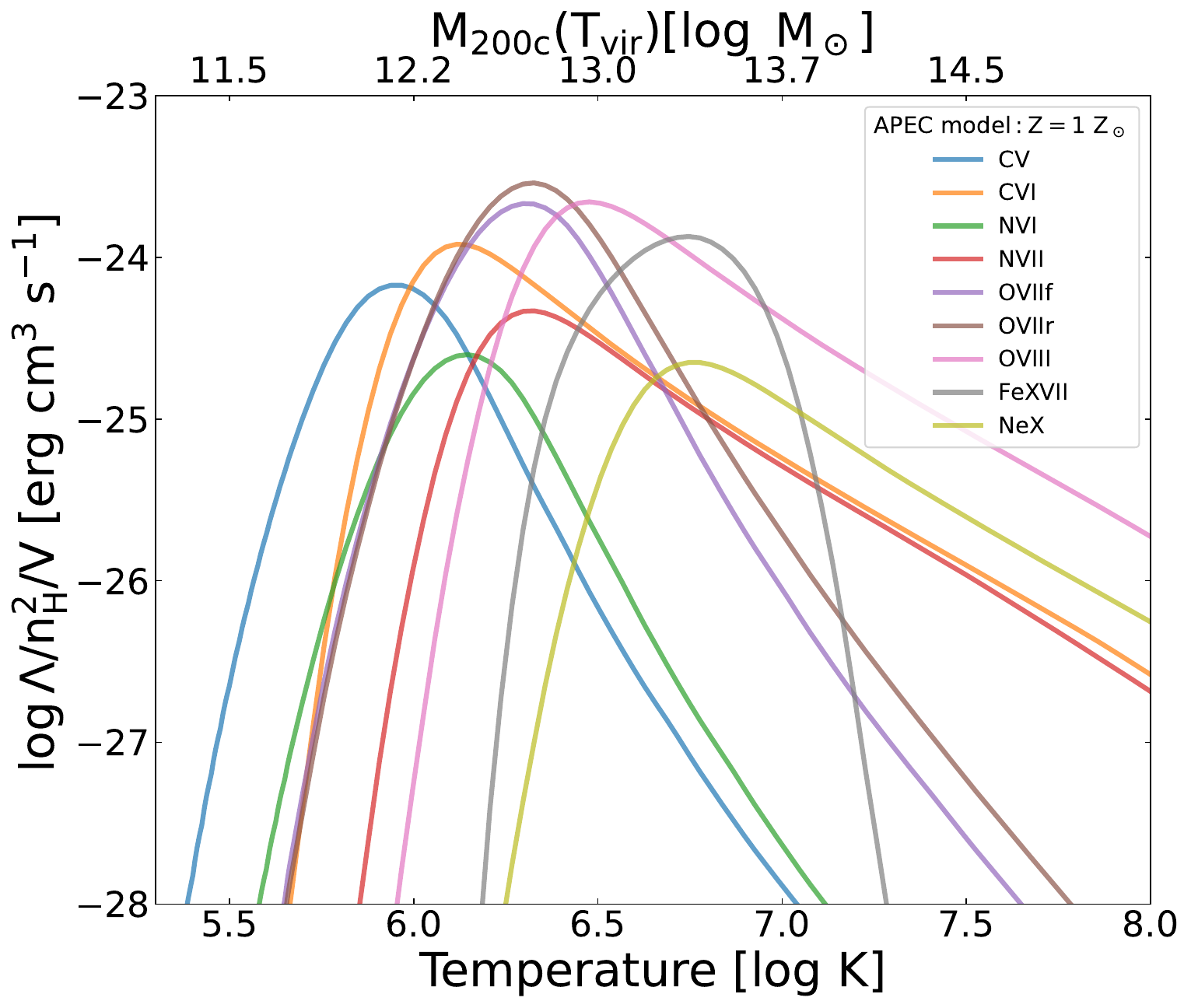}
  \caption{Emissivity as a function temperature of 9 prominent lines listed in Table~\ref{tab:my_lines}. The line emissivity is computed assuming an APEC model with solar elemental abundances. The upper x-axis specifies the corresponding ${\rm M_{200c}}$ derived based on its relation with the virial temperature: ${\rm T_{vir}\simeq (\mu m_p G M_{200c})/(2k_B R_{200c})}$, where ${\rm \mu \simeq0.59}$ is the mean molecular weight, ${\rm m_p}$ is the proton mass, G is the gravitational constant, and ${\rm k_B}$ is the Boltzmann constant.}
  \label{fig:9lines}
\end{figure}

For the analysis in this paper, we focus on simulated galaxies with stellar mass $\MSTARS>10^{10}\, \MSUN$, and we select only galaxies which are the centrals of their host halos, i.e. excluding satellite galaxies. Throughout, we measure $\MSTARS$ within twice the half-stellar mass radius (${\rm R_e}$). There are a few thousand galaxies at $z=0$ above this mass limit in each of the simulated volumes. 

\begin{figure*}
  \centering
  \includegraphics[width=0.98\textwidth]{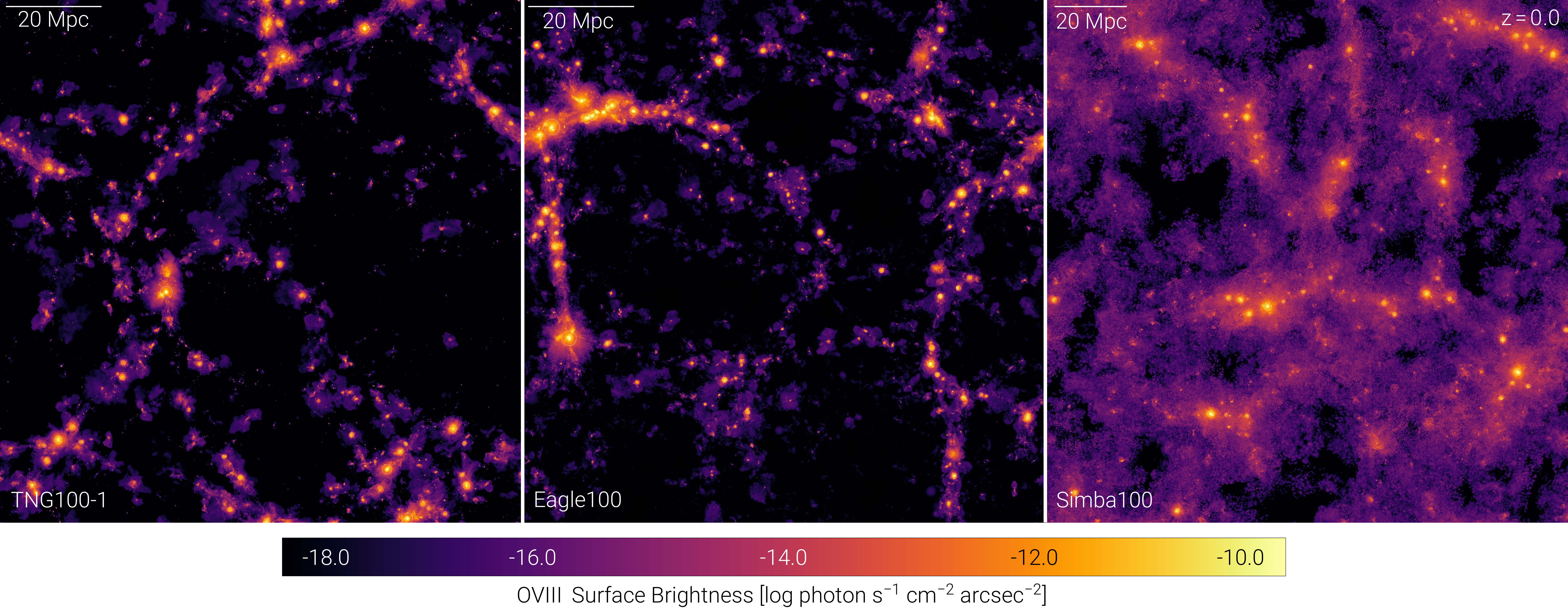}
  \caption{OVIII 18.9709\AA\, emission maps on large, intergalactic scales from three different cosmological hydrodynamical simulations at $z=0$. In each case we show the entire simulated volume across, which is comparable for all three simulations: $\sim100$ Mpc for EAGLE, $\sim110$ Mpc for TNG100, and $\sim147$ Mpc for SIMBA. The projection depth in each case is 15 per cent of the box-side length, a few tens of Mpc.}
  \label{fig:1}
\end{figure*} 

As motivated in Section~\ref{sec:intro}, we particularly aim to quantify the differences in the gas properties around galaxies at the Milky Way and Andromeda mass scale ($\MSTARS = 10^{10.5-11.2}\, \MSUN$). This range encompasses the stellar mass estimates, including 2-$\sigma$ and systematic uncertainties, of our Milky Way and Andromeda \citep[see ][for discussion]{pillepich.etal.2023}, and thus represents a mass scale of reference and importance. Furthermore, $\MSTARS \sim 10^{10-11}\, \MSUN$ is the range where the galaxy population transitions from being dominated by star-forming galaxies to quenched systems, in both the real Universe and in the models \citep[][and references therein]{donnari.etal.2021b}. As a result, in all three models, the physical properties of the circumgalactic gas around galaxies at this mass scale encode information about the effects and functioning of feedback energy injection from SMBHs. We elect to use a galaxy selection based on galaxy stellar mass because this is accessible via observations. As a result, differences among simulations will be, at least partially, driven also by underlying differences in the stellar-to-halo mass relation.

We hence focus on the physical and observable properties of CGM gas. Throughout this paper, by CGM we mean all non-starforming gas within a 3-dimensional distance from a galaxy center of ${\rm [0.1-1]} \RVIR$, where $\RVIR$ is the radius within which the average density is 200 times the present cosmological critical density. While the outer limit is approximately the virial radius of haloes, the inner limit ($>0.1 \RVIR$) is larger than twice the stellar effective radius for our sample, and is adopted to avoid the hot interstellar medium, whose physical interpretation requires more care. For the CGM in the considered radial range of ${\rm [0.1-1]\RVIR}$, we note that the inclusion or exclusion of the star-forming gas produces negligible differences in the obtained results. {In all cases we include gas within each friends-of-friends (FoF) halo as determined by the halo finding algorithm, which is identical in all analyzed simulations.

\subsection{Computation of the gas thermodynamical properties and intrinsic X-ray emission}
\label{sec:computation}

For each galaxy, we extract CGM properties predicted by the simulations (density, temperature and metallicity) by neglecting star-forming gas cells/particles, e.g. with non-zero instantaneous star formation rate, but without imposing any additional cut on temperature or phase. Unless otherwise stated, gas temperature and metallicity are mass weighted and are meant to convey the true physical state of the gas. 

To connect to X-ray observations, we compute the intrinsic emission in the soft band from the diffuse gas, without accounting for observational effects nor telescope responses and without adding the possible {\it contamination} from point-like sources such as X-ray binaries and AGNs. In particular, we focus on either the full X-ray spectrum in the $[0.3-1.3]$ or $[0.3-2]$ keV bands (continuum + emission lines) or on a few selected soft X-ray emission lines from key metal species, as listed in Table~\ref{tab:my_lines}.

X-ray emission is computed using the simulated physical properties of the gas on a cell-by-cell basis. We pre-calculate emissivity tables using a single-temperature APEC model (version 3.0.9) implemented in the XSPEC package (\citealt{smith.etal.2001}). The APEC model assumes an optically-thin plasma in collisional ionization equilibrium (CIE). We use the density, temperature, and elemental abundances of each gas cell/particle to derive its emission. For the solar abundance values we employ \cite{anders.grevesse.1989}. Our calculations exclude the effects of photo-ionization, due to the presence of UV/X-ray photons, local or otherwise. We also focus on non-resonant lines so that we can safely neglect X-ray resonant scattering \citep{gilfanov.etal.1987, zhuravleva.etal.2013, nelson.etal.2023}. In Fig.~\ref{fig:9lines}, we show the emissivity as a function of temperature for the 9 lines listed in Table~\ref{tab:my_lines} assuming an APEC model. The emissivity of selected lines peaks in the temperature range of ${\rm T\sim10^{5.5-7.0}}$ K, which corresponds to virial temperatures of halos across a broad mass range of ${\rm M_{200c}\sim10^{11.5-13.7}M_\odot}$.

When quoting the emission from specific metal lines, the emission is extracted from an energy range of $\sim0.4$ eV, while mock spectra are shown at a resolution of $\sim2$ eV. The total X-ray emission from a galaxy, from its CGM or a portion thereof, is obtained by summing up the contribution of all gas cells/particles in the given region.

\begin{figure}
  \centering
  \includegraphics[width=0.48\textwidth]{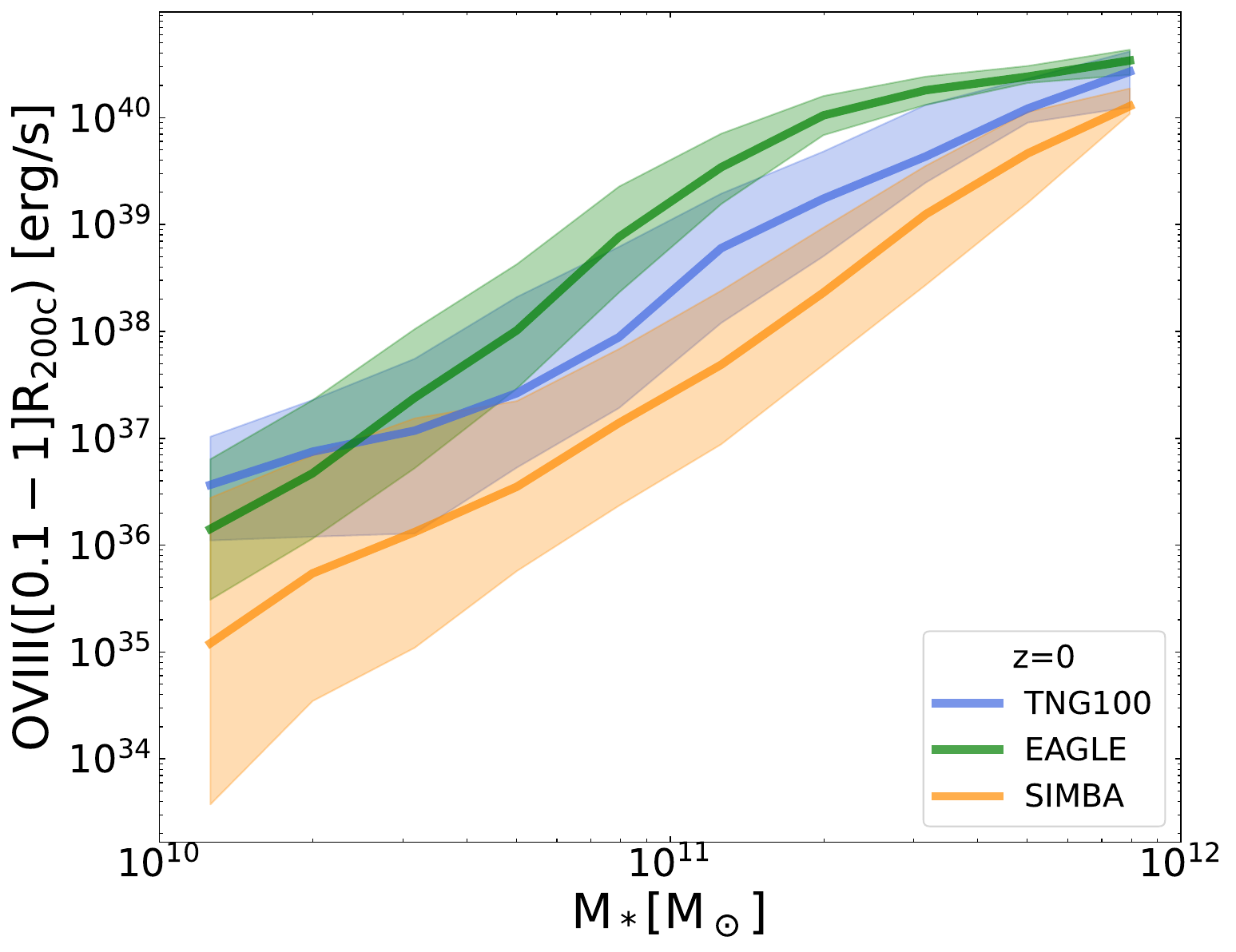}
  \caption{Total OVIII 18.9709\AA\, luminosity of the CGM as a function of galaxy stellar mass according to the TNG100, EAGLE, and SIMBA simulations at $z=0$. The solid lines represent the median across galaxies, while the shaded areas denote the 16th-84th percentiles in bins of galaxy mass.}
  \label{fig:2}
\end{figure}

\begin{figure*}
  \centering
  \includegraphics[width=0.9\textwidth]{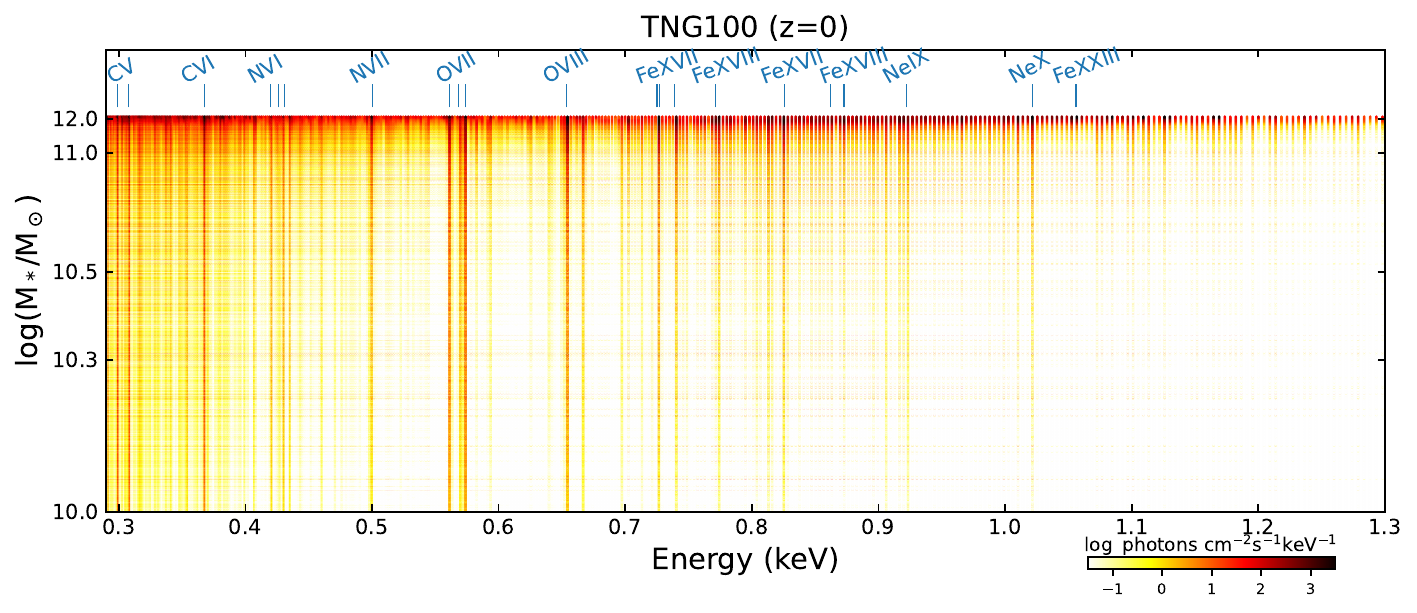}
  \includegraphics[width=0.75\textwidth]{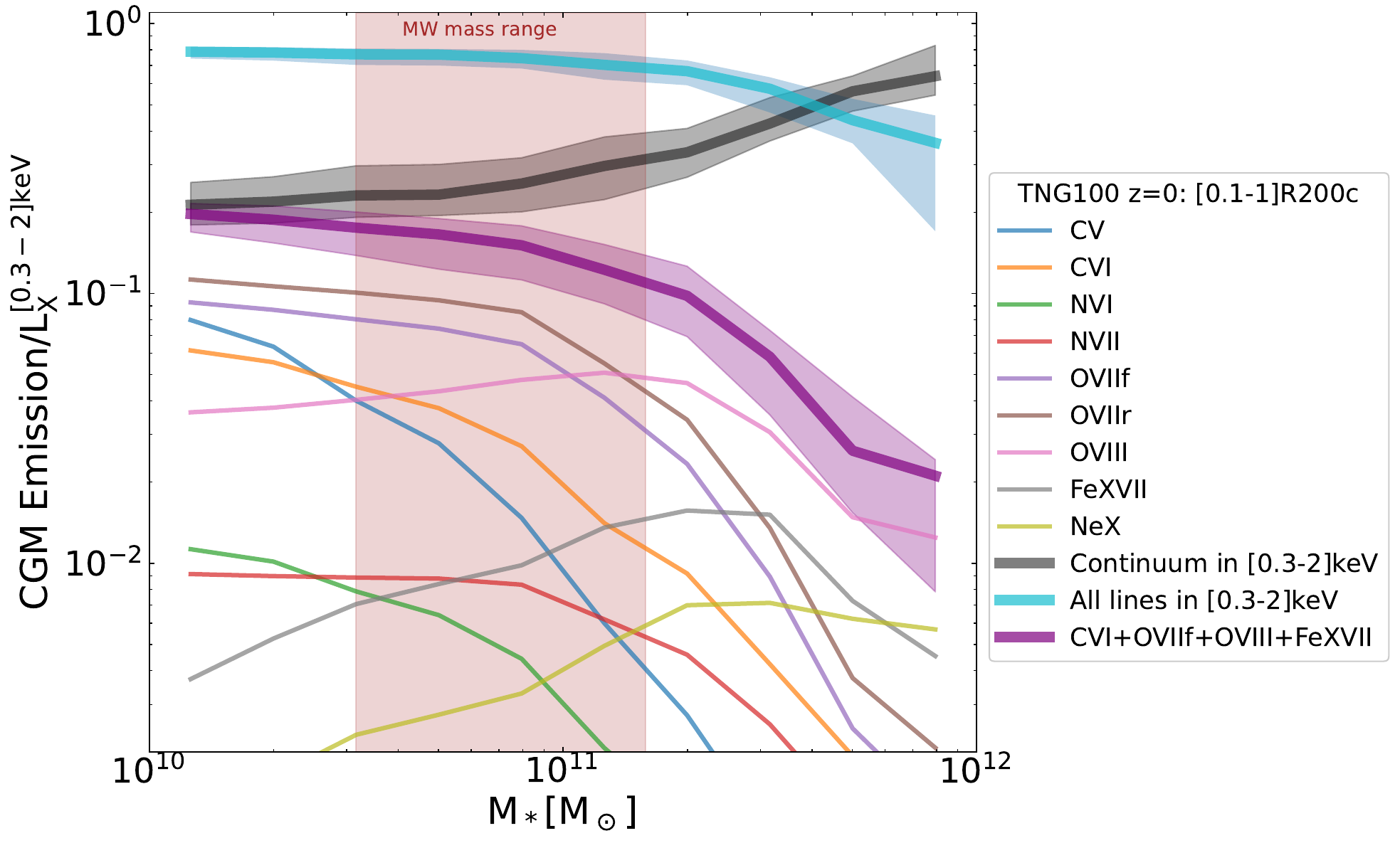}
  \caption{Soft X-ray spectra of the CGM around galaxies predicted by the cosmological galaxy simulation TNG100 and the relative contributions of emission lines to the continuum. {\it Top:} We show for each TNG100 galaxy at a given mass the emission from metal lines in the ${\rm [0.3-1.3]}$ keV band to highlight prominent lines emitted by the CGM: one galaxy, one row; spectral resolution of $\sim2$ eV; each energy bin color-coded based on intensity. It should be noted that the y-axis, representing stellar mass, is not up to scale as rows representing galaxies are layered on top of each other for the sake of effective visualisation. {\it Bottom:} We quantify the contribution of individual lines (thin color curves) and of all of them (thick blue curve) to the total X-ray soft broad band (e.g. ${\rm [0.3-2]}$ keV), in comparison to the continuum (thick gray curve). The shaded area around the thick curves represents the ${\rm 16^{th}-84^{th}}$percentile envelope. Both line and broad band emissions are measured within the radial range ${\rm [0.1-1]R_{200c}}$, i.e. from the CGM. In the CGM of galaxies with stellar mass below a few $10^{11}\,\MSUN$, the soft X-ray emission is by far dominated by metal lines.}
  \label{fig:intrinsic_spectra}
\end{figure*}

%%%%%%%%%%%%%%%%%%%%%%%%% RESULTS %%%%%%%%%%%%%%%%%%%%%%%%%%%%%%%

\section{Predictions for soft X-ray line emission of the CGM at z=0}
\label{section3}

\subsection{From large spatial scales to the haloes}
\label{overview}

Fig.~\ref{fig:1} shows the large-scale distribution of OVIII line emission from TNG100, EAGLE, and SIMBA at redshift $z=0$. The three simulations consistently predict that, as expected, most of the OVIII emission concentrates within gravitationally-bound DM haloes, marked by the bright regions in the maps, where most hot gas is also found. However, they differ significantly from each other on how distant from the halo centers the emission extends. As clearly visible, the SIMBA simulation predicts significantly more extended OVIII emission outside of haloes in comparison to TNG100 and EAGLE. This is qualitatively consistent with the finding that different feedback models lead to different baryon re-distribution within and beyond haloes: the `closure radius' in SIMBA is much larger than in the other considered models (\citealt{ayromlou.nelson.pillepich.2022}), because of the differently far-reaching effects of feedback \citep[see also][]{borrow.etal.2020,sorini.2022}.

To better quantify the OVIII emission from haloes, in Fig.~\ref{fig:2} we show the OVIII luminosity as a function of central galaxy stellar mass for the three simulations. The luminosity is measured within the radial range ${\rm [0.1-1]} \RVIR$ for each galaxy, representing the integrated luminosity of the CGM. Solid curves denote the median across galaxies and the shaded areas visualize the galaxy-to-galaxy variation at fixed galaxy stellar mass. Across the considered mass range, the OVIII CGM luminosity predicted by SIMBA is consistently lower than the other two simulations, especially at the low-mass end. The differences between TNG100 and EAGLE depend on galaxy stellar mass, and is larger for high-mass haloes: for galaxies more massive than the Milky Way, EAGLE predicts OVIII brighter haloes than TNG100, whereas the opposite is true at the lower-mass end. The predictions of Fig.~\ref{fig:2} are the result of the different physical properties of the CGM gas and hence depend on the typical density, temperature and Oxygen abundance in the gaseous haloes and on how these change, on average, with halo mass. Overall, CGM properties depend on the combined effects of hierarchical assembly of structures and feedback, which unfold differently in the different simulations and differently so in galaxies with different halo masses, SMBHs, etc.

\subsection{Most important soft X-ray emission lines in the CGM}
\label{overview}

The OVIII line at 18.9709\AA\, of the previous Section is only one of the prominent metal lines at soft X-ray wavelengths that characterize the CGM of normal galaxies according to current cosmological simulations. 

\begin{figure*}
  \centering
  \includegraphics[width=0.99\textwidth]{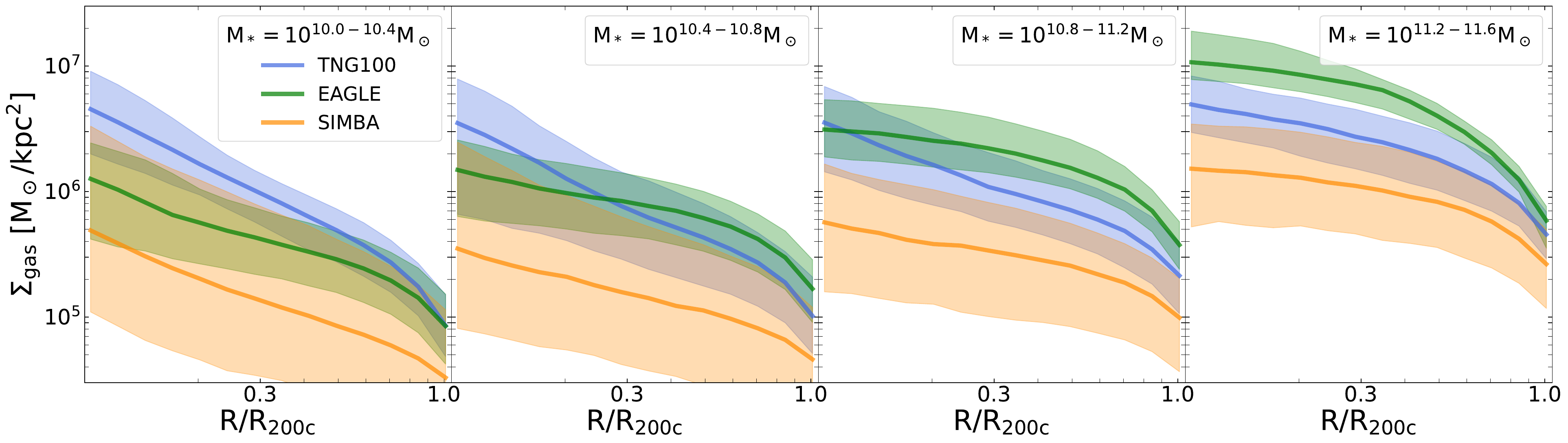}
  \includegraphics[width=0.99\textwidth]{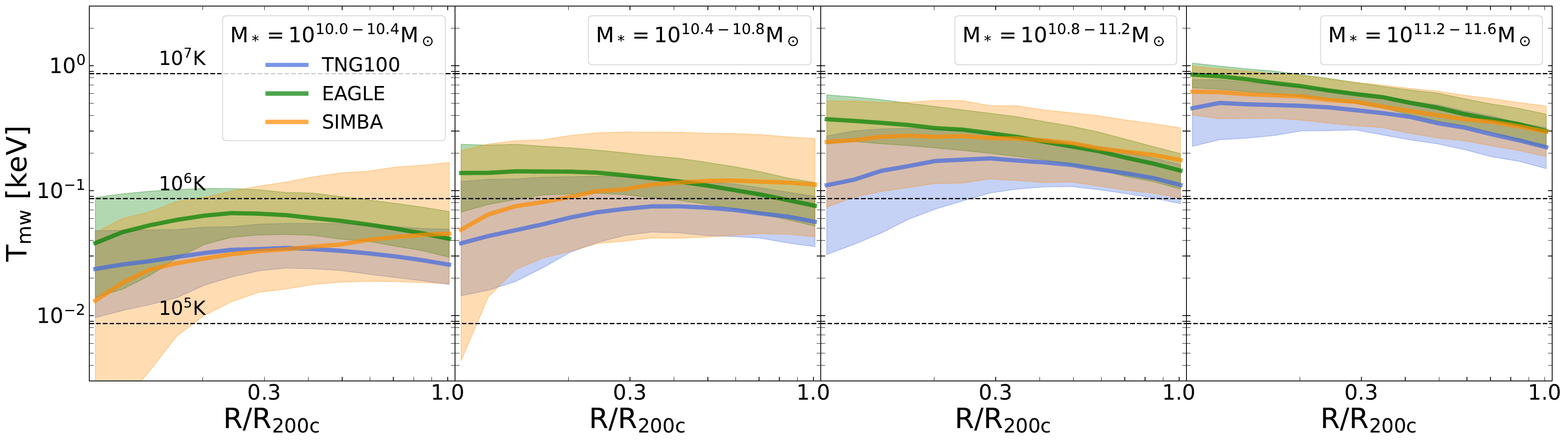}
   \includegraphics[width=0.99\textwidth]{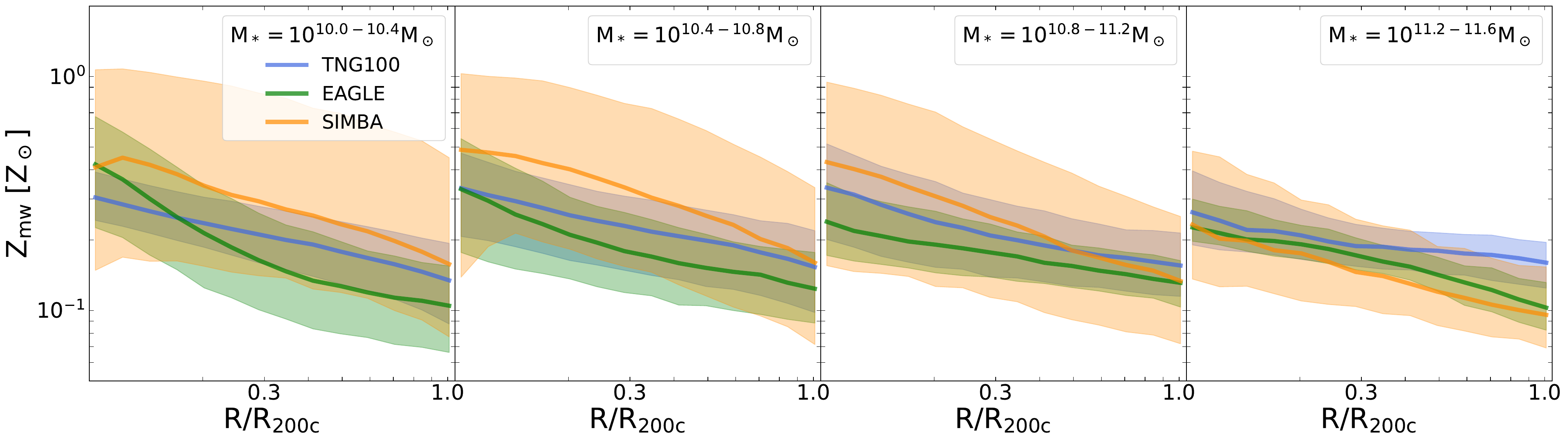}
  \caption{2D projected radial profiles of the CGM thermodynamical and metal content across galaxy stellar mass ranges (different columns) in TNG100, EAGLE, and SIMBA. From {\it top} to {\it bottom} panels we show the gas column density (${\rm \Sigma_{gas}}$), gas mass-weighted temperature (${\rm T_{mw}}$), and gas mass-weighted metallicity (${\rm Z_{mw}}$). The solid lines represent the population median, whereas the shaded areas show the 16th and 84th percentiles, i.e. quantify the galaxy-to-galaxy variation.} 
  \label{fig:3}
\end{figure*}

Fig.~\ref{fig:intrinsic_spectra}, top panel, shows the soft X-ray spectra from the CGM of all TNG100 galaxies in the $10^{10-12}\,\MSUN$ stellar mass range. Each horizontal row in the figure represents one galaxy and its spectrum, whereby emission at each energy is shown via the colorbar. Across the considered mass range, there are several prominent metal lines, whose intensity generally depend on galaxy mass. For example, according to TNG100, high-energy lines such as FeXVIII or NeX mostly appear in massive galaxies ($\MSTARS\gtrsim11.5 \, \MSUN$), while lower-energy lines such as FeXVII, OVIII, or OVII (triplets) are visible virtually at all considered mass scale. This finding is expected as high-energy ions are primarily associated with high-temperature CGM, as indicated by Fig.~\ref{fig:9lines}, which is more abundant in massive galaxies. In low-mass galaxies the CGM temperature is not high enough to effectively ionise highly-charged ions, limiting their line emission. A qualitatively-similar picture also holds in EAGLE and SIMBA, albeit with different quantitative predictions, as seen in Fig.~\ref{fig:2} for the case of OVIII.

\begin{figure*}
  \centering
  \includegraphics[width=0.95\textwidth]{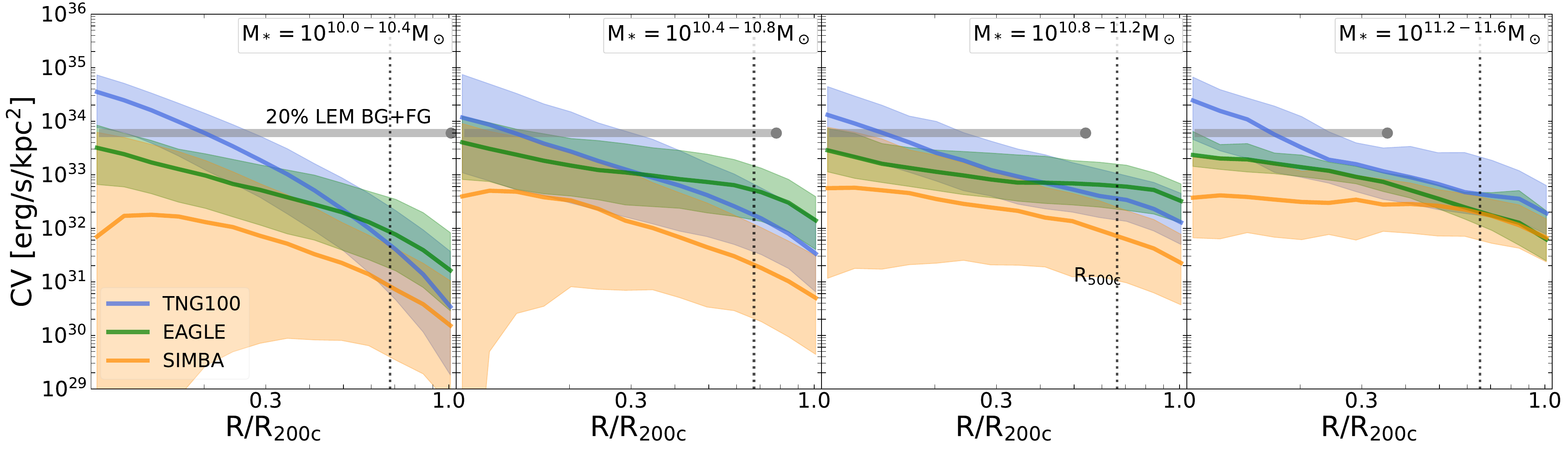}
  \includegraphics[width=0.95\textwidth]
  {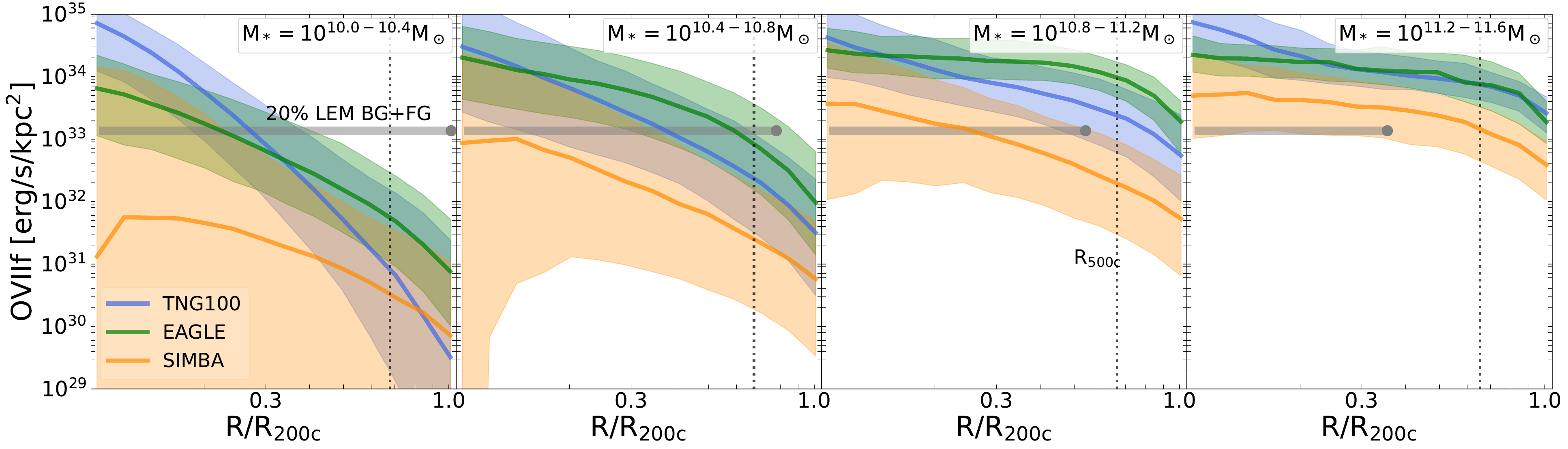}
  \includegraphics[width=0.95\textwidth]{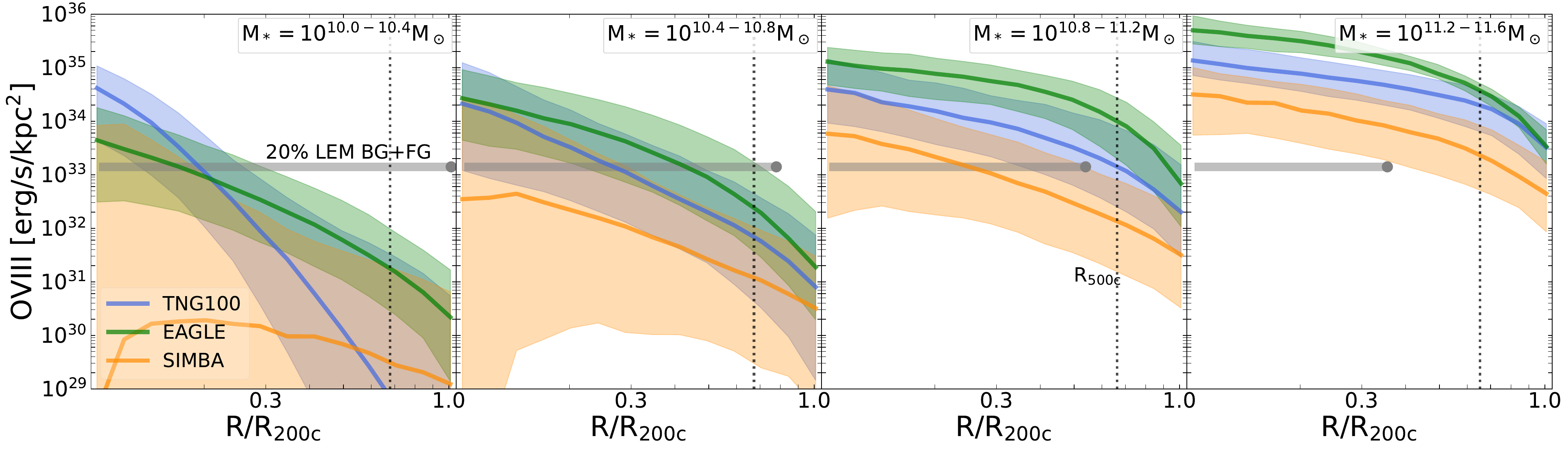}
  \includegraphics[width=0.95\textwidth]{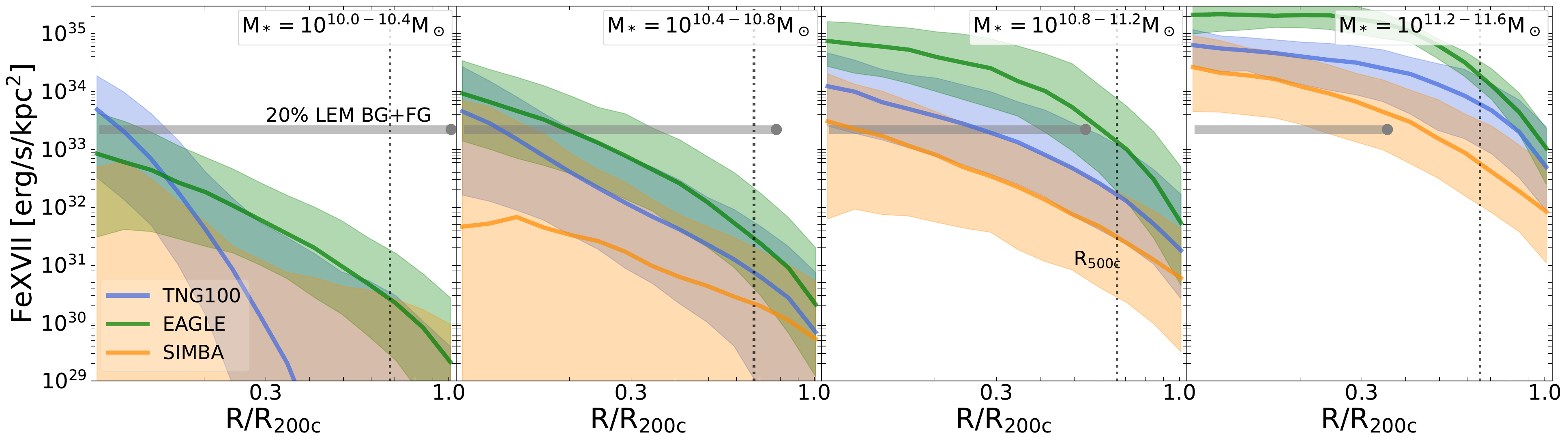}
  \caption{Similar to Fig.~\ref{fig:3} but for the surface brightness profile of a selected sample of emitting lines in the three simulations. From {\it top} to {\it bottom} we show profiles of: CV, OVIIf, OVIII, and FeXVII (intrinsic emission from the diffuse gas, as per Section~\ref{sec:computation}). The horizontal grey lines represent the 20 per cent level of the LEM background plus foreground (see text for more details), and their horizontal extension specifies the area covered by LEM field of view at $z=0.01$ (with radius given in $\RVIR$). For context, the median $\RVIR$ of the depicted TNG galaxies reads 188, 245, 354, 537 kpc, from left to right.}
\label{fig:4}
\end{figure*}

In the bottom panel of Fig.~\ref{fig:intrinsic_spectra} we show that the soft X-ray emission from the CGM of galaxies with stellar mass below a few $10^{11}\,\MSUN$ is strongly dominated by metal lines. This is quantified with TNG100, but holds also for EAGLE and SIMBA. The contribution of nine prominent metal lines to the CGM X-ray emission relative to the broad-band emission (in ${\rm [0.3-2]}$ keV) is given as a function of galaxy stellar mass: this varies from a few percent to over 10 per cent in the case of the resonant OVII line, the brightest line below ${\MSTARS = 10^{11}M_\odot}$ among the considered lines. Low-energy lines such as CVI, NVII, OVII are more prominent in low-mass galaxies (${\MSTARS\sim10^{10}M_\odot}$), whereas the contribution of higher-energy lines such as OVIII, FeXVII, or NeX peaks at a higher mass scale ($\gtrsim10^{11}M_\odot$). Notably, in galaxies with stellar mass in the MW-mass range the combined emission from CVI, OVII(f), OVIII, and FeXVII lines approaches the level of the continuum contribution, accounting for nearly $20\%$ of the total broad-band emission. These lines are promising targets for LEM detection of the CGM in MW-mass range (See Section~\ref{sec:narrow_band_obs}).

%%%%%%%%%%%%%%%%%%%%%%%%%%%%%%%%%%%%%%%%%%%%%%%%%%%%%%%%%%%%%

\subsection{Radial dependence}
\label{mass_dependence}

\subsubsection{Physical properties of the gas.}

We turn to radial trends, and inspect the thermodynamics as well as metal content of the CGM predicted by TNG100, EAGLE and SIMBA. In Fig.~\ref{fig:3} we show the radial profiles of the hot gas surface density, temperature, and metallicity in 4 different stellar mass bins, from left to right. Overall, the three simulations produce significant diversity in the radial profiles of gas properties across the considered galactic mass range (${10^{10}<\MSTARS/\MSUN<10^{11.6}}$), most notably in gas column density. The SIMBA simulation consistently has lower gas column density profiles compared to the other two simulations across the considered radial and mass range. For instance, at the MW-mass range (${\MSTARS=10^{10.8-11.2}\, \MSUN}$) and at the radius ${\rm r=0.5R_{200c}}$, SIMBA ${\rm \Sigma_{gas}}$ profile is below TNG100 (EAGLE) by a factor of $\sim3$ ($\sim6$). Comparing TNG100 and EAGLE, the former has a larger gas column density profile at the low-mass end ($\MSTARS<10^{10.4}\, \MSUN$) but smaller densities at the high-mass end ($\MSTARS>10^{11.2}\, \MSUN$). At the Milky Way-mass range, the TNG100 density profile is comparable within the galaxy central region ($r<0.1R_{200c}$), but smaller at larger radii, in comparison to the EAGLE profile. Concerning the shape i.e. slope, TNG100 produces significantly steeper profiles compared to EAGLE and SIMBA, in particular at the low-mass end (${\MSTARS<10^{10.8}\, \MSUN}$). 

For the mass-weighted temperature profiles, the differences between the three simulations are less significant than in gas column density, especially at large radii and at the high-mass end where the three simulations produce consistent profiles. In low-mass galaxies (below the MW-mass range) and within small radii (${\rm r\lesssim0.5 R_{200c}}$), EAGLE has flatter temperature profiles compared to the other simulations. For the mass-weighted metallicity profiles, TNG100 and EAGLE profiles are broadly consistent within the sample uncertainty, whereas SIMBA has more metal-enriched CGM than the other two simulations (within a factor of 3) for galaxies with ${\MSTARS} < {\rm 10^{11.0}M_\odot}$. 

It is worth noticing that SIMBA produces a significantly larger scatter in gas properties than the other two simulations, especially at the low-mass end. For example, the average scatter in the SIMBA gas column density at the MW-mass range is about 0.5 dex in comparison to 0.1-0.2 dex in TNG100 or EAGLE. 

We note that while the three simulations are designed to reproduce realistic samples of galaxies, the degree to which they do so differs. In addition, they predict considerable variation in in the thermodynamics and metal content of the CGM, due in large part to their distinctive models for stellar and SMBH feedback. To zeroth order, stellar feedback serves as the dominant channel in galaxies with mass substantially below the MW mass range, whereas SMBH feedback dominates in galaxies within or above the MW mass range (see e.g. \citealt{weinberger.etal.2018}). 

\subsubsection{Observable properties of the gas.}

The diversity in the gas properties across the three simulations is reflected in the line emission profiles. In Fig.~\ref{fig:4} we show the surface brightness radial profiles for a selected sample of four metal lines from low to high energy: CV, OVIIf, OVIII, and FeXVII. The results for other lines in the soft band ([0.3-1.3] keV) are qualitatively similar. These lines, especially OVIIf, OVIII, and FeXVII, are crucial for probing the CGM in nearby galaxies (e.g. ${z\sim0.01}$) through LEM observations (see Section \ref{sec:narrow_band_obs} for more details). Overall, SIMBA predicts significantly less line emission compared to TNG100 and EAGLE, which are more similar depending on galaxy mass. Moreover, at the low-mass end the line emission profiles in TNG100 are significantly steeper than the EAGLE profiles. These results emphasize the observational manifestation of the diversity in the gas column density profiles predicted by the three simulations. 

In Fig. \ref{fig:4}, we also illustrate the LEM detectability threshold for three metal lines: OVIIf, OVIII, and FeXVII. To do so, we show the $20$ per cent level of the background+foreground (BG+FG), and use it as an approximate detection threshold.\footnote{We refer to a companion paper by \cite{schellenberger.etal.2023} for a systematic study on the LEM detectability.} The LEM BG+FG level is derived assuming that each galaxy is observed at $z=0.01$, and the extent of the horizontal lines indicate the galactocentric distance covered by the LEM field of view, which is $30$ arcmin. For galaxies more massive than MW/M31 (${\MSTARS\gtrsim10^{11.2}M_\odot}$), a single LEM pointing will cover a spherical region out to ${\rm r\lesssim0.4 R_{200c}}$, whereas for less massive galaxies (${\MSTARS\lesssim10^{11.0}M_\odot}$), LEM can comfortably cover the CGM out to ${\rm R_{500c}}$ or even $\RVIR$ in a single pointing. Detectability for LEM depends on various factors, such as the line energy, the considered mass range, and the simulation model. For instance, for galaxies in the MW/M31 mass range (middle columns of Figs.~\ref{fig:3} and \ref{fig:4}), SIMBA predicts that LEM could detect the CGM emission in OVIIf or OVIII only out to ${\rm \sim0.3 R_{200c}}$. On the other hand, TNG100 and EAGLE predict that LEM will detect OVIIf or OVIII line emission out to ${\rm \sim0.6-0.7 R_{200c}}$, which is similar to ${\rm R_{500c}}$ for galaxies at this mass range.

%%%%%%%%%%%%%%%%%%%%%%%% Intrinsic Scatter %%%%%%%%%%%%%%%%%%%%%%%%%%%%%

\subsection{Dichotomy between star-forming versus quiescent galaxies in X-ray line emission}
\label{lx_dichotomy}

\begin{figure*}
  \centering
  \includegraphics[width=0.9\textwidth]{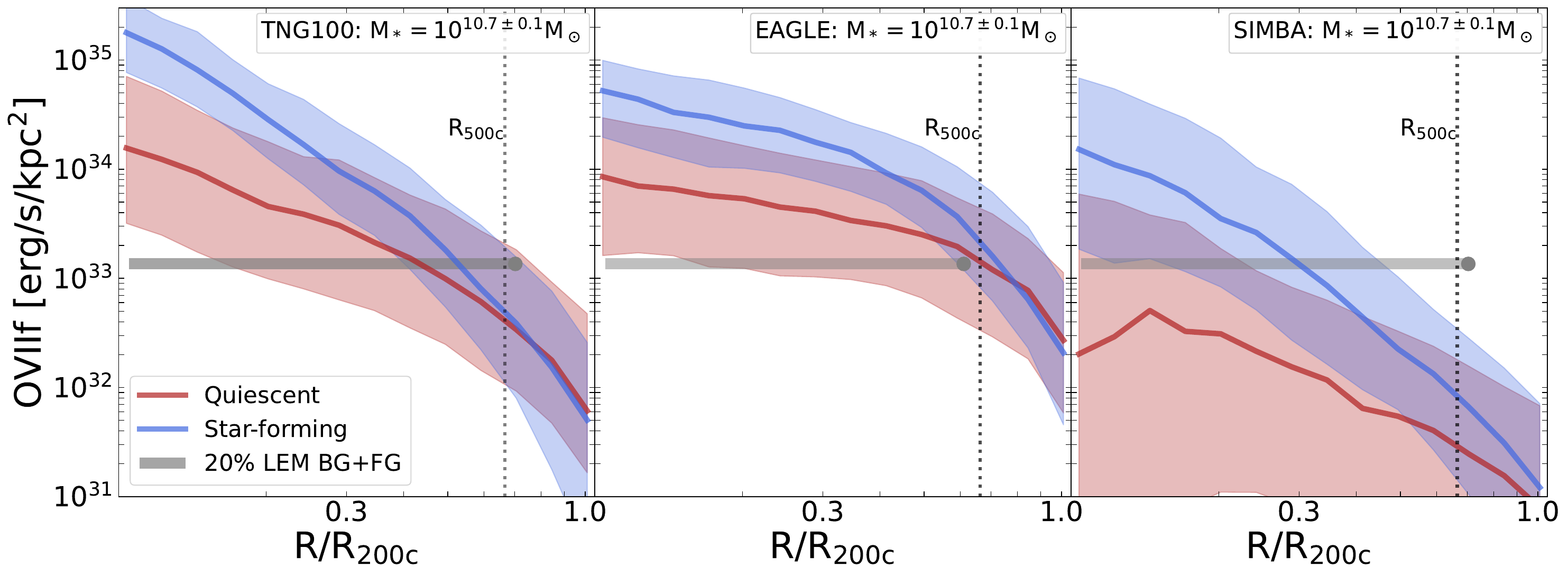}
  \includegraphics[width=0.9\textwidth]{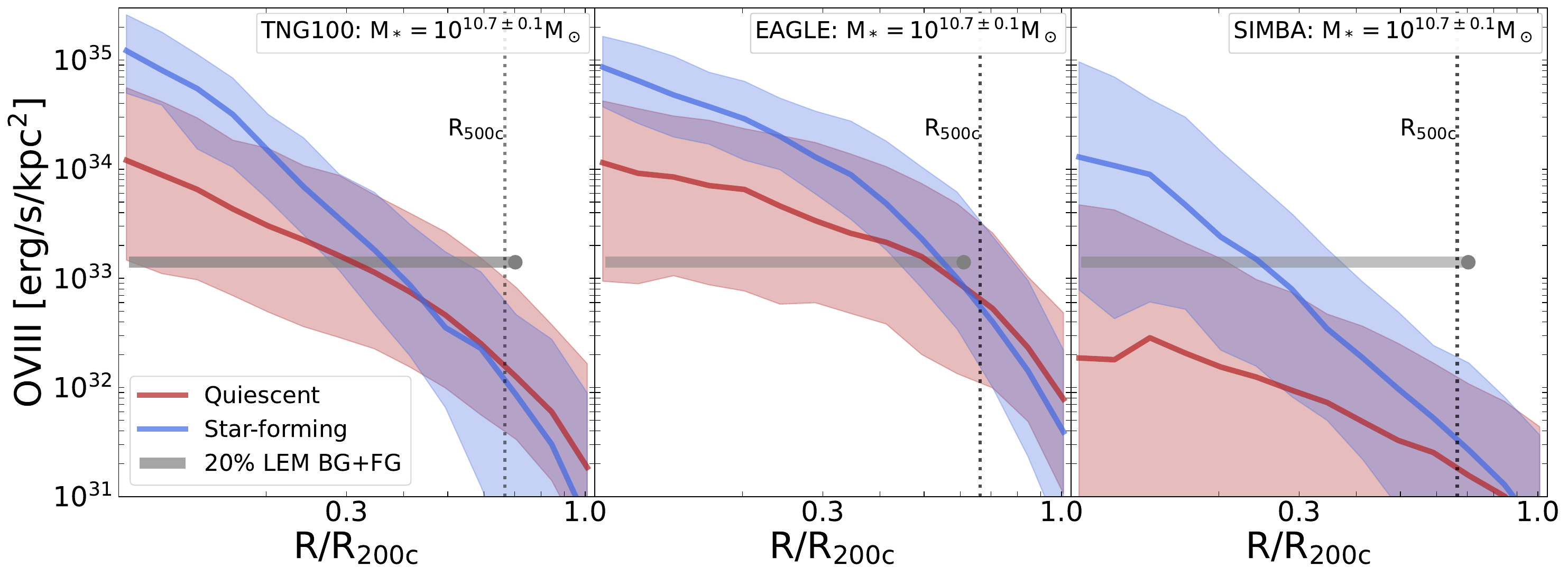}
  \includegraphics[width=0.9\textwidth]{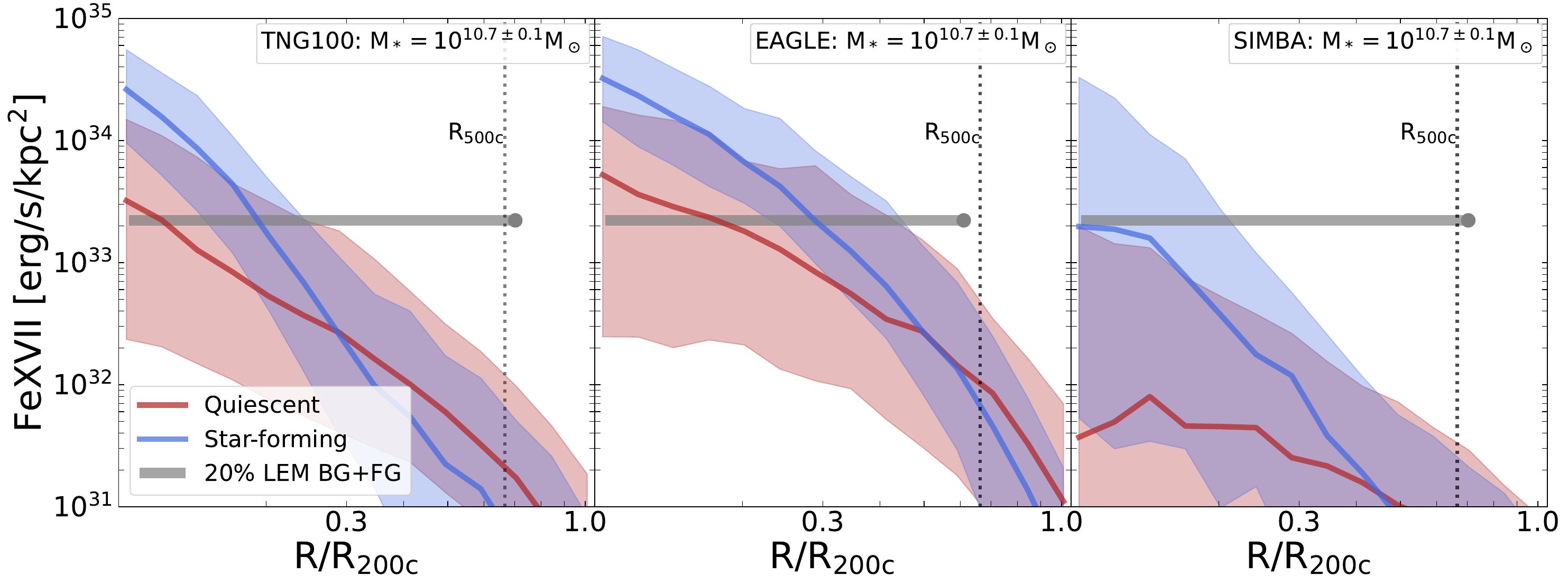}
  \caption{Diversity between star-forming and quiescent galaxies at the MW-mass range (${\MSTARS=10^{10.7\pm0.1}\, \MSUN}$) according to TNG100, EAGLE and SIMBA at $z=0$. From {\it top} to {\it bottom}, we show the intrinsic surface brightness profiles of OVIIf, OVIII, and FeXVII line emission, respectively. Galaxies are split based on their star-formation status: star-forming (blue) and quiescent (red) galaxies depending on their ${\rm sSFR}$ with respect to quenched threshold ${\rm< 10^{-11}yr^{-1}}$. The solid curves denote the median profile of each subgroup, and the shaded areas represent the 16th-84th percentile envelope. Despite the vastly different implementations of SMBH feedback, all considered galaxy formation models predict that, at this mass range, the (inner) CGM of star-forming galaxies is brighter than that around quiescent galaxies, in each of the specific soft X-ray emission lines we consider. This is because, according to the models, quiescent galaxies exhibit overall less dense, less enriched, but hotter halo gas in comparison to star-forming galaxies (see Fig.~\ref{fig:cgm_sf_q}).}
  \label{fig:6b}
\end{figure*}

\begin{figure*}
  \centering
  \includegraphics[width=0.9\textwidth]{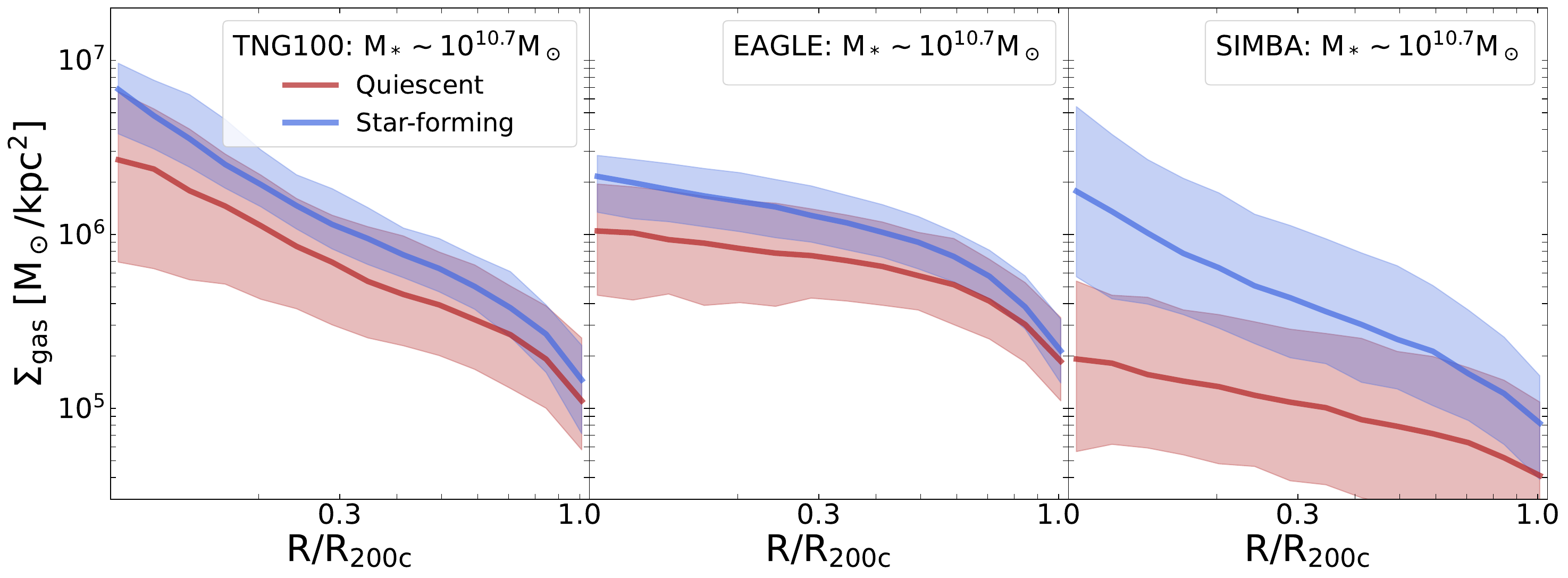}
  \includegraphics[width=0.9\textwidth]{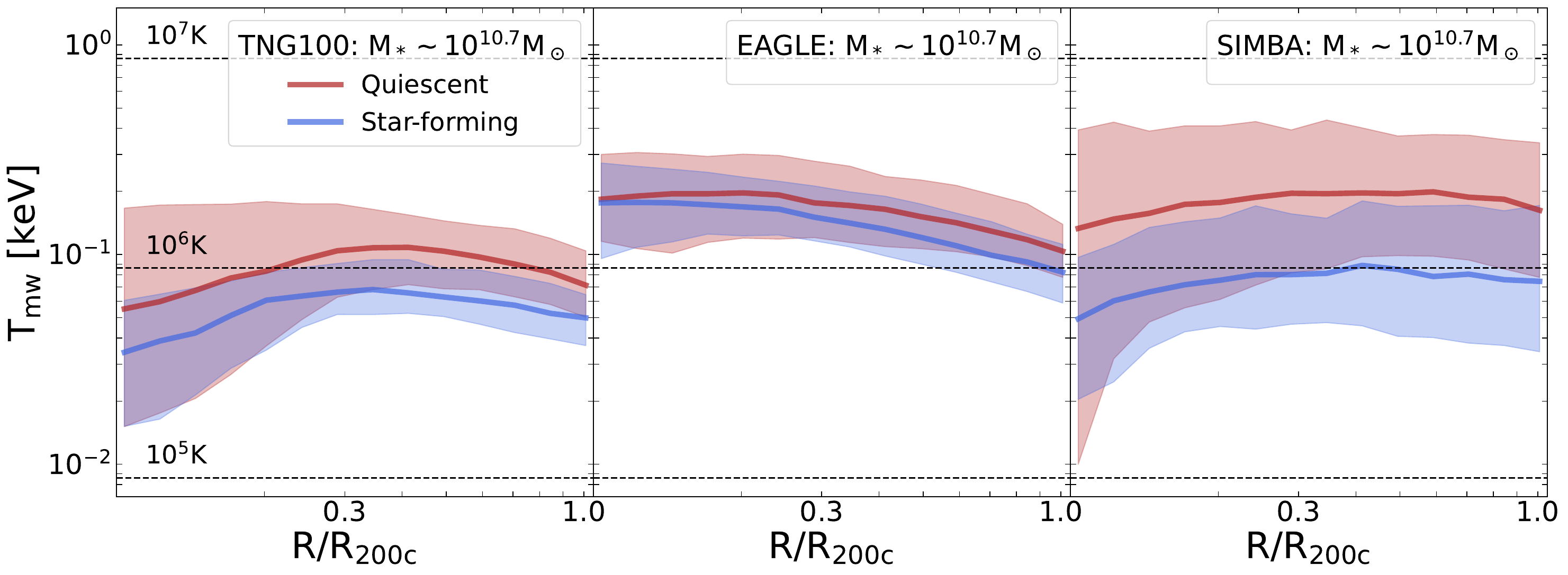}
  \includegraphics[width=0.9\textwidth]{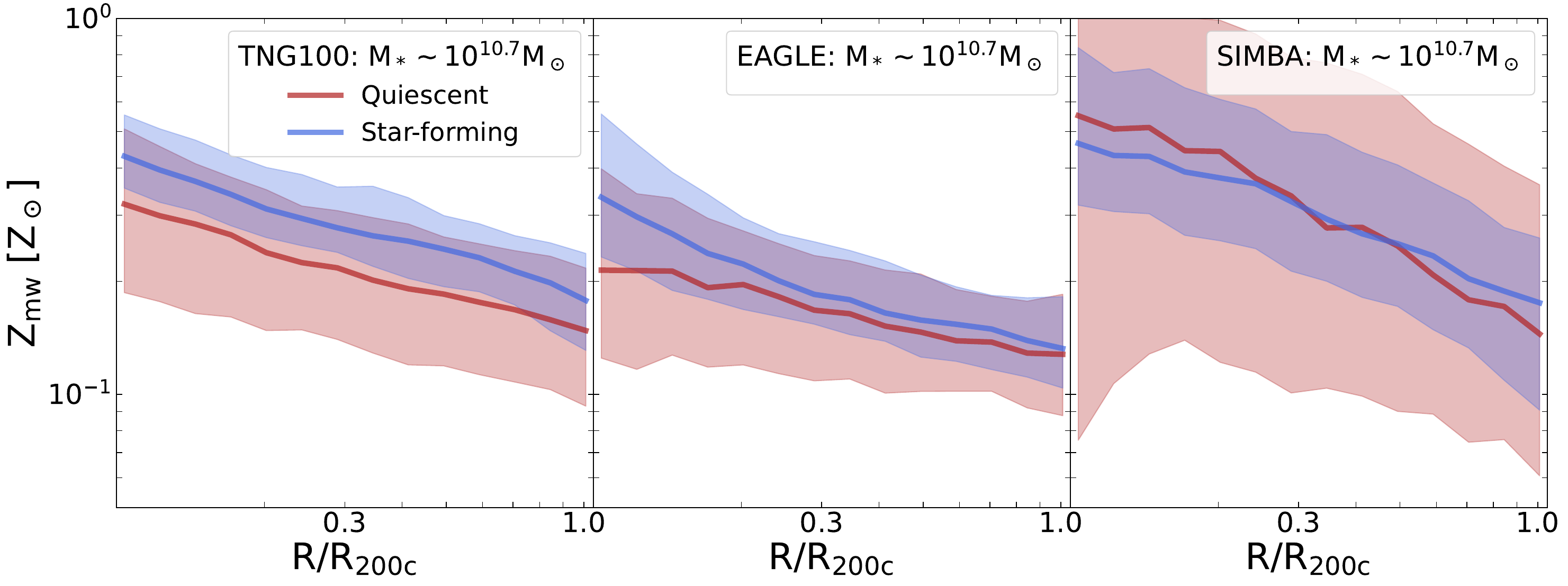}
  \caption{CGM properties in star-forming versus quiescent galaxies in the three simulations: gas column density ({\it top}), gas mass-weighted temperature ({\it middle}), gas mass-weighted metallicity ({\it bottom}).}
  \label{fig:cgm_sf_q}
\end{figure*}

We now investigate the diversity between star-forming and quiescent galaxies. To do so we restrict our study to a sample of galaxies within a narrow stellar mass range: $\MSTARS=10^{10.7\pm0.1}\, \MSUN$, which corresponds approximately the MW-mass range. The number of simulated galaxies within this MW-mass range amounts to 410 for TNG, 240 for EAGLE, and 775 for SIMBA.At this mass scale, galaxies can be both star-forming as well as quenched in all three simulations, as well as in the Universe \citep[e.g.][and references therein]{donnari.etal.2021b}. To identify these two classes we use the specific star formation rate ${\rm sSFR=SFR/\MSTARS}$, where SFR is measured within the aperture of ${\rm 2R_e}$ and ${\MSTARS}$ is the stellar mass within the same aperture. The galaxies are subdivided into 2 subsamples: star-forming versus quiescent galaxies, with the division lines specified as ${\rm sSFR=10^{-11}yr^{-1}}$. At the redshift z=0 the exact definition of `quenched' is unimportant. Different tracers for SFR e.g. instantaneous vs. averaged over the last 100-200 million years result in negligible differences in quenched fractions \citep{donnari.etal.2021b}.

In Fig.~\ref{fig:6b} we show the X-ray line emission from the CGM for star-forming versus quiescent galaxies across the three simulations. After dividing the sample into the two populations, we find that the star-forming population is significantly more luminous in line emission compared to the quenched counterparts \cite{truong.etal.2020,oppenheimer.etal.2020}. Remarkably, at the MW/M31 transitional mass scale, star-forming galaxies are X-ray brighter than quenched ones for each line we consider and for all three simulations, despite the vastly different implementations of SMBH feedback and of the quenching mechanisms. However, there are also more subtle quantitative differences. The difference between the two galaxy populations is most significant in SIMBA, and in OVIIf. For OVIIf in particular, the three simulations agree that the dichotomy is strongest at small radii, for example at ${\rm r\sim0.1R_{200c}}$ the difference between star-forming and quiescent OVIIf emission is 1-2 orders of magnitudes. However, they differ in the extent out to which the dichotomy persists. TNG100 and EAGLE predict that the dichotomy can be seen out to ${\rm r\sim0.7-0.8\ R_{200c}}$, while in SIMBA the dichotomy exists all the way out to $\RVIR$.  

\subsubsection{{\bf Physical origin of the CGM X-ray dichotomy between star-forming and quiescent galaxies.}}

To identify the physical origin of this dichotomy, we quantify thermodynamical and metallicity differences between the two galaxy populations. In Fig.~\ref{fig:cgm_sf_q} we show the physical properties of CGM gas that are responsible for the dichotomy in X-ray emission between star-forming and quiescent galaxies. The figure displays, from top to bottom, radial profiles of gas mass surface density, mass-weighted temperature, and mass-weighted metallicity. There are a number of qualitative similarities among the three simulations at fixed galaxy stellar mass: i) star-forming galaxies contain more CGM gas compared to quiescent galaxies; ii) their CGM temperature is lower than of quiescent counterparts; and iii) their CGM is slightly more metal enriched compared to the quenched galaxies on average, although this feature is less clear in the SIMBA simulation. These trends suggest that the line-emission dichotomy between star-forming versus quiescent galaxies is mainly due to the offset in the gas content between the two populations. The difference in metallicity can in principle also increase line emission from star-forming galaxies, thereby partly explaining the emission dichotomy, however the density effect dominates.  

\begin{figure*}
  \centering
    \includegraphics[width=0.99\textwidth]{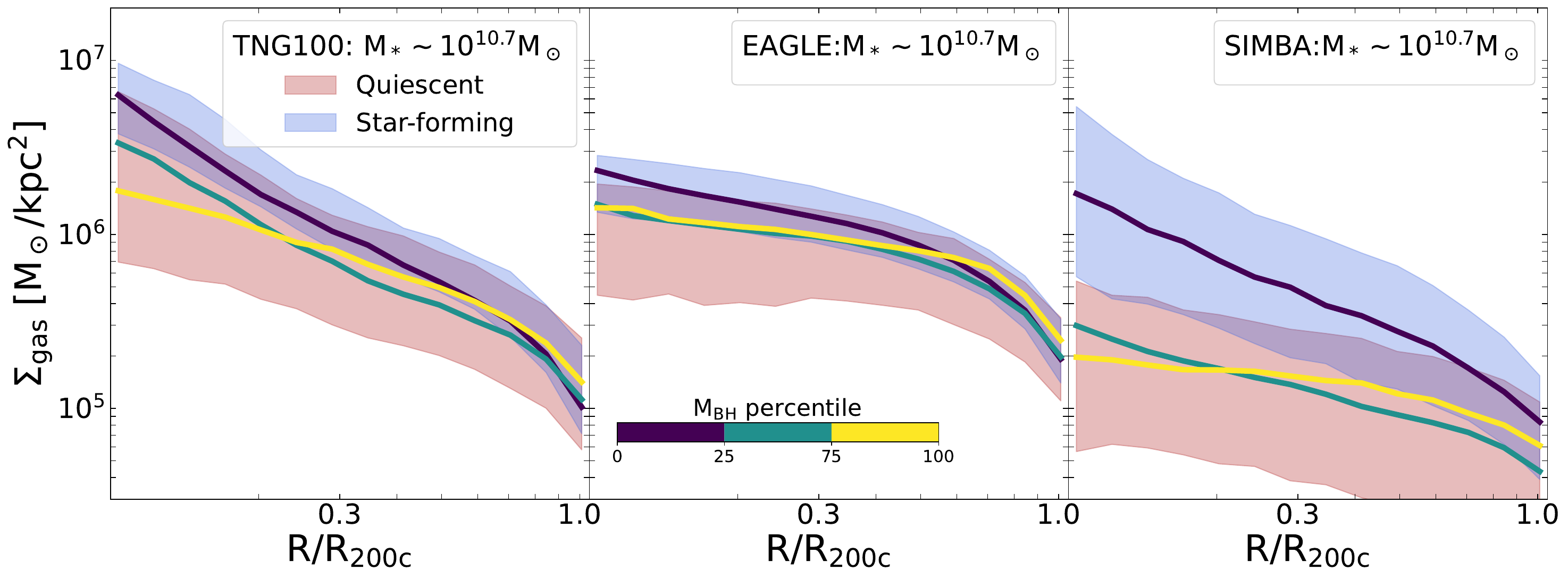}
    \includegraphics[width=0.69\textwidth]{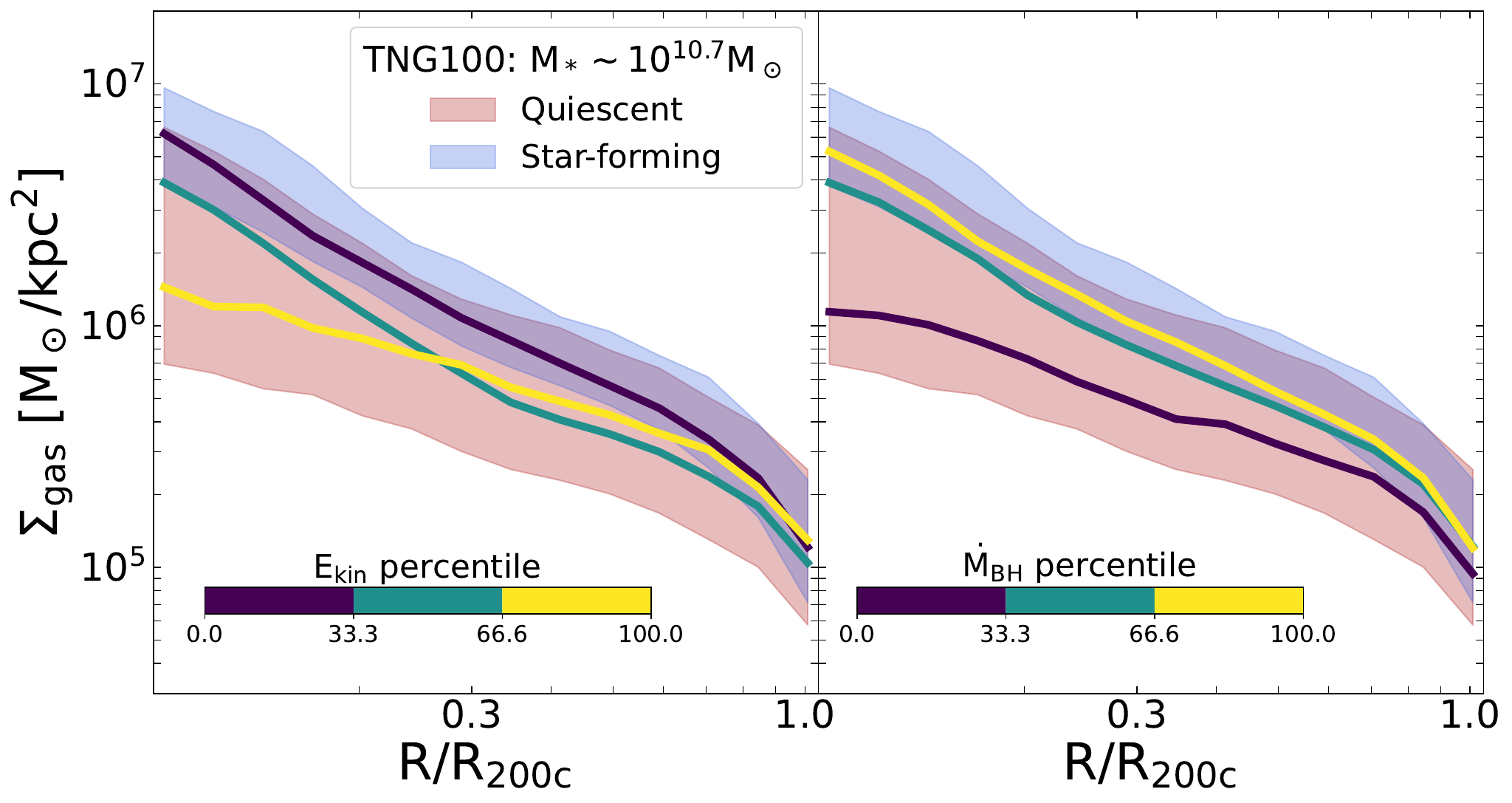}
    \caption{Dependence of the gas column density on various properties of the central SMBH. The results are shown for a subset of galaxies within a narrow MW-range mass bin: ${\MSTARS=10^{10.7\pm0.1}\, \MSUN}$. For each property, the galaxies are subdivided into three subgroups depending on the percentile of their SMBH property (see the color bar), and the solid lines are the subgroup median profiles. Shaded areas represent the 16th-84th percentile envelopes of the star-forming and quiescent populations as per Fig.~\ref{fig:6b}. {\it Top:} Variation of the gas column density profile as a function of SMBH mass, with each simulation shown in a separate panel. {\it Bottom:} Dependence of the gas column density on the accumulated kinetic feedback energy (${\rm E_{kin}}$) and the instantaneous accretion rate (${\rm \dot{M}_{BH}}$) onto the SMBH in TNG100. Notwithstanding the large galaxy-to-galaxy variation, galaxies with more massive SMBHs and with larger relative amounts of feedback in SMBH kinetic mode are surrounded by less dense gaseous haloes.}
    \label{fig:SMBH_Dens}
\end{figure*}

Several studies have reported that SMBH feedback is the primary mechanism for quenching star formation in both TNG100 and EAGLE simulations (e.g. \citealt{davies.etal.2020, zinger.etal.2020, terrazas.etal.2020}). Quenching is chiefly driven by ejection of gas from the central region of galaxies caused by feedback from SMBH, which leads to a depletion of gas in quiescent galaxies compared to their star-forming counterparts in the transitional mass range \citep[e.g.][]{nelson.etal.2018b, ramesh.etal.2023a}. To further explore this phenomenon, we investigate the impact of SMBH feedback on CGM gas content in the three simulations by examining the relationship between the gas content and central SMBH mass, which serves as a proxy for the amount of feedback energy injected in surrounding environment. The key idea for interpreting the diagnostics below is that, in all three simulations, galaxies would continue to form stars in the absence of SMBH feedback \citep[e.g.][]{weinberger.etal.2017}. Thus, the differences in the physical and observable properties of the CGM between quiescent and star-forming galaxies are due to the impact of SMBH feedback. We refer the reader to previous investigations about the interrelation between the star formation with SMBH feedback (\citealt{davies.etal.2020, terrazas.etal.2020}), the effects of SMBH feedback on the CGM properties (\citealt{nelson.etal.2018b, davies.etal.2019, oppenheimer.etal.2020a, zinger.etal.2020}), and how the X-ray emission of the gaseous halo is impacted by different models for SMBH and stellar feedback \citep[e.g.][]{truong.etal.2020, truong.etal.2021, ayromlou.nelson.pillepich.2022}.

In the top row of Fig.\ref{fig:SMBH_Dens}, we show the radial profile of the gas column density for the sample of MW-like galaxies, dividing it into three subsets based on percentiles of SMBH mass. Overplotted are also the loci of star-forming and quiescent galaxies as per Fig.~\ref{fig:6b}. We find that galaxies with higher ${\rm M_{BH}}$ percentiles have lower density profiles compared to those with smaller SMBH masses, indicating an anti-correlation between the CGM gas content and the SMBH feedback activity integrated across time, of which SMBH mass is a proxy. The significance and the radial extent of this anti-correlation varies among the three simulations. In the TNG100 and EAGLE simulations the anti-correlation is mostly found within ${\rm 0.3-0.5R_{200c}}$, while in SIMBA it extends all the way to $\RVIR$. All this is consistent with the finding that SMBH mass (and not SMBH accretion rate, see below) is unanimously the most predictive parameter of central galaxy quenching at all epochs, in both current galaxy simulations and in observations \citep{piotrowska.etal.2022, bluck.etal.2023}.

The bottom row of Fig. \ref{fig:SMBH_Dens} explores how halo gas column density varies with the amount of feedback energy that is injected by the central SMBH into the surrounding gas, for TNG100 (see a similar quantification in Figure 8 of \citealt{truong.etal.2020}). On the left panel, we show the dependence of the gas column density profile on the total amount of kinetic feedback energy that is injected by the central SMBH over its lifetime. TNG100 galaxies that have a larger amount of accumulated SMBH feedback energy have lower densities, i.e. contain less CGM gas. On the right panel, we show the dependence of the gas column density profiles on the instantaneous accretion rate onto SMBH. In the case of the instantaneous SMBH accretion rate (or luminosity), the trend is reversed: this is expected in the context of the considered simulations, as in the TNG model (as well as in SIMBA) SMBHs in the high accretion rate mode (yellow) exert less efficient feedback (thermal mode) in comparison to the low accretion rate mode i.e. kinetic wind mode (purple). That is, 
%neither the total energy released across all feedback modes, nor 
the instantaneous accretion rate is not expected to explain the rarefaction of halo gas column density for quenched galaxies. Rather, it is the energy injected in the kinetic wind mode, which is able to efficiently couple to the surrounding gas, which impacts and thus correlates with global CGM properties \citep[see also][]{pillepich.etal.2021,ramesh.etal.2023b}.

\begin{figure*}
  \centering
  \includegraphics[width=0.99\textwidth]{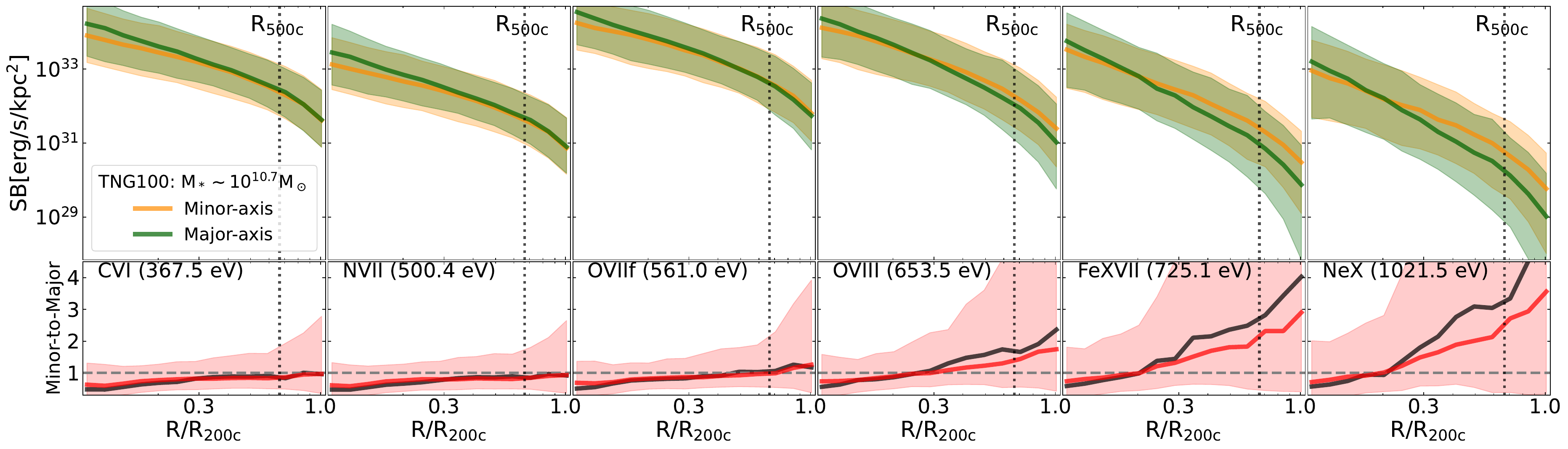}
  \includegraphics[width=0.99\textwidth]{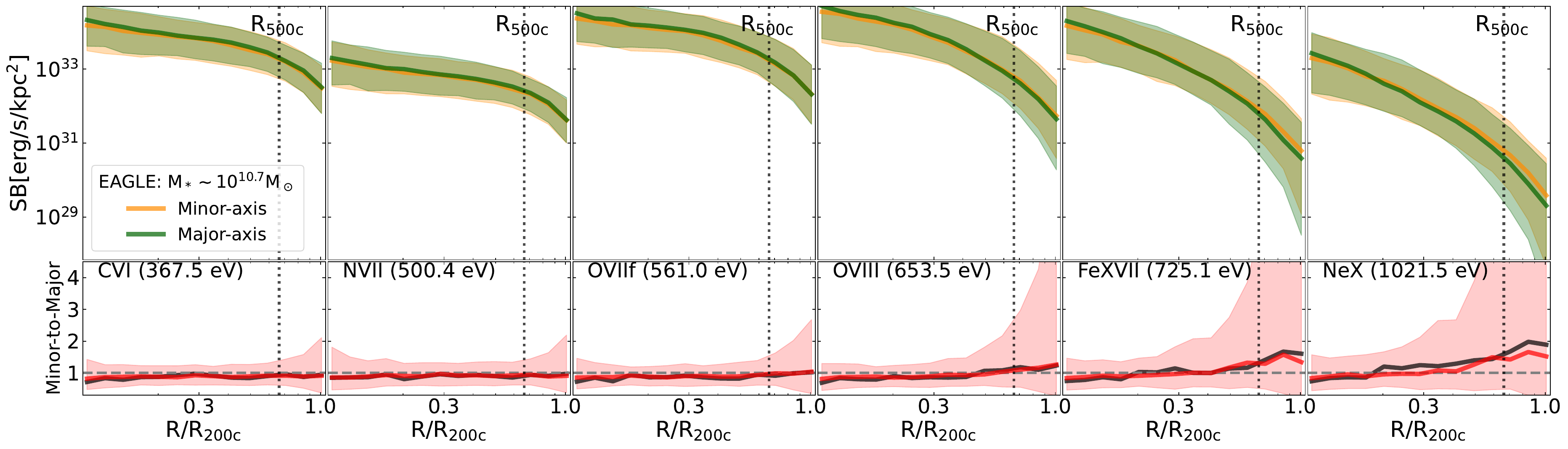}
  \includegraphics[width=0.99\textwidth]{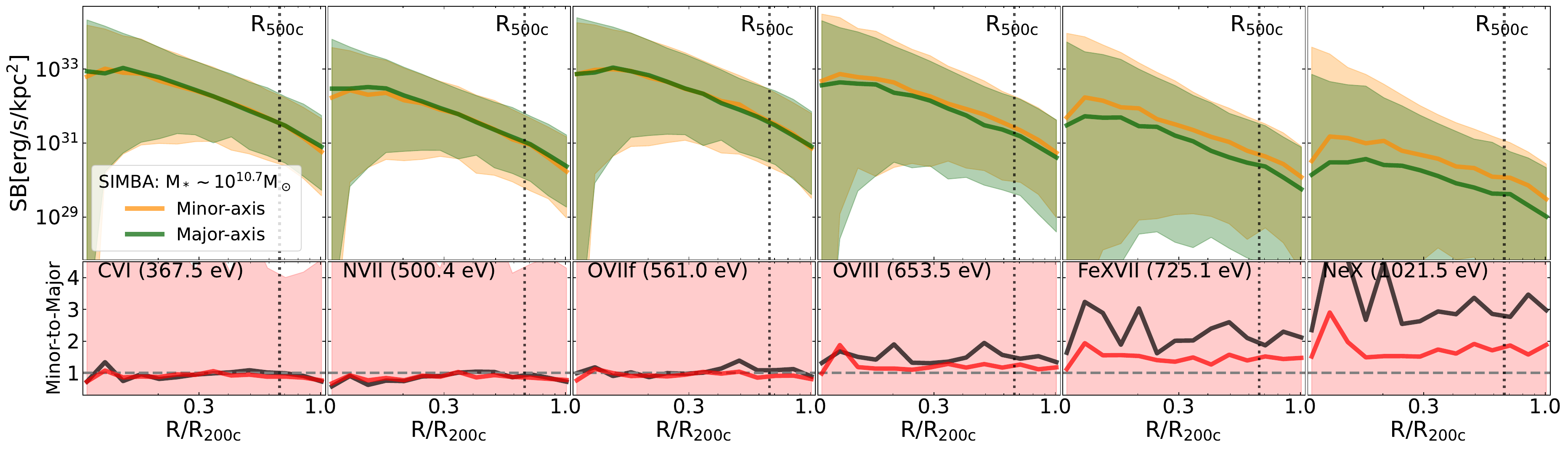}  
  \caption{Predictions for the surface brightness profiles measured along the minor axis versus those along the major axis of galaxies for the three simulations ({\it top} to {\it bottom}). The comparison is shown for various lines from low to high energy ranges ({\it left} to {\it right}). In each comparison plot, the top panel displays the median profiles and their associated 16th-84th percentile galaxy-to-galaxy variation, while the bottom panel shows the ratio of the minor to major median profiles shown in the top panel (black curve), as well as the median profile of galaxy-by-galaxy ratio profiles (red curve) associated by the 16th-84th percentile envelope (red-shaded area). The signals are predicted for galaxies in the MW mass range: ${\MSTARS=10^{10.7\pm0.1}\, \MSUN}$.}
  \label{fig:9}
\end{figure*}

Overall, our results suggest the picture that SMBH feedback has an ejective effect on the CGM, as has been suggested and demonstrated in earlier studies of TNG100 and EAGLE simulations. This effect is also observed in the SIMBA simulation, highlighting a commonality. Despite the differences in their implementations of SMBH growth and feedback, the three simulations all produce a large-scale ejective effect via SMBH feedback. This leads to the dichotomy between star-forming and quiescent galaxies in X-ray line emission. This prediction is in line with previous studies that found a dichotomy in broad band X-ray emission for the inner halos in TNG100 \citep{truong.etal.2020} and EAGLE \citep{oppenheimer.etal.2020}, which also manifests as a dichotomy in OVI column density i.e. tracing the warm-hot phase of the CGM \citep{nelson.etal.2018b}.

%%%%%%%%%%%%%%%%%%%%%%%% Azimuthal %%%%%%%%%%%%%%%%%%%%%%%%%%%%%

\subsection{Azimuthal dependence}
\label{azimuthal_dependence}

We now explore the azimuthal distribution of the CGM properties, which has been found to exhibit striking anisotropic features in a number of recent simulation studies \citep{peroux.etal.2020, truong.etal.2021b,nica.etal.2022}. In general, there is a difference in the CGM physical properties along the minor versus major axes, i.e. as a function of azimuthal angle with respect to the orientation of the stellar body of the central galaxy. We evaluate the angular anisotropy of X-ray line emission with a sample of simulated galaxies with stellar masses within the MW mass range, $\MSTARS=10^{10.7\pm0.1}\, \MSUN$. Surface brightness profiles of CGM line emission are computed for separate regions along the minor and major axis. The minor axis is defined based on the total angular momentum of stellar particles within twice the half-stellar mass radius (${\rm R_e}$), and the CGM is subdivided into evenly-sized quadrants, each spanning an angular degree of 90. These quadrants are delineated along the minor and major axes respectively. The surface brightness profiles are computed along each axis using the gas elements from the two quadrants situated along the corresponding axis.

\begin{figure*}
  \centering
  \includegraphics[width=0.99\textwidth]{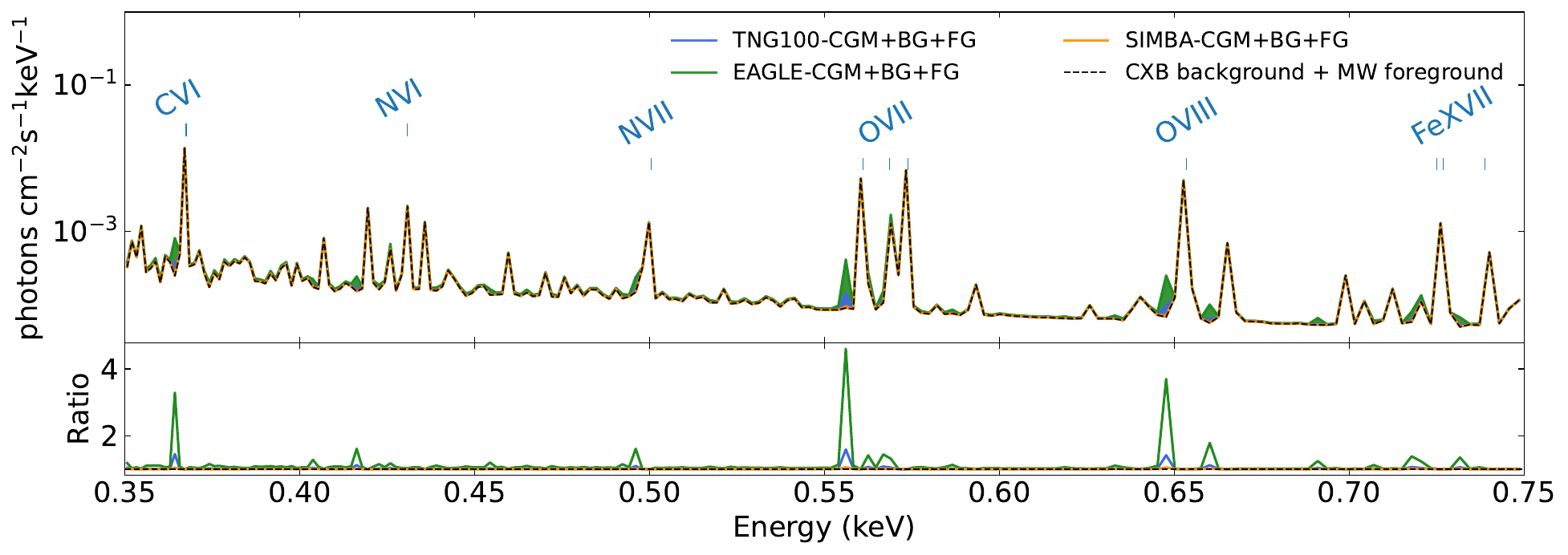}
  \caption{Separation of CGM signals originating from external galaxies (z=0.01) and the emission from the Milky-Way's own CGM. {\it Top:} Median CGM spectra of simulated galaxies in the MW-mass range ($\MSTARS=10^{[10.6-10.8]}\, \MSUN$) shown in a small segment of the X-ray soft band ($[0.35-0.75]$ keV) in comparison with the emission spectrum from the CXB background (BG) and MW foreground (FG). {\it Bottom: } The ratio of the total signals to the background+foreground. The CGM spectra are extracted from a thin annulus at $r=0.5R_{200c}$. The spectra are created with the spectral resolution of $\sim2$ eV.}
  \label{fig:shift_spectra}
\end{figure*}

In Fig.~\ref{fig:9} we present a comparison between the surface brightness profiles computed along the minor axis versus those computed along the major axis in the three simulations. All three simulations predict that the emission in low-energy lines, such as CV, NVII, OVIIf, is largely isotropic, while the emission in higher-energy lines, including OVIII, FeXVII and NeX, is enhanced along the minor axis. However, the level of this CGM angular dependence  varies quantitatively among the three simulations. For instance, TNG100 and SIMBA predict that the emission in FeXVII at ${\rm R_{500c}}$ is significantly higher along the minor axis -- the minor-to-major ratio is a factor of $\sim 3$ -- while the corresponding ratio in EAGLE is approximately half that value. 

The increased emission of high-energy lines along the minor axis may be due to the higher CGM temperature along that axis, caused by SMBH feedback \citep{truong.etal.2021b}. The outflows driven by SMBH feedback can evacuate, heat, and metal enrich the gas along the way, and these outflows occur preferentially along the minor axis, where the outflows propagate along the path of least resistance \citep{nelson.etal.2019a}. However, the predicted level of the anisotropy among the simulations varies, and we speculate this is due to two main factors. First, the efficiency of SMBH feedback in driving outflows: the TNG100 and SIMBA models for SMBH feedback appear to be more efficient in this regard than EAGLE model, which is more efficient in heating up the gas at small radii (e.g ${\rm r<0.3R_{200c}}$, as shown in Fig. \ref{fig:3}). Second, the degree of collimation of SMBH-driven outflows: SIMBA injects SMBH feedback energy explicitly along a bi-polar direction, which is parallel to the angular momentum of the gas particles in the neighbour to the central SMBH (see \citealt{dave.etal.2019} for a more detailed description). Whereas TNG100 and EAGLE inject all (time-averaged) SMBH energy isotropically. Together with the morphological distribution of gas in the surroundings of SMBHs, this undoubtedly affects the degree of outflow collimation, and hence the degree of anisotropy arising in the CGM.

Finally, we also detect an inverse anisotropy of greater emission along the major axis, in the interior (${\rm r\lesssim0.3 R_{200c}}$) of the TNG100 halos, and to a lesser extent in the EAGLE and SIMBA halos. The major axis anisotropy in TNG100 is most pronounced for lower energy lines, implying an augmented presence of cooler gas in this direction.  \citet{nica.etal.2022} found that EAGLE star-forming galaxies above $M_*=10^{10.7}\MSUN$ exhibit a major axis enhancement of the X-ray emission in the soft band ([0.2-1.5] keV) inside 30 kpc, which corresponds to $0.1 R_{200c}$, and is only weakly apparent just above this radius. This elevated major axis emission may be associated with the accretion of $\geq 10^6$ K gas onto disc galaxies \citep{hafen.etal.2022}, or the overall density structure of the inner disk \citep{nelson.etal.2021}, and is explored in more detail for LEM in (\citealt{zuhone.etal.2023}).

%%%%%%%%%%%%%%%%%%%%%%%%%%%%%%%%%%%%%%%%%%%%%%%%%%%%%%%%%%%%%%%%%%%%%%%%%%%%%%%%%%%%%%%%%%%%%%%%

\section{Observability considerations}
\label{sec:obs}

To potentially distinguish between the model differences highlighted previously, we proceed to explore the prospects of observationally detecting X-ray line emission with the LEM mission. Compared to other planned X-ray missions with microcalorimeters onboard, such as XRISM and ATHENA, LEM has a significantly larger field of view and a greater grasp, which is defined as the product of the effective area and the field of view, making it ideal for probing physical properties of the CGM in nearby galaxies \citep{kraft.etal.2022}. Our objective is not to conduct a comprehensive assessment of the LEM capabilities, which would necessitate analysis of LEM mock observations (\citealt{schellenberger.etal.2023}). Rather, we aim to point out potential observations that LEM can make to illuminate the physics of feedback as traced through CGM observables.

\subsection{Observing galaxies in narrow X-ray bands}
\label{sec:narrow_band_obs}

To illustrate the advantage of observing the CGM via line emission, we compare the intrinsic spectrum of an extragalactic CGM with the emission spectrum of the Milky Way CGM. The latter is the main foreground component that contaminates the intrinsic CGM emission of an extragalactic galaxy. 

\begin{figure*}
  \centering
  \includegraphics[width=0.85\textwidth, trim={0cm, 15cm, 0cm, 0cm}]{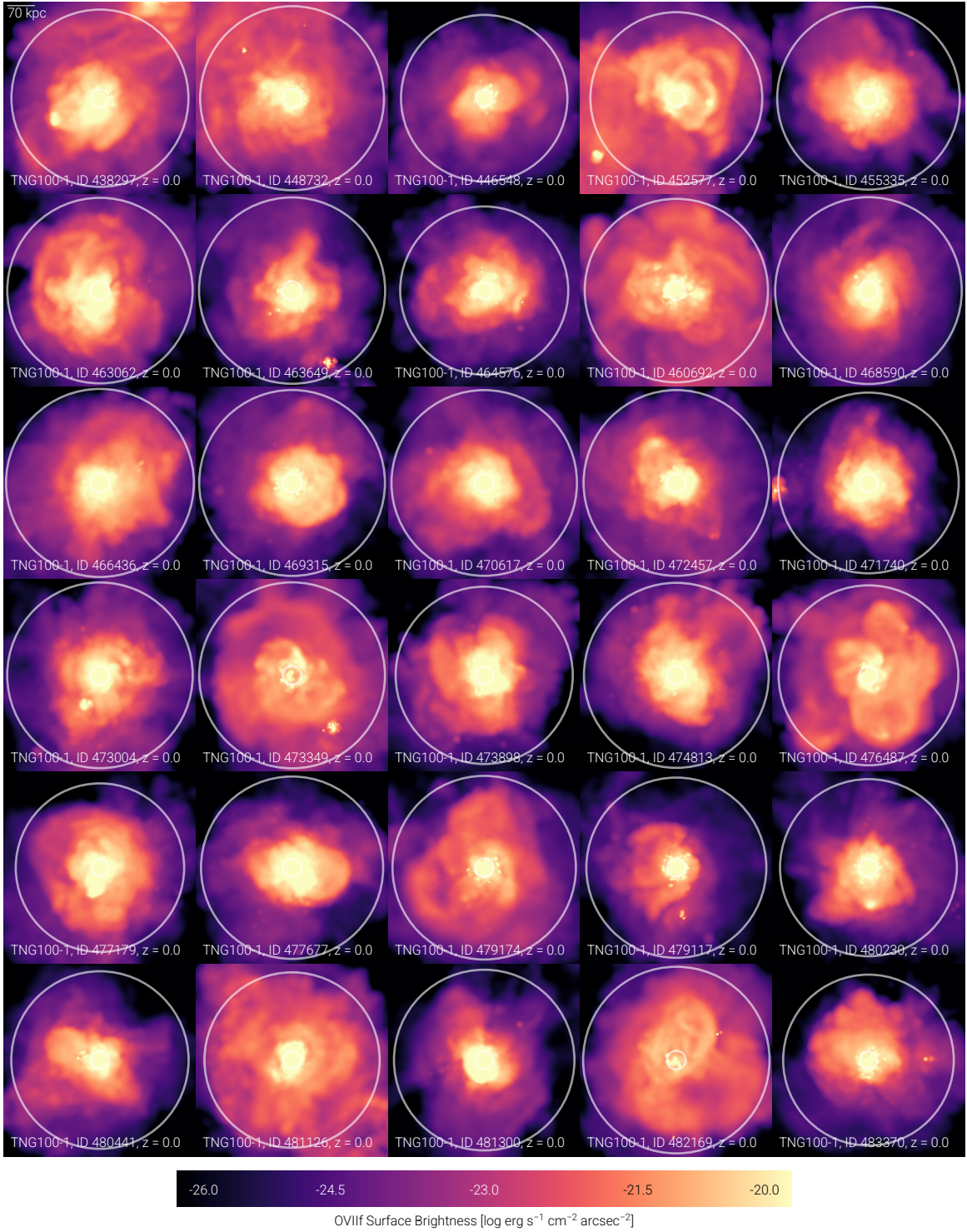}
  \caption{OVIIf surface brightness maps for a sample of star-forming galaxies with MW-like mass from the TNG100 simulation, selected to be blue based on their (g-r) optical colors. The size of each map is 500 kpc per side. The white circles specify the radii of ${\rm 0.1R_{200c}}$ and ${\rm R_{200c}}$.}
  \label{fig:7}
\end{figure*}

\begin{figure*}
  \centering
  \includegraphics[width=0.85\textwidth, trim={0cm, 15cm, 0cm, 0cm}]{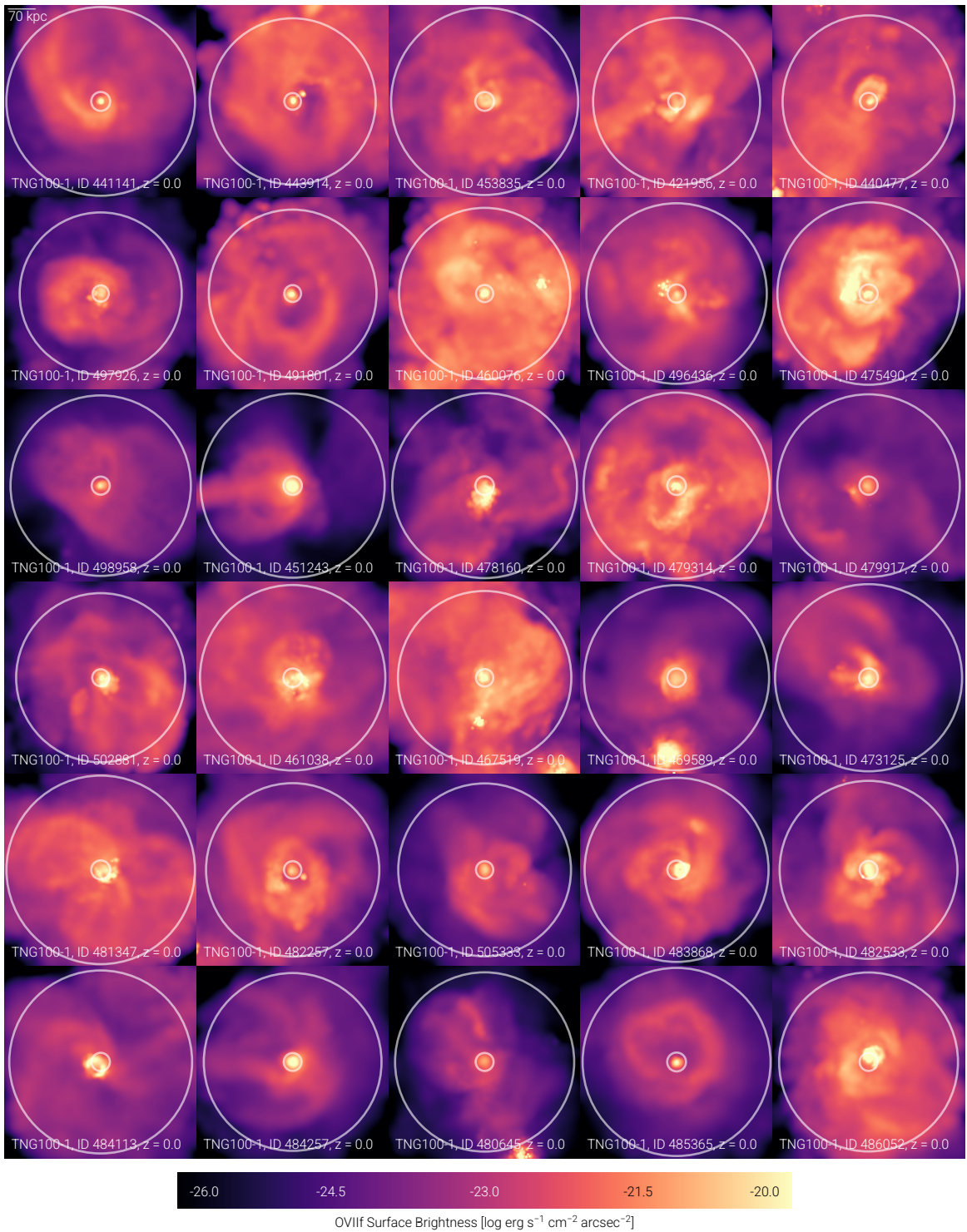}
  \caption{As in Fig.~\ref{fig:7} but for TNG100 quiescent (red) galaxies, selected by their optical (g-r) colors. OVIIf surface brightness maps, each 500 kpc across. The white circles specify the radii of ${\rm 0.1R_{200c}}$ and ${\rm R_{200c}}$.} 
  \label{fig:8}
\end{figure*}

In Fig.~\ref{fig:shift_spectra} we show X-ray spectra of different components: the cosmic X-ray background (CXB) + Milky-Way foreground, and the intrinsic CGM emission from galaxies of MW-like mass from TNG100, EAGLE, and SIMBA. The CXB background is estimated assuming emission from a power-law spectral source with a spectral index of 1.47 (see e.g. \citealt{lehmer.etal.2012}). The MW foreground emission is modelled as a sum of 2 thermal components: i) the local hot bubble as a thermal plasma with ${\rm k_BT\sim0.99}$ keV and metallicity ${z\sim1.0\ Z_\odot}$; ii) the hot halo as a 2-temperature plasma with ${\rm k_BT\sim0.225}$ keV and ${\rm k_BT\sim0.7}$ keV. Both are assumed to have a metallicity of ${\rm Z\sim1.0\ Z_\odot}$ (see e.g.  \citealt{McCammon.etal.2002} and \citealt{bluem.etal.2022}). The CGM emission of external galaxies is taken from a thin annulus at $r=(0.5\pm0.1)R_{\rm 200c}$ and computed for a sample of simulated galaxies with stellar mass $\MSTARS=10^{10.6-10.8}\, \MSUN$ placed at a redshift ${z=0.01}$. Of these, we show the sample-median spectra of their CGM emission predicted by each of the three simulations plus the foreground (MW-CGM) and background (CXB) emission. The total emission is largely dominated by the CXB background and the MW-CGM foreground across the soft X-ray band. However, thanks to the line redshifting, the intrinsic CGM emission is visible at certain lines including CVI, OVIIf, OVIII, and FeXVII (denoted by shaded areas in the plot). The strength of those lines varies between the simulations. For instance, as expected from the previous figures, the OVIIf strength in the EAGLE simulation is the highest and it is larger than the CXB background+MW-CGM foreground level by a factor of $\sim4$. The TNG100 line strength is about $50$ per cent of the background+foreground level. SIMBA produces almost no significant OVIIf line compared to the background+foreground level (a few percent). This result implies that the detectability of CGM in emission via metal lines strongly depends on the underlying simulation model.

\begin{figure*}
  \centering
  \includegraphics[width=0.99\textwidth]{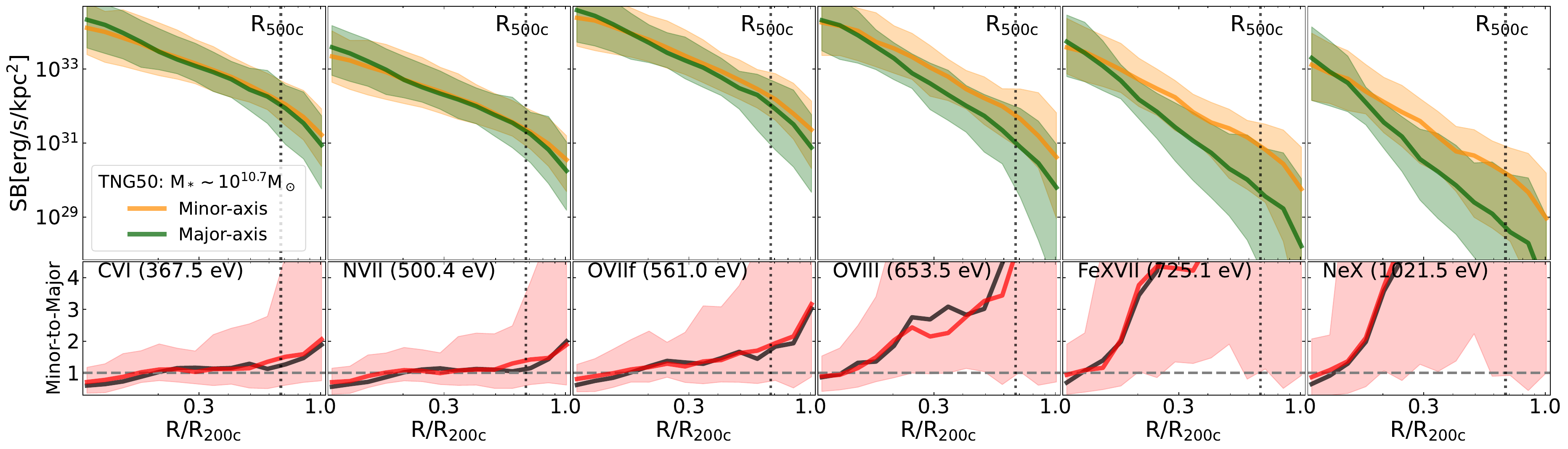}
  \caption{The surface brightness profiles of various lines measured along the minor axis versus those along the major axis of TNG50 galaxies in the MW-mass range. The plot format and notations are identical to those in Fig. \ref{fig:9}.}
  \label{fig:tng50}
\end{figure*}

In this illustrating exercise our primary focus is to study galaxies at a redshift of z=0.01, where the redshifting of lines starts to separate few of them from the MW foreground. Observations at higher redshifts could reveal additional lines. For instance, observations of galaxies at z=0.035 could unveil the entire OVII triplets. However, this would be at the expense of a corresponding decrease in surface brightness. On the other hand, observing galaxies at lower redshifts would limit the number of detectable lines. Nonetheless, such scenario may permit studies of the inner part of the CGM, since its X-ray emission could be bright enough to overcome the MW foreground. Such observations would be useful to study the dynamical and chemical properties of the inner CGM as well as its interplay with the galactic disks.

\subsection{The dichotomy between star-forming and quiescent galaxies}

As shown in Section~\ref{lx_dichotomy}, there is a predicted dichotomy between star-forming versus quiescent galaxies in intrinsic X-ray line emission. With a future large field of view X-ray telescope like LEM, this prediction can be observationally tested via surface brightness mapping on a sample of galaxies each imaged individually. For illustration, in Fig.~\ref{fig:7} and Fig.~\ref{fig:8} we show surface brightness maps of OVIIf for a set of 30 star-forming and 30 quiescent TNG100 galaxies, respectively. Overall, it is visually clear from the maps that in star-forming galaxies, the OVII emission is remarkably higher than quiescent counterparts. More importantly, the maps clearly show that while star-forming galaxies have centrally peaked distributions of OVIII abundance, the quiescent counterparts exhibit much less concentrated distributions. The contrast between the two populations is a visual reflection of the ejective effect of SMBH feedback at $100$-kpc scales, as discussed in Section~\ref{lx_dichotomy}.

\begin{figure*}
  \centering
  \includegraphics[width=0.99\textwidth]{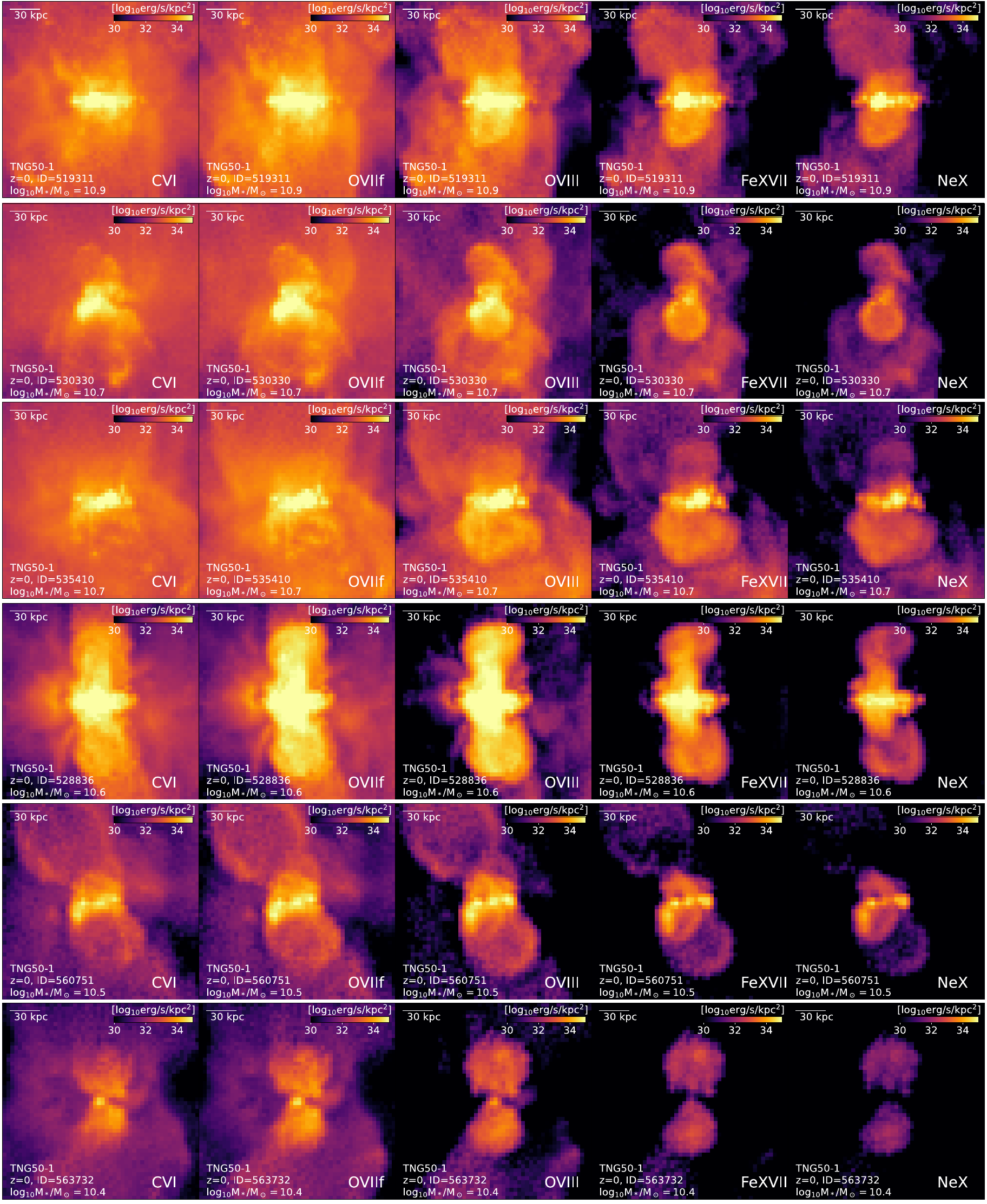}
  \caption{Azimuthal dependence of the X-ray line emission from the CGM for six simulated MW-like galaxy in the TNG50 simulation (one row, one galaxy). The size of each map is 200 kpc per side, pixels are 4 kpc per side, and all the maps are oriented edge-on with respect to the galactic disk, placed horizontally at the center. From {\it left} to {\it right}, we show the intrinsic emission of six lines in energy-increasing order: CVI, OVIIf, OVIII, FeXVII, and NeX.}
  \label{fig:13}
\end{figure*}

\subsection{Detecting X-ray bubbles in the CGM}

The result presented in Section \ref{azimuthal_dependence} suggests that the CGM exhibits a notable degree of anisotropy in its line emission, with an increasing level of angular dependence at higher energies. More specifically, the three simulations predict that the emission of heavy-element lines, such as FeXVII and NeX, is more pronounced along the minor axis compared to the major axis of the central galaxy's stellar body. We note that this result is a statistical finding, that is manifest at the galaxy population level. In this Section, our objective is to explore how the anisotropic signals might be observed at the individual galaxy level. For this purpose, we utilise simulated galaxies taken from TNG model, as it predicts the most robust signal among the three simulation especially at extended radii. Additionally, we employ the TNG50 simulation \citep{nelson.etal.2019b, pillepich.etal.2019} due to its superior resolution among TNG simulations (2.5 times better spatial resolution than TNG100), making it ideally suited for examining the spatial distribution of the CGM.

To explore whether this anisotropic signal can be seen also for individual galaxies, we again use a sample of MW-like galaxies with stellar masses ${\MSTARS\sim10^{10.7} M_\odot}$, in this case from the TNG50 simulation. We first examine, as shown in Fig.~\ref{fig:tng50}, the azimuthal dependence of the averaged radial profiles of lines emission in those TNG50 MW-like galaxies, an exercise similar to the one presented in Section~\ref{azimuthal_dependence}. Analogous to TNG100, TNG50 predicts qualitatively consistent anisotropic signals in the lines emission, displaying enhanced emission along the minor axis compared to the major axis, which becomes increasingly pronounced with higher energy lines. We note, however, that the quantitative level of anisotropy in TNG50, indicated by the minor-to-major ratio, is larger than in TNG100 by a factor of 2 or 3 depending on the lines, despite the significantly larger scatter in TNG50. The quantitative difference between the predicted anisotropy levels in TNG50 and TNG100 could be explained by: i) The smaller sample in TNG50, which returns 58 galaxies at the chosen MW-mass range compared to 410 galaxies in TNG100; ii) The resolution difference between the two simulations; or the combination of both. This, however, lies beyond the scope of the current study.

In Fig.~\ref{fig:13} we show X-ray surface brightness maps of a selection of 6 TNG50 MW-like galaxies that exhibit bubbles in their CGM reminiscent of the eROSITA/Fermi bubbles in the Milky Way \citep{predehl.etal.2020}. A comprehensive study of X-ray bubbles in TNG50 can be found in \cite{pillepich.etal.2021}. The maps are presented for an edge-on orientation and displayed in various lines, from low to high energy. Each map covers an area of $200$ kpc per side, which falls well within the LEM field of view when observing a galaxy at redshift $z=0.01$. It is evident from the maps that the bubbles appear to be more prominent at progressively high energy lines. The existence of the bubbles is a strong manifestation of the anisotropic signal in CGM line emission, but at the individual galaxy level. However, the presence of an anisotropy signal does not necessarily imply the existence of bubbles, as the latter requires a sufficiently strong outflow to create shock fronts, which are the defining boundaries where a discontinuity in gas temperature and other gas properties occurs (see e.g. Figure 6 in \citealt{pillepich.etal.2021}).

\begin{figure*}
  \centering
  \includegraphics[width=0.36\textwidth]{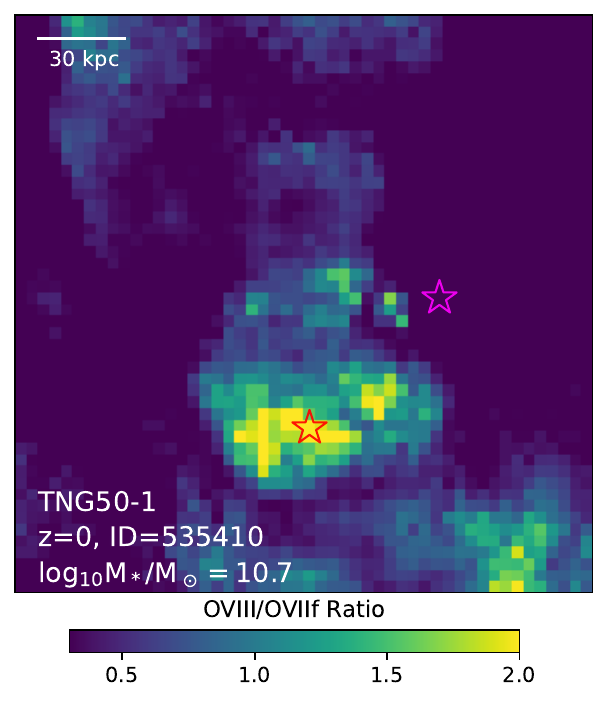}
  \includegraphics[width=0.52\textwidth]{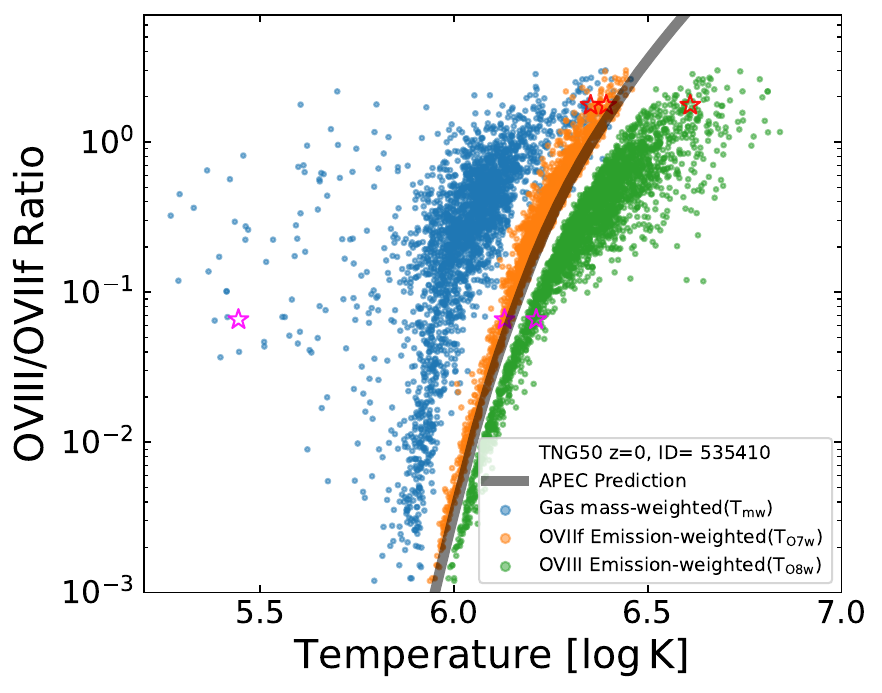}
  \includegraphics[width=0.49\textwidth]{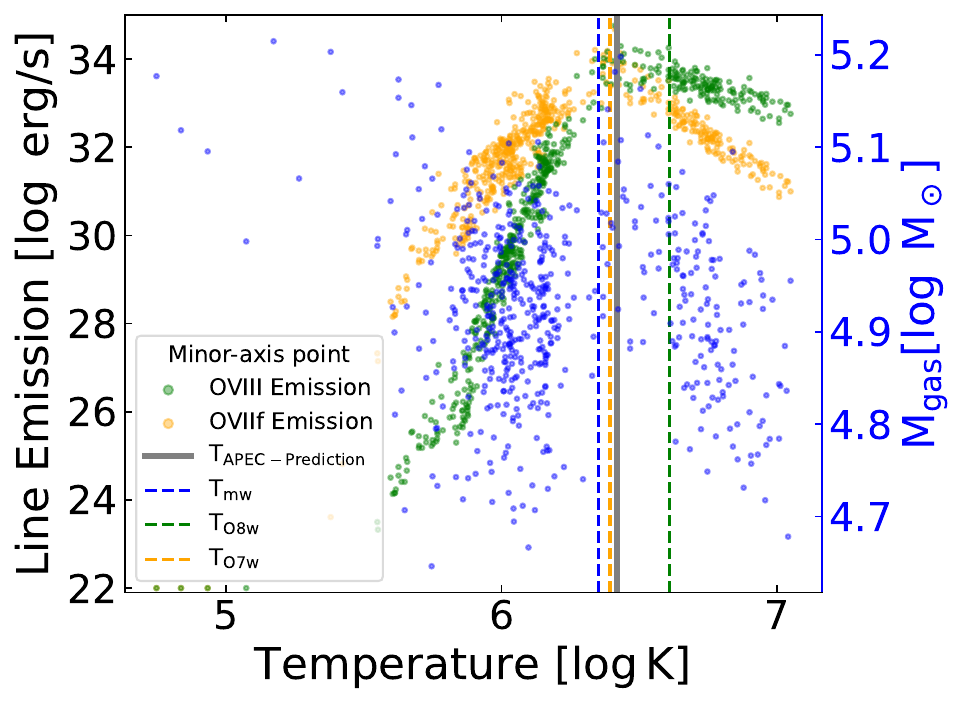}
  \includegraphics[width=0.49\textwidth]{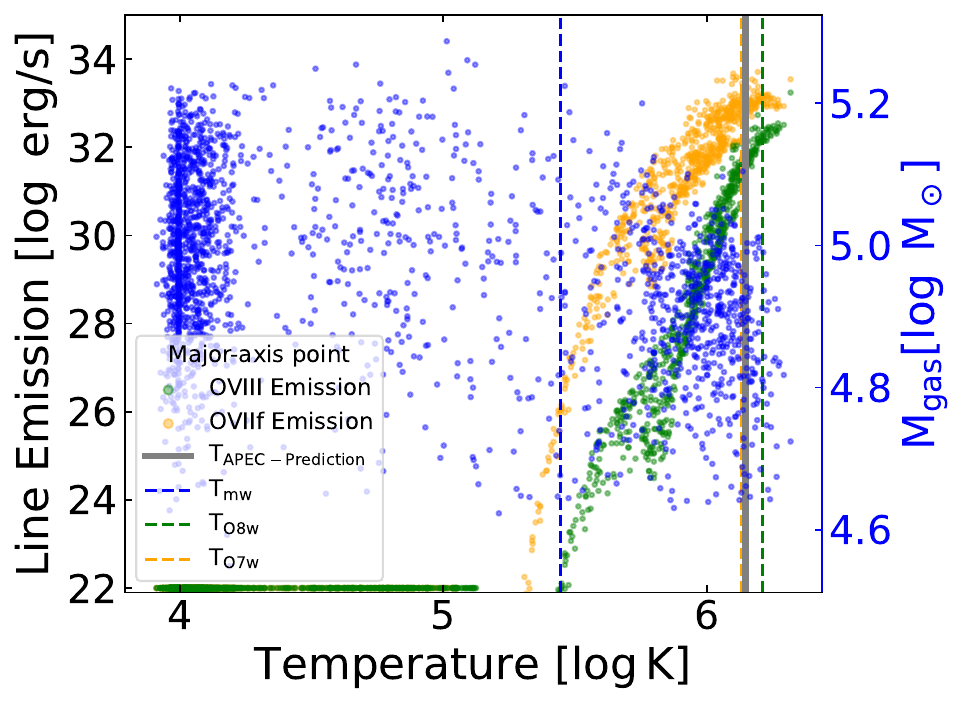}
  \caption{Relationship between the OVIII/OVIIf line ratio and various estimates of the CGM temperature for a single TNG50 MW-like galaxy. {\it Top left:} The OVIII/OVIIf ratio map, spanning an area of 200 kpc per side, with pixels of 4 kpc a side, and with an edge-on orientation, i.e. the galactic disk is oriented horizontally in the center. The pink and red stars mark two specific regions, each with an area of $\rm{(10kpc)}^2$, where we extract gas cells for further analysis below. {\it Top right:} The line ratio-temperature diagram for individual pixels in the top-left map, in which the pixels temperature are obtained from various mass/emission-weighted estimates (blue, orange, and green), and the solid gray line is the prediction from the 1T APEC model. The stars specify the location on this diagram for the two extracted regions, and from left-to-right order they represent the gas mass-weighted, OVIIf emission-weighted, and OVIII emission-weighted temperatures. {\it Bottom row:} In the two extracted regions, the relationship between OVIIf and OVIII emission, and mass, as a function of gas temperature. One region is along the minor axis ({\it left}), while one is along the major axis ({\it right}). For each plot, the data points represent the gas cells belonging to the extracted region, and the vertical lines denote various average temperatures of the region. Points on the x-axis correspond to gas cells with zero line emission in either the OVIIf or OVIII line.}
  \label{fig:5}
\end{figure*}

The observation of X-ray bubbles beyond our own Milky Way \citep{predehl.etal.2020} will be an important target for LEM \citep{kraft.etal.2022}. The presence of bubbles provides direct, spatially resolved evidence for galactic feedback processes impacting the CGM. The morphology and extent of bubbles encodes information on the physical nature of feedback processes operating within the CGM (see also \citealt{pillepich.etal.2021} for a comprehensive discussion). Therefore, the detection (or non-detection) of these X-ray bubbles represents a critical test for current models of galaxy formation.

\subsection{Inferring gas temperature from line ratios}

As soft X-ray lines of different transition energies depend on the temperature of the emitting gas, the ratios of different emission lines constraints the thermal properties of the CGM. That is, emission line ratios are an observable proxy for temperature (e.g. \citealt{kaastra.etal.2008, chakraborty.etal.2020, chakraborty.etal.2022}). 

To study this probe, we take the single TNG50 galaxy for which X-ray emission maps were shown in Fig.~\ref{fig:13} (ID 535410, third row from the top). In the top left panel of Fig.~\ref{fig:5} we present the map of the OVIII/OVIIf ratio for this galaxy, which is oriented edge-on and covers an area similar to that of the maps in Fig.~\ref{fig:13}. It is apparent that the line ratio map qualitatively resembles the emission maps in heavy-element lines such as FeXVII and NeX. Specifically, there are dome-like features located above and below the galactic disk, within which the line ratio is significantly higher than surrounding values. Additionally, we find that the ratio map correlates strongly with the map of intrinsic mass-weighted gas temperature (not shown). 

To examine how accurately the line ratio reflects the CGM temperature, we show in the top right panel of Fig.~\ref{fig:5} the line ratio-temperature diagram for all individual pixels of the map. We weight by gas mass ${\rm T_{mw}}$ (blue), OVIIf emission (orange), and OVIII emission (green). In addition, the predicted relation between the OVIII/OVIIf ratio and gas temperature based on the APEC 1-temperature model is shown as the solid gray line in the plot. The plot reveals that the APEC-predicted temperature, which is derived from the OVIII/OVIIf ratio, overestimates the gas mass-weighted temperature, and underestimates the OVIII emission-weighted temperature, by roughly $0.2$ dex in each case. On the other hand, the APEC-predicted temperature tracks the OVIIf emission-weighted estimate reasonably well. This result arises from the multi-temperature nature of the CGM, also within small probed areas, by biasing ${\rm T_{mw}}$ low, whereas the line ratio OVIII/OVIIf is mainly sensitive to the gas component responsible for most of the OVIIf emission.

In the bottom panels of Fig.~\ref{fig:5} we show the line emission (orange and green), as well as gas mass (blue), versus temperature for gas cells obtained from the two extraction regions indicated by star symbols on the ratio map, one along the minor axis and the other along the major axis, each region covers a (10kpc x 10kpc x 400kpc) volume}. We find that the difference between the gas mass and line emission distributions determines the ability of the APEC-predicted temperatures to recover the mass-weighted values. In the selected minor-axis region (bottom left panel), the line emission distribution, both in OVIIf and OVIII, peaks around a temperature similar to that of the gas mass distribution. As a consequence, the APEC-predicted temperature is quite close to the mass-weighted value (within $0.1$ dex). On the other hand, for the major axis region (bottom right panel), the gas mass distribution is skewed toward lower temperatures (${\rm T\sim 10^{4}K}$) resulting in a much lower mass-weighted temperature than the APEC-predicted or emission-weighted temperatures. 

To conclude, this exercise demonstrates the potential of deriving fairly precise measurements of the underlying physical temperature of the X-ray emitting CGM through the extraction of individual line ratios in spatially-resolved regions. The temperature of the hot phase CGM, as characterised by emission-weighted estimates, can be well recovered within 0.2 dex. It is worth noting that the results in this Section are obtained based on the assumption that the CGM is in the state of collisional ionisation equilibrium and without considering potential impacts of photoionisation. Studies on non-equilibrium ionisation effects, e.g. \cite{gnat.sternberg.2007,oppenheimer.schaye.2013, sarkar.etal.2022}, have indicated that the cooling efficiency of both OVII and OVIII ion species could be significantly impacted depending on the specific background radiation present.

This implies that in cases of non-equilibrium ionisation, a CIE-based derived temperature based on the OVIII/OVIIf ratio could potentially be misestimated. We postpone a quantitative investigation of the implications of non-equilibrium ionisation as well as photoionisation on the line diagnostics of CGM temperature for a future work.

%%%%%%%%%%%%%%%%%%%%%%%%%%%%%%%%%%%%%%%%%%%%%%%%%%%%%%%%%%%%%%%%%%%%%%%%%%%%%%%%

\section{Summary and conclusions}
\label{sec:conclusion}

Future X-ray imaging microcalorimeter missions with high spatial and spectral resolution such as the Line Emission Mapper (LEM) will, for the first time, detect and characterize the hot circumgalcatic medium (CGM) of galaxies via X-ray metal line emission. In this paper, we utilise three modern cosmological hydrodynamical simulations of galaxy formation (TNG100, EAGLE, and SIMBA) to make predictions for the spatial distribution of soft band X-ray metal line emission in the CGM of galaxies over a wide range of stellar masses ${\rm 10^{10-11.6}\,M_\odot}$ at $z=0$. We focus on the effects of supermassive black hole (SMBH) feedback on the CGM, the observability of the signal, and its ability to constrain the physics of galactic feedback and gaseous atmospheres. Our main results are:

\begin{enumerate}
    \item The three simulations show significant diversity in the thermodynamical properties of the CGM, as well as in the radial profiles of X-ray line emission (Fig.~\ref{fig:3}, \ref{fig:4}). SIMBA predicts the lowest overall emission, due to lower CGM gas content than in TNG100 or EAGLE. This reflects the different efficiencies and ejective strengths of the SMBH feedback models across these three simulations.
    
    \item For galaxies at the Milky-Way mass scale, all three simulations consistently indicate a clear correlation between CGM line emission and star formation activity: star-forming galaxies are significantly more luminous compared to their quiescent counterparts (Fig.~\ref{fig:6b}) over a broad radial range up to ${\rm R_{500c}}$. This dichotomy between star-forming versus quiescent galaxies is caused by SMBH feedback that expels gas from the central regions of galaxies and halos as the quenching process proceeds (Fig. \ref{fig:SMBH_Dens}).
    
    \item At the population level, the three simulations predict that the spatial distribution of CGM line emission is largely anisotropic, differing along the major versus minor axis directions with respect to the orientation of the central galaxy. This anisotropy is maximal for high-energy lines, such as FeXVII or NeX: emission is enhanced along the minor axis (Fig.~\ref{fig:9}). However, the magnitude of the angular signal varies quantitatively between the simulations: TNG100 predicts the strongest signal, while EAGLE predicts the weakest one. 
\end{enumerate}

Finally, we show that future X-ray missions will provide sufficient spectral resolution (${\rm \sim eV}$) to disentangle an extragalactic CGM signal from the Milky-Way foreground, thanks to the line redshifitng (Fig.~\ref{fig:shift_spectra}). In particular, the currently planned LEM mission, which is designed to have a large field of view and advanced spectral resolution \citep{kraft.etal.2022}, will have excellent mapping capability to explore many physical aspects of the CGM:
    
\begin{enumerate}
    \item {\bf Star-forming versus quiescent dichotomy.} By mapping the surface brightness in X-ray line emission around star-forming and quiescent galaxies at similar masses, LEM will test the dichotomy between the two populations predicted to exist out to large radii ($\sim100-200$ kpc) by these cosmological simulations ({Fig. \ref{fig:7}, \ref{fig:8}}).
    \item {\bf Detection of X-ray bubbles.} The anisotropic signal observed at the population level may manifest as X-ray bubbles around individual galaxies at the Milky-Way mass scale, as predicted by TNG. LEM will detect, map, and measure the physical properties of these bubbles for the first time. The whole extent of the bubbles and inner CGM ($\sim40-50$ kpc) are comfortably within its single pointing field of view (Fig. \ref{fig:13}).
    \item {\bf Mapping the CGM thermal structure.} The ratio between emission lines, such as OVIII/OIIf, offers a powerful diagnostic of CGM temperature (Fig.~\ref{fig:5}). The ratio serves as a good proxy for the emission-weighted temperature. 
\end{enumerate}

Although the cosmological simulations studied in this paper employ different physical models for SMBH feedback, they predict a number of qualitative similarities for the impact of SMBH feedback on the CGM. First, gas is physically ejected. Feedback can expel gas out of galaxies and even their halos, particularly as a result of the quenching of star formation in massive galaxies. Second, halo gas experiences anisotropic outflows. Feedback-driven outflows tend to move and heat gas preferentially along the minor axis of galaxies. This angular anistropy is reflected in observable X-ray line emission. The next generation of imaging microcalorimeter-based X-ray missions, including LEM, will provide new opportunities to study the CGM through its narrow-band emission. These will reveal the fundamental connection between galactic feedback and the circumgalactic medium, a key missing link in our understanding of the cosmic baryon cycle and thus galaxy evolution as a whole. 

\section*{Data Availability}

The IllustrisTNG simulations are publicly available and accessible at \url{www.tng-project.org/data} \citep{nelson.etal.2019a}. Similarly, data from the Illustris and EAGLE projects are available at \url{www.illustris-project.org} \citep{nelson15} and e.g. \url{eagle.strw.leidenuniv.nl} \citep{Mcalpine.etal.2016}, respectively. The SIMBA data is publicly available at \url{simba.roe.ac.uk}. We note that for this study we have not used the original versions of EAGLE and SIMBA, but instead have used versions thereof re-processed into a TNG-like format, enabling an apples-to-apples comparison. Data directly related to this publication and its figures are available on request from the corresponding author.

\section*{Acknowledgements}

AP and NT thank Elad Zinger for useful conversations. NT thanks Nastasha Wijers for useful discussions. DN acknowledges funding from the Deutsche Forschungsgemeinschaft (DFG) through an Emmy Noether Research Group (grant number NE 2441/1-1) and NT and AP acknowledge funding by the Deutsche Forschungsgemeinschaft (DFG, German Research Foundation) -- Project-ID 138713538 -- SFB 881 (``The Milky Way System'', subproject A01). NW is supported by the GACR grant 21-13491X. AB, WF acknowledge support from the Smithsonian Institution and the Chandra High Resolution
Camera Project through NASA contract NAS8-03060. The material is based upon work supported by NASA under award number 80GSFC21M0002. The primary TNG simulations were carried out with compute time granted by the Gauss Centre for Supercomputing (GCS) under Large-Scale Projects GCS-ILLU and GCS-DWAR on the GCS share of the supercomputer Hazel Hen at the High Performance Computing Center Stuttgart (HLRS). Additional simulations and analyses had been carried out on the Isaac machine of the Max Planck Institute for Astronomy (MPIA) and on the other systems at the Max Planck Computing and Data Facility (MPCDF). 

\bibliographystyle{mnras}
\bibliography{refs}

\begin{thebibliography}{}
\makeatletter
\relax
\def\mn@urlcharsother{\let\do\@makeother \do\$\do\&\do\#\do\^\do\_\do\%\do\~}
\def\mn@doi{\begingroup\mn@urlcharsother \@ifnextchar [ {\mn@doi@}
  {\mn@doi@[]}}
\def\mn@doi@[#1]#2{\def\@tempa{#1}\ifx\@tempa\@empty \href
  {http://dx.doi.org/#2} {doi:#2}\else \href {http://dx.doi.org/#2} {#1}\fi
  \endgroup}
\def\mn@eprint#1#2{\mn@eprint@#1:#2::\@nil}
\def\mn@eprint@arXiv#1{\href {http://arxiv.org/abs/#1} {{\tt arXiv:#1}}}
\def\mn@eprint@dblp#1{\href {http://dblp.uni-trier.de/rec/bibtex/#1.xml}
  {dblp:#1}}
\def\mn@eprint@#1:#2:#3:#4\@nil{\def\@tempa {#1}\def\@tempb {#2}\def\@tempc
  {#3}\ifx \@tempc \@empty \let \@tempc \@tempb \let \@tempb \@tempa \fi \ifx
  \@tempb \@empty \def\@tempb {arXiv}\fi \@ifundefined
  {mn@eprint@\@tempb}{\@tempb:\@tempc}{\expandafter \expandafter \csname
  mn@eprint@\@tempb\endcsname \expandafter{\@tempc}}}

\bibitem[\protect\citeauthoryear{{Anders} \& {Grevesse}}{{Anders} \&
  {Grevesse}}{1989}]{anders.grevesse.1989}
{Anders} E.,  {Grevesse} N.,  1989, \mn@doi [\gca]
  {10.1016/0016-7037(89)90286-X}, \href
  {https://ui.adsabs.harvard.edu/abs/1989GeCoA..53..197A} {53, 197}

\bibitem[\protect\citeauthoryear{{Anderson}, {Bregman}  \& {Dai}}{{Anderson}
  et~al.}{2013}]{anderson.etal.2013}
{Anderson} M.~E.,  {Bregman} J.~N.,   {Dai} X.,  2013, \mn@doi [\apj]
  {10.1088/0004-637X/762/2/106}, \href
  {https://ui.adsabs.harvard.edu/abs/2013ApJ...762..106A} {762, 106}

\bibitem[\protect\citeauthoryear{{Anderson}, {Gaspari}, {White}, {Wang}  \&
  {Dai}}{{Anderson} et~al.}{2015}]{anderson.etal.2015}
{Anderson} M.~E.,  {Gaspari} M.,  {White} S. D.~M.,  {Wang} W.,   {Dai} X.,
  2015, \mn@doi [\mnras] {10.1093/mnras/stv437}, \href
  {https://ui.adsabs.harvard.edu/abs/2015MNRAS.449.3806A} {449, 3806}

\bibitem[\protect\citeauthoryear{{Ayromlou}, {Nelson}  \&
  {Pillepich}}{{Ayromlou} et~al.}{2022}]{ayromlou.nelson.pillepich.2022}
{Ayromlou} M.,  {Nelson} D.,   {Pillepich} A.,  2022, arXiv e-prints, \href
  {https://ui.adsabs.harvard.edu/abs/2022arXiv221107659A} {p. arXiv:2211.07659}

\bibitem[\protect\citeauthoryear{{Babyk}, {McNamara}, {Nulsen}, {Hogan},
  {Vantyghem}, {Russell}, {Pulido}  \& {Edge}}{{Babyk}
  et~al.}{2018}]{babyk.etal.2018}
{Babyk} I.~V.,  {McNamara} B.~R.,  {Nulsen} P.~E.~J.,  {Hogan} M.~T.,
  {Vantyghem} A.~N.,  {Russell} H.~R.,  {Pulido} F.~A.,   {Edge} A.~C.,  2018,
  \mn@doi [\apj] {10.3847/1538-4357/aab3c9}, \href
  {https://ui.adsabs.harvard.edu/abs/2018ApJ...857...32B} {857, 32}

\bibitem[\protect\citeauthoryear{{Barret} et~al.,}{{Barret}
  et~al.}{2016}]{Barret.etal.2016}
{Barret} D.,  et~al., 2016, in {den Herder} J.-W.~A.,  {Takahashi} T.,
  {Bautz} M.,  eds,  Society of Photo-Optical Instrumentation Engineers (SPIE)
  Conference Series Vol. 9905, Space Telescopes and Instrumentation 2016:
  Ultraviolet to Gamma Ray. p. 99052F (\mn@eprint {arXiv} {1608.08105}),
  \mn@doi{10.1117/12.2232432}

\bibitem[\protect\citeauthoryear{{Barret} et~al.,}{{Barret}
  et~al.}{2018}]{barret.etal.2018}
{Barret} D.,  et~al., 2018, in {den Herder} J.-W.~A.,  {Nikzad} S.,
  {Nakazawa} K.,  eds,  Society of Photo-Optical Instrumentation Engineers
  (SPIE) Conference Series Vol. 10699, Space Telescopes and Instrumentation
  2018: Ultraviolet to Gamma Ray. p. 106991G (\mn@eprint {arXiv} {1807.06092}),
  \mn@doi{10.1117/12.2312409}

\bibitem[\protect\citeauthoryear{{Bertone}, {Schaye}, {Dalla Vecchia}, {Booth},
  {Theuns}  \& {Wiersma}}{{Bertone} et~al.}{2010}]{bertone.etal.2010}
{Bertone} S.,  {Schaye} J.,  {Dalla Vecchia} C.,  {Booth} C.~M.,  {Theuns} T.,
   {Wiersma} R. P.~C.,  2010, \mn@doi [\mnras]
  {10.1111/j.1365-2966.2010.16932.x}, \href
  {https://ui.adsabs.harvard.edu/abs/2010MNRAS.407..544B} {407, 544}

\bibitem[\protect\citeauthoryear{{Bluck}, {Piotrowska}  \& {Maiolino}}{{Bluck}
  et~al.}{2023}]{bluck.etal.2023}
{Bluck} A. F.~L.,  {Piotrowska} J.~M.,   {Maiolino} R.,  2023, \mn@doi [\apj]
  {10.3847/1538-4357/acac7c}, \href
  {https://ui.adsabs.harvard.edu/abs/2023ApJ...944..108B} {944, 108}

\bibitem[\protect\citeauthoryear{{Bluem} et~al.,}{{Bluem}
  et~al.}{2022}]{bluem.etal.2022}
{Bluem} J.,  et~al., 2022, \mn@doi [\apj] {10.3847/1538-4357/ac8662}, \href
  {https://ui.adsabs.harvard.edu/abs/2022ApJ...936...72B} {936, 72}

\bibitem[\protect\citeauthoryear{{Bogd{\'a}n} \& {Gilfanov}}{{Bogd{\'a}n} \&
  {Gilfanov}}{2011}]{bogdan.etal.2011}
{Bogd{\'a}n} {\'A}.,  {Gilfanov} M.,  2011, \mn@doi [\mnras]
  {10.1111/j.1365-2966.2011.19608.x}, \href
  {https://ui.adsabs.harvard.edu/abs/2011MNRAS.418.1901B} {418, 1901}

\bibitem[\protect\citeauthoryear{{Bogd{\'a}n} \& {Goulding}}{{Bogd{\'a}n} \&
  {Goulding}}{2015}]{Bogdan.Goulding.2015}
{Bogd{\'a}n} {\'A}.,  {Goulding} A.~D.,  2015, \mn@doi [\apj]
  {10.1088/0004-637X/800/2/124}, \href
  {https://ui.adsabs.harvard.edu/abs/2015ApJ...800..124B} {800, 124}

\bibitem[\protect\citeauthoryear{{Bogd{\'a}n} et~al.,}{{Bogd{\'a}n}
  et~al.}{2013}]{bogdan.etal.2013}
{Bogd{\'a}n} {\'A}.,  et~al., 2013, \mn@doi [\apj]
  {10.1088/0004-637X/772/2/97}, \href
  {https://ui.adsabs.harvard.edu/abs/2013ApJ...772...97B} {772, 97}

\bibitem[\protect\citeauthoryear{{Bogd{\'a}n} et~al.,}{{Bogd{\'a}n}
  et~al.}{2015}]{bogdan.etal.2015}
{Bogd{\'a}n} {\'A}.,  et~al., 2015, \mn@doi [\apj]
  {10.1088/0004-637X/804/1/72}, \href
  {https://ui.adsabs.harvard.edu/abs/2015ApJ...804...72B} {804, 72}

\bibitem[\protect\citeauthoryear{{Bogd{\'a}n}, {Bourdin}, {Forman}, {Kraft},
  {Vogelsberger}, {Hernquist}  \& {Springel}}{{Bogd{\'a}n}
  et~al.}{2017}]{bogdan.etal.2017}
{Bogd{\'a}n} {\'A}.,  {Bourdin} H.,  {Forman} W.~R.,  {Kraft} R.~P.,
  {Vogelsberger} M.,  {Hernquist} L.,   {Springel} V.,  2017, \mn@doi [\apj]
  {10.3847/1538-4357/aa9523}, \href
  {https://ui.adsabs.harvard.edu/abs/2017ApJ...850...98B} {850, 98}

\bibitem[\protect\citeauthoryear{{Bogdan} et~al.,}{{Bogdan}
  et~al.}{2023}]{bogdan.etal.2023}
{Bogdan} A.,  et~al., 2023, \mn@doi [arXiv e-prints]
  {10.48550/arXiv.2306.05449}, \href
  {https://ui.adsabs.harvard.edu/abs/2023arXiv230605449B} {p. arXiv:2306.05449}

\bibitem[\protect\citeauthoryear{{Borrow}, {Angl{\'e}s-Alc{\'a}zar}  \&
  {Dav{\'e}}}{{Borrow} et~al.}{2020}]{borrow.etal.2020}
{Borrow} J.,  {Angl{\'e}s-Alc{\'a}zar} D.,   {Dav{\'e}} R.,  2020, \mn@doi
  [\mnras] {10.1093/mnras/stz3428}, \href
  {https://ui.adsabs.harvard.edu/abs/2020MNRAS.491.6102B} {491, 6102}

\bibitem[\protect\citeauthoryear{{Bregman}, {Anderson}, {Miller},
  {Hodges-Kluck}, {Dai}, {Li}, {Li}  \& {Qu}}{{Bregman}
  et~al.}{2018}]{bregman.etal.2018}
{Bregman} J.~N.,  {Anderson} M.~E.,  {Miller} M.~J.,  {Hodges-Kluck} E.,  {Dai}
  X.,  {Li} J.-T.,  {Li} Y.,   {Qu} Z.,  2018, \mn@doi [\apj]
  {10.3847/1538-4357/aacafe}, \href
  {https://ui.adsabs.harvard.edu/abs/2018ApJ...862....3B} {862, 3}

\bibitem[\protect\citeauthoryear{{Byrohl} \& {Nelson}}{{Byrohl} \&
  {Nelson}}{2022}]{byrohl.2022}
{Byrohl} C.,  {Nelson} D.,  2022, \mn@doi [arXiv e-prints]
  {10.48550/arXiv.2212.08666}, \href
  {https://ui.adsabs.harvard.edu/abs/2022arXiv221208666B} {p. arXiv:2212.08666}

\bibitem[\protect\citeauthoryear{{Chadayammuri}, {Bogd{\'a}n}, {Oppenheimer},
  {Kraft}, {Forman}  \& {Jones}}{{Chadayammuri}
  et~al.}{2022}]{chadayammuri.etal.2022}
{Chadayammuri} U.,  {Bogd{\'a}n} {\'A}.,  {Oppenheimer} B.~D.,  {Kraft} R.~P.,
  {Forman} W.~R.,   {Jones} C.,  2022, \mn@doi [\apjl]
  {10.3847/2041-8213/ac8936}, \href
  {https://ui.adsabs.harvard.edu/abs/2022ApJ...936L..15C} {936, L15}

\bibitem[\protect\citeauthoryear{{Chakraborty}, {Ferland}, {Chatzikos},
  {Guzm{\'a}n}  \& {Su}}{{Chakraborty} et~al.}{2020}]{chakraborty.etal.2020}
{Chakraborty} P.,  {Ferland} G.~J.,  {Chatzikos} M.,  {Guzm{\'a}n} F.,   {Su}
  Y.,  2020, \mn@doi [\apj] {10.3847/1538-4357/abaaac}, \href
  {https://ui.adsabs.harvard.edu/abs/2020ApJ...901...69C} {901, 69}

\bibitem[\protect\citeauthoryear{{Chakraborty}, {Ferland}, {Chatzikos},
  {Fabian}, {Bianchi}, {Guzm{\'a}n}  \& {Su}}{{Chakraborty}
  et~al.}{2022}]{chakraborty.etal.2022}
{Chakraborty} P.,  {Ferland} G.~J.,  {Chatzikos} M.,  {Fabian} A.~C.,
  {Bianchi} S.,  {Guzm{\'a}n} F.,   {Su} Y.,  2022, \mn@doi [\apj]
  {10.3847/1538-4357/ac7eb9}, \href
  {https://ui.adsabs.harvard.edu/abs/2022ApJ...935...70C} {935, 70}

\bibitem[\protect\citeauthoryear{{Comparat} et~al.,}{{Comparat}
  et~al.}{2022}]{comparat.etal.2022}
{Comparat} J.,  et~al., 2022, arXiv e-prints, \href
  {https://ui.adsabs.harvard.edu/abs/2022arXiv220105169C} {p. arXiv:2201.05169}

\bibitem[\protect\citeauthoryear{{Crain} et~al.,}{{Crain}
  et~al.}{2015}]{crain.etal.2015}
{Crain} R.~A.,  et~al., 2015, \mn@doi [\mnras] {10.1093/mnras/stv725}, \href
  {https://ui.adsabs.harvard.edu/abs/2015MNRAS.450.1937C} {450, 1937}

\bibitem[\protect\citeauthoryear{{Cui} et~al.,}{{Cui}
  et~al.}{2020}]{cui.etal.2011}
{Cui} W.,  et~al., 2020, in {den Herder} J.-W.~A.,  {Nikzad} S.,   {Nakazawa}
  K.,  eds,  Society of Photo-Optical Instrumentation Engineers (SPIE)
  Conference Series Vol. 11444, Space Telescopes and Instrumentation 2020:
  Ultraviolet to Gamma Ray. p. 114442S (\mn@eprint {arXiv} {2101.05587}),
  \mn@doi{10.1117/12.2560871}

\bibitem[\protect\citeauthoryear{{Dav{\'e}}, {Angl{\'e}s-Alc{\'a}zar},
  {Narayanan}, {Li}, {Rafieferantsoa}  \& {Appleby}}{{Dav{\'e}}
  et~al.}{2019}]{dave.etal.2019}
{Dav{\'e}} R.,  {Angl{\'e}s-Alc{\'a}zar} D.,  {Narayanan} D.,  {Li} Q.,
  {Rafieferantsoa} M.~H.,   {Appleby} S.,  2019, \mn@doi [\mnras]
  {10.1093/mnras/stz937}, \href
  {https://ui.adsabs.harvard.edu/abs/2019MNRAS.486.2827D} {486, 2827}

\bibitem[\protect\citeauthoryear{{Davies}, {Crain}, {McCarthy}, {Oppenheimer},
  {Schaye}, {Schaller}  \& {McAlpine}}{{Davies}
  et~al.}{2019}]{davies.etal.2019}
{Davies} J.~J.,  {Crain} R.~A.,  {McCarthy} I.~G.,  {Oppenheimer} B.~D.,
  {Schaye} J.,  {Schaller} M.,   {McAlpine} S.,  2019, \mn@doi [\mnras]
  {10.1093/mnras/stz635}, \href
  {https://ui.adsabs.harvard.edu/abs/2019MNRAS.485.3783D} {485, 3783}

\bibitem[\protect\citeauthoryear{{Davies}, {Crain}, {Oppenheimer}  \&
  {Schaye}}{{Davies} et~al.}{2020}]{davies.etal.2020}
{Davies} J.~J.,  {Crain} R.~A.,  {Oppenheimer} B.~D.,   {Schaye} J.,  2020,
  \mn@doi [\mnras] {10.1093/mnras/stz3201}, \href
  {https://ui.adsabs.harvard.edu/abs/2020MNRAS.491.4462D} {491, 4462}

\bibitem[\protect\citeauthoryear{{Donnari}, {Pillepich}, {Nelson}, {Marinacci},
  {Vogelsberger}  \& {Hernquist}}{{Donnari} et~al.}{2021}]{donnari.etal.2021b}
{Donnari} M.,  {Pillepich} A.,  {Nelson} D.,  {Marinacci} F.,  {Vogelsberger}
  M.,   {Hernquist} L.,  2021, \mn@doi [\mnras] {10.1093/mnras/stab1950}, \href
  {https://ui.adsabs.harvard.edu/abs/2021MNRAS.506.4760D} {506, 4760}

\bibitem[\protect\citeauthoryear{{Furlong} et~al.,}{{Furlong}
  et~al.}{2015}]{furlong.etal.2015}
{Furlong} M.,  et~al., 2015, \mn@doi [\mnras] {10.1093/mnras/stv852}, \href
  {https://ui.adsabs.harvard.edu/abs/2015MNRAS.450.4486F} {450, 4486}

\bibitem[\protect\citeauthoryear{{Gilfanov}, {Syunyaev}  \&
  {Churazov}}{{Gilfanov} et~al.}{1987}]{gilfanov.etal.1987}
{Gilfanov} M.~R.,  {Syunyaev} R.~A.,   {Churazov} E.~M.,  1987, Soviet
  Astronomy Letters, \href
  {https://ui.adsabs.harvard.edu/abs/1987SvAL...13....3G} {13, 3}

\bibitem[\protect\citeauthoryear{{Gnat} \& {Sternberg}}{{Gnat} \&
  {Sternberg}}{2007}]{gnat.sternberg.2007}
{Gnat} O.,  {Sternberg} A.,  2007, \mn@doi [\apjs] {10.1086/509786}, \href
  {https://ui.adsabs.harvard.edu/abs/2007ApJS..168..213G} {168, 213}

\bibitem[\protect\citeauthoryear{{Goulding} et~al.,}{{Goulding}
  et~al.}{2016}]{goulding.etal.2016}
{Goulding} A.~D.,  et~al., 2016, \mn@doi [\apj] {10.3847/0004-637X/826/2/167},
  \href {https://ui.adsabs.harvard.edu/abs/2016ApJ...826..167G} {826, 167}

\bibitem[\protect\citeauthoryear{{Hafen} et~al.,}{{Hafen}
  et~al.}{2022}]{hafen.etal.2022}
{Hafen} Z.,  et~al., 2022, \mn@doi [\mnras] {10.1093/mnras/stac1603}, \href
  {https://ui.adsabs.harvard.edu/abs/2022MNRAS.514.5056H} {514, 5056}

\bibitem[\protect\citeauthoryear{{Hopkins}}{{Hopkins}}{2015}]{hopkins.2015}
{Hopkins} P.~F.,  2015, \mn@doi [\mnras] {10.1093/mnras/stv195}, \href
  {https://ui.adsabs.harvard.edu/abs/2015MNRAS.450...53H} {450, 53}

\bibitem[\protect\citeauthoryear{{Kaastra}, {Paerels}, {Durret}, {Schindler}
  \& {Richter}}{{Kaastra} et~al.}{2008}]{kaastra.etal.2008}
{Kaastra} J.~S.,  {Paerels} F.~B.~S.,  {Durret} F.,  {Schindler} S.,
  {Richter} P.,  2008, \mn@doi [\ssr] {10.1007/s11214-008-9310-y}, \href
  {https://ui.adsabs.harvard.edu/abs/2008SSRv..134..155K} {134, 155}

\bibitem[\protect\citeauthoryear{{Kauffmann}, {White}  \&
  {Guiderdoni}}{{Kauffmann} et~al.}{1993}]{kauffmann.etal.1993}
{Kauffmann} G.,  {White} S.~D.~M.,   {Guiderdoni} B.,  1993, \mn@doi [\mnras]
  {10.1093/mnras/264.1.201}, \href
  {https://ui.adsabs.harvard.edu/abs/1993MNRAS.264..201K} {264, 201}

\bibitem[\protect\citeauthoryear{{Kim} \& {Fabbiano}}{{Kim} \&
  {Fabbiano}}{2015}]{kim.fabbiano.2015}
{Kim} D.-W.,  {Fabbiano} G.,  2015, \mn@doi [\apj]
  {10.1088/0004-637X/812/2/127}, \href
  {https://ui.adsabs.harvard.edu/abs/2015ApJ...812..127K} {812, 127}

\bibitem[\protect\citeauthoryear{{Kraft} et~al.,}{{Kraft}
  et~al.}{2022}]{kraft.etal.2022}
{Kraft} R.,  et~al., 2022, \mn@doi [arXiv e-prints]
  {10.48550/arXiv.2211.09827}, \href
  {https://ui.adsabs.harvard.edu/abs/2022arXiv221109827K} {p. arXiv:2211.09827}

\bibitem[\protect\citeauthoryear{{Lakhchaura} et~al.,}{{Lakhchaura}
  et~al.}{2018}]{lakhchaura.etal.2018}
{Lakhchaura} K.,  et~al., 2018, \mn@doi [\mnras] {10.1093/mnras/sty2565}, \href
  {https://ui.adsabs.harvard.edu/abs/2018MNRAS.481.4472L} {481, 4472}

\bibitem[\protect\citeauthoryear{{Lehmer} et~al.,}{{Lehmer}
  et~al.}{2012}]{lehmer.etal.2012}
{Lehmer} B.~D.,  et~al., 2012, \mn@doi [\apj] {10.1088/0004-637X/752/1/46},
  \href {https://ui.adsabs.harvard.edu/abs/2012ApJ...752...46L} {752, 46}

\bibitem[\protect\citeauthoryear{{Li} \& {Wang}}{{Li} \&
  {Wang}}{2013}]{li.wang.2013}
{Li} J.-T.,  {Wang} Q.~D.,  2013, \mn@doi [\mnras] {10.1093/mnras/sts183},
  \href {https://ui.adsabs.harvard.edu/abs/2013MNRAS.428.2085L} {428, 2085}

\bibitem[\protect\citeauthoryear{{Li}, {Bregman}, {Wang}, {Crain}, {Anderson}
  \& {Zhang}}{{Li} et~al.}{2017}]{li.etal.2017}
{Li} J.-T.,  {Bregman} J.~N.,  {Wang} Q.~D.,  {Crain} R.~A.,  {Anderson} M.~E.,
    {Zhang} S.,  2017, \mn@doi [\apjs] {10.3847/1538-4365/aa96fc}, \href
  {https://ui.adsabs.harvard.edu/abs/2017ApJS..233...20L} {233, 20}

\bibitem[\protect\citeauthoryear{{Liu} et~al.,}{{Liu}
  et~al.}{2019}]{liu.etal.2019}
{Liu} W.,  et~al., 2019, \mn@doi [\apj] {10.3847/1538-4357/aafdfc}, \href
  {https://ui.adsabs.harvard.edu/abs/2019ApJ...872...39L} {872, 39}

\bibitem[\protect\citeauthoryear{{Marinacci} et~al.,}{{Marinacci}
  et~al.}{2018}]{marinacci.etal.2018}
{Marinacci} F.,  et~al., 2018, \mn@doi [\mnras] {10.1093/mnras/sty2206}, \href
  {https://ui.adsabs.harvard.edu/abs/2018MNRAS.480.5113M} {480, 5113}

\bibitem[\protect\citeauthoryear{{McAlpine} et~al.,}{{McAlpine}
  et~al.}{2016}]{Mcalpine.etal.2016}
{McAlpine} S.,  et~al., 2016, \mn@doi [Astronomy and Computing]
  {10.1016/j.ascom.2016.02.004}, \href
  {https://ui.adsabs.harvard.edu/abs/2016A&C....15...72M} {15, 72}

\bibitem[\protect\citeauthoryear{{McCammon} et~al.,}{{McCammon}
  et~al.}{2002}]{McCammon.etal.2002}
{McCammon} D.,  et~al., 2002, \mn@doi [\apj] {10.1086/341727}, \href
  {https://ui.adsabs.harvard.edu/abs/2002ApJ...576..188M} {576, 188}

\bibitem[\protect\citeauthoryear{{Mineo}, {Gilfanov}  \& {Sunyaev}}{{Mineo}
  et~al.}{2012}]{mineo.etal.2012}
{Mineo} S.,  {Gilfanov} M.,   {Sunyaev} R.,  2012, \mn@doi [\mnras]
  {10.1111/j.1365-2966.2012.21831.x}, \href
  {https://ui.adsabs.harvard.edu/abs/2012MNRAS.426.1870M} {426, 1870}

\bibitem[\protect\citeauthoryear{{Naiman} et~al.,}{{Naiman}
  et~al.}{2018}]{naiman.etal.2018}
{Naiman} J.~P.,  et~al., 2018, \mn@doi [\mnras] {10.1093/mnras/sty618}, \href
  {https://ui.adsabs.harvard.edu/abs/2018MNRAS.477.1206N} {477, 1206}

\bibitem[\protect\citeauthoryear{{Nardini}, {Wang}, {Fabbiano}, {Elvis},
  {Pellegrini}, {Risaliti}, {Karovska}  \& {Zezas}}{{Nardini}
  et~al.}{2013}]{nardini.etal.2013}
{Nardini} E.,  {Wang} J.,  {Fabbiano} G.,  {Elvis} M.,  {Pellegrini} S.,
  {Risaliti} G.,  {Karovska} M.,   {Zezas} A.,  2013, \mn@doi [\apj]
  {10.1088/0004-637X/765/2/141}, \href
  {https://ui.adsabs.harvard.edu/abs/2013ApJ...765..141N} {765, 141}

\bibitem[\protect\citeauthoryear{{Nelson} et~al.,}{{Nelson}
  et~al.}{2015}]{nelson15}
{Nelson} D.,  et~al., 2015, \mn@doi [Astronomy and Computing]
  {10.1016/j.ascom.2015.09.003}, \href
  {https://ui.adsabs.harvard.edu/abs/2015A&C....13...12N} {13, 12}

\bibitem[\protect\citeauthoryear{{Nelson} et~al.,}{{Nelson}
  et~al.}{2018a}]{nelson.etal.2018}
{Nelson} D.,  et~al., 2018a, \mn@doi [\mnras] {10.1093/mnras/stx3040}, \href
  {https://ui.adsabs.harvard.edu/abs/2018MNRAS.475..624N} {475, 624}

\bibitem[\protect\citeauthoryear{{Nelson} et~al.,}{{Nelson}
  et~al.}{2018b}]{nelson.etal.2018b}
{Nelson} D.,  et~al., 2018b, \mn@doi [\mnras] {10.1093/mnras/sty656}, \href
  {https://ui.adsabs.harvard.edu/abs/2018MNRAS.477..450N} {477, 450}

\bibitem[\protect\citeauthoryear{{Nelson} et~al.,}{{Nelson}
  et~al.}{2019a}]{nelson.etal.2019a}
{Nelson} D.,  et~al., 2019a, \mn@doi [Computational Astrophysics and Cosmology]
  {10.1186/s40668-019-0028-x}, \href
  {https://ui.adsabs.harvard.edu/abs/2019ComAC...6....2N} {6, 2}

\bibitem[\protect\citeauthoryear{{Nelson} et~al.,}{{Nelson}
  et~al.}{2019b}]{nelson.etal.2019b}
{Nelson} D.,  et~al., 2019b, \mn@doi [\mnras] {10.1093/mnras/stz2306}, \href
  {https://ui.adsabs.harvard.edu/abs/2019MNRAS.490.3234N} {490, 3234}

\bibitem[\protect\citeauthoryear{{Nelson}, {Byrohl}, {Peroux}, {Rubin}  \&
  {Burchett}}{{Nelson} et~al.}{2021}]{nelson.etal.2021}
{Nelson} D.,  {Byrohl} C.,  {Peroux} C.,  {Rubin} K. H.~R.,   {Burchett} J.~N.,
   2021, \mn@doi [\mnras] {10.1093/mnras/stab2177}, \href
  {https://ui.adsabs.harvard.edu/abs/2021MNRAS.507.4445N} {507, 4445}

\bibitem[\protect\citeauthoryear{{Nelson} et~al.,}{{Nelson}
  et~al.}{2023}]{nelson.etal.2023}
{Nelson} D.,  et~al., 2023, \mn@doi [\mnras] {10.1093/mnras/stad1195}, \href
  {https://ui.adsabs.harvard.edu/abs/2023MNRAS.522.3665N} {522, 3665}

\bibitem[\protect\citeauthoryear{{Nica}, {Oppenheimer}, {Crain}, {Bogd{\'a}n},
  {Davies}, {Forman}, {Kraft}  \& {ZuHone}}{{Nica}
  et~al.}{2022}]{nica.etal.2022}
{Nica} A.,  {Oppenheimer} B.~D.,  {Crain} R.~A.,  {Bogd{\'a}n} {\'A}.,
  {Davies} J.~J.,  {Forman} W.~R.,  {Kraft} R.~P.,   {ZuHone} J.~A.,  2022,
  \mn@doi [\mnras] {10.1093/mnras/stac2020}, \href
  {https://ui.adsabs.harvard.edu/abs/2022MNRAS.517.1958N} {517, 1958}

\bibitem[\protect\citeauthoryear{{Oppenheimer} \& {Schaye}}{{Oppenheimer} \&
  {Schaye}}{2013}]{oppenheimer.schaye.2013}
{Oppenheimer} B.~D.,  {Schaye} J.,  2013, \mn@doi [\mnras]
  {10.1093/mnras/stt1043}, \href
  {https://ui.adsabs.harvard.edu/abs/2013MNRAS.434.1043O} {434, 1043}

\bibitem[\protect\citeauthoryear{{Oppenheimer} et~al.,}{{Oppenheimer}
  et~al.}{2020a}]{oppenheimer.etal.2020a}
{Oppenheimer} B.~D.,  et~al., 2020a, \mn@doi [\mnras] {10.1093/mnras/stz3124},
  \href {https://ui.adsabs.harvard.edu/abs/2020MNRAS.491.2939O} {491, 2939}

\bibitem[\protect\citeauthoryear{{Oppenheimer} et~al.,}{{Oppenheimer}
  et~al.}{2020b}]{oppenheimer.etal.2020}
{Oppenheimer} B.~D.,  et~al., 2020b, \mn@doi [\apjl]
  {10.3847/2041-8213/ab846f}, \href
  {https://ui.adsabs.harvard.edu/abs/2020ApJ...893L..24O} {893, L24}

\bibitem[\protect\citeauthoryear{{P{\'e}roux}, {Nelson}, {van de Voort},
  {Pillepich}, {Marinacci}, {Vogelsberger}  \& {Hernquist}}{{P{\'e}roux}
  et~al.}{2020}]{peroux.etal.2020}
{P{\'e}roux} C.,  {Nelson} D.,  {van de Voort} F.,  {Pillepich} A.,
  {Marinacci} F.,  {Vogelsberger} M.,   {Hernquist} L.,  2020, \mn@doi [\mnras]
  {10.1093/mnras/staa2888}, \href
  {https://ui.adsabs.harvard.edu/abs/2020MNRAS.499.2462P} {499, 2462}

\bibitem[\protect\citeauthoryear{{Pillepich} et~al.,}{{Pillepich}
  et~al.}{2018a}]{pillepich.etal.2018}
{Pillepich} A.,  et~al., 2018a, \mn@doi [\mnras] {10.1093/mnras/stx2656}, \href
  {https://ui.adsabs.harvard.edu/abs/2018MNRAS.473.4077P} {473, 4077}

\bibitem[\protect\citeauthoryear{{Pillepich} et~al.,}{{Pillepich}
  et~al.}{2018b}]{pillepich.etal.2018b}
{Pillepich} A.,  et~al., 2018b, \mn@doi [\mnras] {10.1093/mnras/stx3112}, \href
  {https://ui.adsabs.harvard.edu/abs/2018MNRAS.475..648P} {475, 648}

\bibitem[\protect\citeauthoryear{{Pillepich} et~al.,}{{Pillepich}
  et~al.}{2019}]{pillepich.etal.2019}
{Pillepich} A.,  et~al., 2019, \mn@doi [\mnras] {10.1093/mnras/stz2338}, \href
  {https://ui.adsabs.harvard.edu/abs/2019MNRAS.490.3196P} {490, 3196}

\bibitem[\protect\citeauthoryear{{Pillepich}, {Nelson}, {Truong}, {Weinberger},
  {Martin-Navarro}, {Springel}, {Faber}  \& {Hernquist}}{{Pillepich}
  et~al.}{2021}]{pillepich.etal.2021}
{Pillepich} A.,  {Nelson} D.,  {Truong} N.,  {Weinberger} R.,  {Martin-Navarro}
  I.,  {Springel} V.,  {Faber} S.~M.,   {Hernquist} L.,  2021, arXiv e-prints,
  \href {https://ui.adsabs.harvard.edu/abs/2021arXiv210508062P} {p.
  arXiv:2105.08062}

\bibitem[\protect\citeauthoryear{{Pillepich} et~al.,}{{Pillepich}
  et~al.}{2023}]{pillepich.etal.2023}
{Pillepich} A.,  et~al., 2023, \mn@doi [arXiv e-prints]
  {10.48550/arXiv.2303.16217}, \href
  {https://ui.adsabs.harvard.edu/abs/2023arXiv230316217P} {p. arXiv:2303.16217}

\bibitem[\protect\citeauthoryear{{Piotrowska}, {Bluck}, {Maiolino}  \&
  {Peng}}{{Piotrowska} et~al.}{2022}]{piotrowska.etal.2022}
{Piotrowska} J.~M.,  {Bluck} A. F.~L.,  {Maiolino} R.,   {Peng} Y.,  2022,
  \mn@doi [\mnras] {10.1093/mnras/stab3673}, \href
  {https://ui.adsabs.harvard.edu/abs/2022MNRAS.512.1052P} {512, 1052}

\bibitem[\protect\citeauthoryear{{Predehl} et~al.,}{{Predehl}
  et~al.}{2020}]{predehl.etal.2020}
{Predehl} P.,  et~al., 2020, \mn@doi [\nat] {10.1038/s41586-020-2979-0}, \href
  {https://ui.adsabs.harvard.edu/abs/2020Natur.588..227P} {588, 227}

\bibitem[\protect\citeauthoryear{{Ramesh}, {Nelson}  \& {Pillepich}}{{Ramesh}
  et~al.}{2023a}]{ramesh.etal.2023b}
{Ramesh} R.,  {Nelson} D.,   {Pillepich} A.,  2023a, \mn@doi [\mnras]
  {10.1093/mnras/stad951}, \href
  {https://ui.adsabs.harvard.edu/abs/2023MNRAS.tmp..906R} {}

\bibitem[\protect\citeauthoryear{{Ramesh}, {Nelson}  \& {Pillepich}}{{Ramesh}
  et~al.}{2023b}]{ramesh.etal.2023a}
{Ramesh} R.,  {Nelson} D.,   {Pillepich} A.,  2023b, \mn@doi [\mnras]
  {10.1093/mnras/stac3524}, \href
  {https://ui.adsabs.harvard.edu/abs/2023MNRAS.518.5754R} {518, 5754}

\bibitem[\protect\citeauthoryear{{Sarkar}, {Sternberg}  \& {Gnat}}{{Sarkar}
  et~al.}{2022}]{sarkar.etal.2022}
{Sarkar} K.~C.,  {Sternberg} A.,   {Gnat} O.,  2022, \mn@doi [\apj]
  {10.3847/1538-4357/ac9835}, \href
  {https://ui.adsabs.harvard.edu/abs/2022ApJ...940...44S} {940, 44}

\bibitem[\protect\citeauthoryear{{Schaller}, {Dalla Vecchia}, {Schaye},
  {Bower}, {Theuns}, {Crain}, {Furlong}  \& {McCarthy}}{{Schaller}
  et~al.}{2015}]{schaller.etal.2015}
{Schaller} M.,  {Dalla Vecchia} C.,  {Schaye} J.,  {Bower} R.~G.,  {Theuns} T.,
   {Crain} R.~A.,  {Furlong} M.,   {McCarthy} I.~G.,  2015, \mn@doi [\mnras]
  {10.1093/mnras/stv2169}, \href
  {https://ui.adsabs.harvard.edu/abs/2015MNRAS.454.2277S} {454, 2277}

\bibitem[\protect\citeauthoryear{{Schaye} et~al.,}{{Schaye}
  et~al.}{2015}]{schaye.etal.2015}
{Schaye} J.,  et~al., 2015, \mn@doi [\mnras] {10.1093/mnras/stu2058}, \href
  {https://ui.adsabs.harvard.edu/abs/2015MNRAS.446..521S} {446, 521}

\bibitem[\protect\citeauthoryear{{Schellenberger} et~al.,}{{Schellenberger}
  et~al.}{2023}]{schellenberger.etal.2023}
{Schellenberger} G.,  et~al., 2023, \mn@doi [arXiv e-prints]
  {10.48550/arXiv.2307.01259}, \href
  {https://ui.adsabs.harvard.edu/abs/2023arXiv230701259S} {p. arXiv:2307.01259}

\bibitem[\protect\citeauthoryear{{Smith}, {Brickhouse}, {Liedahl}  \&
  {Raymond}}{{Smith} et~al.}{2001}]{smith.etal.2001}
{Smith} R.~K.,  {Brickhouse} N.~S.,  {Liedahl} D.~A.,   {Raymond} J.~C.,  2001,
  \mn@doi [\apjl] {10.1086/322992}, \href
  {http://adsabs.harvard.edu/abs/2001ApJ...556L..91S} {556, L91}

\bibitem[\protect\citeauthoryear{{Sorini}, {Dav{\'e}}, {Cui}  \&
  {Appleby}}{{Sorini} et~al.}{2022}]{sorini.2022}
{Sorini} D.,  {Dav{\'e}} R.,  {Cui} W.,   {Appleby} S.,  2022, \mn@doi [\mnras]
  {10.1093/mnras/stac2214}, \href
  {https://ui.adsabs.harvard.edu/abs/2022MNRAS.516..883S} {516, 883}

\bibitem[\protect\citeauthoryear{{Springel}}{{Springel}}{2005}]{springel.2005}
{Springel} V.,  2005, \mn@doi [\mnras] {10.1111/j.1365-2966.2005.09655.x},
  \href {https://ui.adsabs.harvard.edu/abs/2005MNRAS.364.1105S} {364, 1105}

\bibitem[\protect\citeauthoryear{{Springel}}{{Springel}}{2010}]{springel.2010}
{Springel} V.,  2010, \mn@doi [\mnras] {10.1111/j.1365-2966.2009.15715.x},
  \href {http://adsabs.harvard.edu/abs/2010MNRAS.401..791S} {401, 791}

\bibitem[\protect\citeauthoryear{{Springel} et~al.,}{{Springel}
  et~al.}{2018}]{springel.etal.2018}
{Springel} V.,  et~al., 2018, \mn@doi [\mnras] {10.1093/mnras/stx3304}, \href
  {https://ui.adsabs.harvard.edu/abs/2018MNRAS.475..676S} {475, 676}

\bibitem[\protect\citeauthoryear{{Strickland}, {Heckman}, {Colbert}, {Hoopes}
  \& {Weaver}}{{Strickland} et~al.}{2004}]{strickland.etal.2004}
{Strickland} D.~K.,  {Heckman} T.~M.,  {Colbert} E. J.~M.,  {Hoopes} C.~G.,
  {Weaver} K.~A.,  2004, \mn@doi [\apjs] {10.1086/382214}, \href
  {https://ui.adsabs.harvard.edu/abs/2004ApJS..151..193S} {151, 193}

\bibitem[\protect\citeauthoryear{{Terrazas} et~al.,}{{Terrazas}
  et~al.}{2020}]{terrazas.etal.2020}
{Terrazas} B.~A.,  et~al., 2020, \mn@doi [\mnras] {10.1093/mnras/staa374},
  \href {https://ui.adsabs.harvard.edu/abs/2020MNRAS.493.1888T} {493, 1888}

\bibitem[\protect\citeauthoryear{{Truong} et~al.,}{{Truong}
  et~al.}{2020}]{truong.etal.2020}
{Truong} N.,  et~al., 2020, \mn@doi [\mnras] {10.1093/mnras/staa685}, \href
  {https://ui.adsabs.harvard.edu/abs/2020MNRAS.494..549T} {494, 549}

\bibitem[\protect\citeauthoryear{{Truong}, {Pillepich}  \& {Werner}}{{Truong}
  et~al.}{2021a}]{truong.etal.2021}
{Truong} N.,  {Pillepich} A.,   {Werner} N.,  2021a, \mn@doi [\mnras]
  {10.1093/mnras/staa3880}, \href
  {https://ui.adsabs.harvard.edu/abs/2021MNRAS.501.2210T} {501, 2210}

\bibitem[\protect\citeauthoryear{{Truong}, {Pillepich}, {Nelson}, {Werner}  \&
  {Hernquist}}{{Truong} et~al.}{2021b}]{truong.etal.2021b}
{Truong} N.,  {Pillepich} A.,  {Nelson} D.,  {Werner} N.,   {Hernquist} L.,
  2021b, \mn@doi [\mnras] {10.1093/mnras/stab2638}, \href
  {https://ui.adsabs.harvard.edu/abs/2021MNRAS.508.1563T} {508, 1563}

\bibitem[\protect\citeauthoryear{{T{\"u}llmann}, {Pietsch}, {Rossa},
  {Breitschwerdt}  \& {Dettmar}}{{T{\"u}llmann}
  et~al.}{2006}]{tullmann.etal.2006}
{T{\"u}llmann} R.,  {Pietsch} W.,  {Rossa} J.,  {Breitschwerdt} D.,   {Dettmar}
  R.~J.,  2006, \mn@doi [\aap] {10.1051/0004-6361:20052936}, \href
  {https://ui.adsabs.harvard.edu/abs/2006A&A...448...43T} {448, 43}

\bibitem[\protect\citeauthoryear{{Tumlinson}, {Peeples}  \& {Werk}}{{Tumlinson}
  et~al.}{2017}]{tumlinson.etal.2017}
{Tumlinson} J.,  {Peeples} M.~S.,   {Werk} J.~K.,  2017, \mn@doi [\araa]
  {10.1146/annurev-astro-091916-055240}, \href
  {https://ui.adsabs.harvard.edu/abs/2017ARA&A..55..389T} {55, 389}

\bibitem[\protect\citeauthoryear{{Veilleux}, {Teng}, {Rupke}, {Maiolino}  \&
  {Sturm}}{{Veilleux} et~al.}{2014}]{veilleux.etal.2014}
{Veilleux} S.,  {Teng} S.~H.,  {Rupke} D.~S.~N.,  {Maiolino} R.,   {Sturm} E.,
  2014, \mn@doi [\apj] {10.1088/0004-637X/790/2/116}, \href
  {https://ui.adsabs.harvard.edu/abs/2014ApJ...790..116V} {790, 116}

\bibitem[\protect\citeauthoryear{{Veilleux}, {Maiolino}, {Bolatto}  \&
  {Aalto}}{{Veilleux} et~al.}{2020}]{veilleux.etal.2020}
{Veilleux} S.,  {Maiolino} R.,  {Bolatto} A.~D.,   {Aalto} S.,  2020, \mn@doi
  [\aapr] {10.1007/s00159-019-0121-9}, \href
  {https://ui.adsabs.harvard.edu/abs/2020A&ARv..28....2V} {28, 2}

\bibitem[\protect\citeauthoryear{{Weinberger} et~al.,}{{Weinberger}
  et~al.}{2017}]{weinberger.etal.2017}
{Weinberger} R.,  et~al., 2017, \mn@doi [\mnras] {10.1093/mnras/stw2944}, \href
  {https://ui.adsabs.harvard.edu/abs/2017MNRAS.465.3291W} {465, 3291}

\bibitem[\protect\citeauthoryear{{Weinberger} et~al.,}{{Weinberger}
  et~al.}{2018}]{weinberger.etal.2018}
{Weinberger} R.,  et~al., 2018, \mn@doi [\mnras] {10.1093/mnras/sty1733}, \href
  {https://ui.adsabs.harvard.edu/abs/2018MNRAS.479.4056W} {479, 4056}

\bibitem[\protect\citeauthoryear{{Werner}, {McNamara}, {Churazov}  \&
  {Scannapieco}}{{Werner} et~al.}{2019}]{werner.etal.2019}
{Werner} N.,  {McNamara} B.~R.,  {Churazov} E.,   {Scannapieco} E.,  2019,
  \mn@doi [\ssr] {10.1007/s11214-018-0571-9}, \href
  {https://ui.adsabs.harvard.edu/abs/2019SSRv..215....5W} {215, 5}

\bibitem[\protect\citeauthoryear{{White} \& {Frenk}}{{White} \&
  {Frenk}}{1991}]{white.frenk.1991}
{White} S. D.~M.,  {Frenk} C.~S.,  1991, \mn@doi [\apj] {10.1086/170483}, \href
  {https://ui.adsabs.harvard.edu/abs/1991ApJ...379...52W} {379, 52}

\bibitem[\protect\citeauthoryear{{White} \& {Rees}}{{White} \&
  {Rees}}{1978}]{white.rees.1979}
{White} S.~D.~M.,  {Rees} M.~J.,  1978, \mn@doi [\mnras]
  {10.1093/mnras/183.3.341}, \href
  {https://ui.adsabs.harvard.edu/abs/1978MNRAS.183..341W} {183, 341}

\bibitem[\protect\citeauthoryear{{Wijers} \& {Schaye}}{{Wijers} \&
  {Schaye}}{2022}]{wijers.schaye.2022}
{Wijers} N.~A.,  {Schaye} J.,  2022, \mn@doi [\mnras] {10.1093/mnras/stac1580},
  \href {https://ui.adsabs.harvard.edu/abs/2022MNRAS.514.5214W} {514, 5214}

\bibitem[\protect\citeauthoryear{{XRISM Science Team}}{{XRISM Science
  Team}}{2020}]{xrism.whitepaper.2020}
{XRISM Science Team} 2020, arXiv e-prints, \href
  {https://ui.adsabs.harvard.edu/abs/2020arXiv200304962X} {p. arXiv:2003.04962}

\bibitem[\protect\citeauthoryear{{Yamasaki}, {Sato}, {Mitsuishi}  \&
  {Ohashi}}{{Yamasaki} et~al.}{2009}]{yamasaki.etal.2009}
{Yamasaki} N.~Y.,  {Sato} K.,  {Mitsuishi} I.,   {Ohashi} T.,  2009, \mn@doi
  [\pasj] {10.1093/pasj/61.sp1.S291}, \href
  {https://ui.adsabs.harvard.edu/abs/2009PASJ...61S.291Y} {61, S291}

\bibitem[\protect\citeauthoryear{{Zhuravleva} et~al.,}{{Zhuravleva}
  et~al.}{2013}]{zhuravleva.etal.2013}
{Zhuravleva} I.,  et~al., 2013, \mn@doi [\mnras] {10.1093/mnras/stt1506}, \href
  {https://ui.adsabs.harvard.edu/abs/2013MNRAS.435.3111Z} {435, 3111}

\bibitem[\protect\citeauthoryear{{Zinger} et~al.,}{{Zinger}
  et~al.}{2020}]{zinger.etal.2020}
{Zinger} E.,  et~al., 2020, \mn@doi [\mnras] {10.1093/mnras/staa2607}, \href
  {https://ui.adsabs.harvard.edu/abs/2020MNRAS.499..768Z} {499, 768}

\bibitem[\protect\citeauthoryear{{ZuHone} et~al.,}{{ZuHone}
  et~al.}{2023}]{zuhone.etal.2023}
{ZuHone} J.~A.,  et~al., 2023, \mn@doi [arXiv e-prints]
  {10.48550/arXiv.2307.01269}, \href
  {https://ui.adsabs.harvard.edu/abs/2023arXiv230701269Z} {p. arXiv:2307.01269}

\bibitem[\protect\citeauthoryear{{van de Voort} \& {Schaye}}{{van de Voort} \&
  {Schaye}}{2013}]{vandeVoort.schaye.2013}
{van de Voort} F.,  {Schaye} J.,  2013, \mn@doi [\mnras]
  {10.1093/mnras/stt115}, \href
  {https://ui.adsabs.harvard.edu/abs/2013MNRAS.430.2688V} {430, 2688}

\makeatother
\end{thebibliography}

%\appendix

%\section{CGM properties in star-forming vs quiescent galaxies}
%\label{appendix_sf_vs_q}

%\begin{figure*}
%  \centering
%  \includegraphics[width=0.9\textwidth]{figures/Fig_Profile_Dens_Profiles_Sf_vs_Quiescent.pdf}
%  \includegraphics[width=0.9\textwidth]{figures/Fig_Profile_Tmw_Profiles_Sf_vs_Quiescent.pdf}
%  \includegraphics[width=0.9\textwidth]{figures/Fig_Profile_Zmw_Profiles_Sf_vs_Quiescent.pdf}
%  \caption{CGM properties in star-forming verus quiescent galaxies: gas column density ({\it top}), gas mass-weighted temperature ({\it middle}), gas mass-weighted metallicity ({\it bottom}).}
%  \label{fig:cgm_sf_q}
%\end{figure*}

%In Fig.~\ref{fig:cgm_sf_q} we show the physical properties of CGM gas that are responsible for the dichotomy in X-ray emission between star-forming and quiescent galaxies (Section \ref{lx_dichotomy}). From top to bottom: radial profiles of gas mass surface density, mass-weighted temperature, and mass-weighted metallicity. In all cases we restrict our galaxy sample to the same MW-mass selection as previously. The three columns show the three simulations separately: TNG100 (left), EAGLE (center), and SIMBA (right).

%In all panels, the star-forming subset is shown with the blue lines and shaded bands, while the quiescent subset is shown in red. Star-forming galaxies, across all three simulations, have higher gas densities, lower temperatures, and higher metallicities throughout their gaseous halos. As discussed in the main text, these differences collectively result in observable differences in X-ray line emission.

\label{lastpage}
\end{document}